 \documentclass[sigconf,10pt]{acmart}
\usepackage[english]{babel}
\renewcommand\footnotetextcopyrightpermission[1]{}

\usepackage{times,color,graphicx,amsmath,array,gensymb,subcaption}
\usepackage{xspace}
\newcommand{\para}[1]{\medskip\noindent\textbf{#1}}

\usepackage{balance, ifthen, multirow}

\usepackage[small,compact]{titlesec}              

\usepackage[font=small,labelfont=bf,textfont=bf,aboveskip=8pt]{caption}
\usepackage{fancyhdr}

\usepackage{enumitem}
\setlist{itemsep=0pt,parsep=0pt,topsep=0pt}

\usepackage{booktabs}
\usepackage{colortbl}
\usepackage{float}                           

\definecolor{placeholderbg}{rgb}{0.85,0.85,0.85}

\usepackage{mathtools}
\usepackage{pifont}

\newenvironment{parafont}{\fontfamily{ptm}\selectfont}{}

\usepackage{url}
\usepackage{cleveref}
\crefname{section}{\S}{\SS}
\crefformat{section}{\S#2#1#3} 
\crefformat{subsection}{\S#2#1#3}
\crefformat{subsubsection}{\S#2#1#3}

\usepackage{listings}
\newcommand\code[1]{\lstinline$#1$}

\lstdefinelanguage{paper}{
 keywords={partition, transform, gather, scatter, apply},
 keywordstyle=\color{blue}\bfseries,
 morekeywords={[2]degrees,branch,commit,v_prev},
 keywordstyle={[2]\color{red}\bfseries},
 morekeywords={[3]if,def,Class,return,else,None,False,True,Array,while,G},
 keywordstyle={[3]\bfseries},
 basicstyle=\small\ttfamily,
 identifierstyle=\color{black},
 sensitive=false,
 comment=[l]{\/\/},
 morecomment=[s]{/*}{*/},
 commentstyle=\color{green}\ttfamily,
 stringstyle=\color{red}\ttfamily,
 breaklines=true,
}
\newcommand{\tightcaption}[1]{\vspace{-7pt}\caption{{\bf \small #1}}
\vspace{-5pt}
}
\lstnewenvironment{python}
{\lstset{
    language=paper,
    frame=tlbr,
    columns=fullflexible,
    postbreak=\mbox{\textcolor{red}{$\hookrightarrow$}\space},
    captionpos=b}}
{}

\usepackage{algorithm}
\usepackage{algpseudocode}

\usepackage{tikz}

\usepackage{epsfig,comment}

\newcommand{\squishlist}{
   \begin{list}{$\bullet$}
    { \setlength{\itemsep}{0pt}      \setlength{\parsep}{3pt}
      \setlength{\topsep}{3pt}       \setlength{\partopsep}{0pt}
      \setlength{\leftmargin}{1.0em} \setlength{\labelwidth}{1em}
      \setlength{\labelsep}{0.5em} } }
\newcommand{\squishend}{
    \end{list}  }

  \newcommand{\mr}[1] {{\textcolor{blue}{MR: {#1}}}}

\usepackage{comment}
\usepackage{soul}
\soulregister\cite7
\soulregister\ref7
\soulregister\pageref7

\usepackage[all]{nowidow}

\begin{document}


\title{\LARGE MadEye: Boosting Live Video Analytics Accuracy with Adaptive Camera Configurations}

\newcommand{\aut}[2]{#1\texorpdfstring{$^{#2}$}{(#2)}}
\author{\aut{Mike Wong}{1}, \aut{Murali Ramanujam}{1}, \aut{Guha Balakrishnan}{2}, \aut{Ravi Netravali}{1}}

\affiliation{\aut{}{1}\textit{Princeton University}\quad
             \aut{}{2}\textit{Rice University}
             \country{}
             }







\begin{abstract}
\vspace{-2pt}
Camera orientations (i.e., rotation and zoom) govern the content that a camera captures in a given scene, which in turn heavily influences the accuracy of live video analytics pipelines. However, existing analytics approaches leave this crucial adaptation knob untouched, instead opting to only alter the way that captured images from fixed orientations are encoded, streamed, and analyzed. We present MadEye, a camera-server system that automatically and continually adapts orientations to maximize accuracy for the workload and resource constraints at hand. To realize this using commodity pan-tilt-zoom (PTZ) cameras, MadEye embeds (1) a search algorithm that rapidly explores the massive space of orientations to identify a fruitful subset at each time, and (2) a novel knowledge distillation strategy to efficiently (with only camera resources) select the ones that maximize workload accuracy. Experiments on diverse workloads show that MadEye boosts accuracy by 2.9-25.7\% for the same resource usage, or achieves the same accuracy with 2-3.7$\times$ lower resource costs.

\vspace{-6pt}





\end{abstract}


\maketitle 
\makeatletter
\def\blfootnote{\xdef\@thefnmark{}\@footnotetext}
\makeatother

\section{Introduction}
\label{s:intro}

Building on the steady growth in camera deployments and advances in deep neural networks (DNNs) for vision tasks (e.g., classification or detection)~\cite{cnn-face-cvpr15, pyramid-network-cvpr17, pedestrian-detection-iccv15, maskrcnn,pytorchyolov3}, live video analytics pipelines have become prevalent. These pipelines operate by continually streaming live video feeds from cameras to processing servers (either edge~\cite{gemel,ekya,VideoKillerApp2,edgenative,AzureStackEdge} or cloud~\cite{videostorm-nsdi17,reducto,dds,chameleon-sigcomm18}), where DNNs are run on incoming frames to produce low latency and highly accurate results for different application queries, i.e., combinations of task, DNN, and object(s) of interest. Key use cases include autonomous driving, footfall tracking, traffic coordination, business analytics, among others~\cite{are-we-ready-for-ai-powered-security-cameras,powering-the-edge-with-ai-in-an-iot-world,smart-mall,traff2,fortune_va_market_report, analyzing-social-distancing, traffic-analysis,sports_va,sports, soccer_tracking}.

Given their practical importance, much research has been devoted to improving both the resource efficiency and accuracy of live video analytics pipelines. Existing solutions include accuracy-aware tuning of inference configuration, encoding, or appearance knobs~\cite{chameleon,videostorm-nsdi17,accmpeg,camtuner}, filtering out redundant content~\cite{dds,reducto,glimpse-sensys15,spatula}, using cheaper model variants~\cite{InFaaSATC21,ekya}, improving job scheduling~\cite{videostorm-nsdi17,nexus,gemel}, and so on. However, all of these works assume that the content observable by cameras is unchangeable, and instead can only be encoded, streamed, or analyzed differently. In essence, they focus on optimizing \emph{fixed, preset} camera deployments.

Unfortunately, the deployment of cameras for analytics is itself a daunting task for operators. Subject to practical constraints (e.g., mounts, power sources), for a scene of interest, operators must determine the number of cameras to deploy and the orientation (i.e., combination of rotation and zoom factor) to use for each. There exist many possible orientations, and altering these decisions requires manual intervention. Yet we find that doing so can be highly fruitful: across different workloads and scenes, dynamically adapting orientations over time can yield accuracy improvements of 21.3-35.3\% (without inflating resource usage) compared to even the \emph{best} fixed-orientation scheme. Further, these wins cannot be reaped by simply deploying more fixed cameras to simultaneously cover more orientations: most orientations are `best' for short total periods of time (median of 6 sec for each 10-min video), drastically hindering the efficiency of such an approach, especially in the resource-constrained settings where video analytics are run~\cite{ekya,shi2016edge,AzureStackEdge,spider}.

An alternative strategy is to leverage PTZ (pan-tilt-zoom) cameras that offer software libraries for tuning orientations, thereby providing a logical approach to capturing the above wins. Indeed, despite existing for nearly two decades, PTZ camera popularity has surged in recent years (global market value of \$3 billion in 2020~\cite{global_ptz_report}) largely due to declining price points that now rival fixed-camera costs~\cite{ptz_camera_cost, cheap_ptz_comparison}. However, multiple challenges complicate their use for live analytics (\S\ref{ss:challenges}). First, queries are highly sensitive, in different ways, to orientation knobs due to their diverse goals (e.g., tasks), inherent model biases (how models perceive scenes and objects), and scene dynamism (where objects are located) -- optimizing orientation tuning for one workload can forego up to 25.1\% of the potential median accuracy wins for another. Second, the `grid' of orientations is large, but the selection space is sparse, with steep accuracy drops from the best orientation(s) to other at any time. Third, the best orientation changes rapidly, e.g., 85\% of changes occur in $\leq$1 sec since the last change.





To overcome these issues, we present \textbf{MadEye}, a camera-server system that automatically and continually adapts PTZ camera orientations to maximize analytics accuracy for the scene and workload at hand. The key insight behind MadEye is that the speed at which commodity PTZ cameras can change orientations (i.e., upwards of 600\degree~per sec with near-instantaneous digital zoom) far outpaces the rate at which applications require analytics results (typically 1-30 frames per second (fps), i.e., every 33-1000 ms). This, in turn, allows MadEye to eschew typical non-stationary multi-armed bandit strategies~\cite{mab-cournot-games, mab-link-adaption, mab-variation-budget} that rely purely on previous explorations to determine orientation importance, in favor of a more informed strategy based on \emph{current} scene content. Concretely, in each timestep (33 ms for 30 fps) and subject to network/compute resource availability, MadEye cameras explore multiple orientations and quickly determine which will maximize workload accuracy and warrant transmission to the backend for full inference. However, realizing this strategy in practice involves addressing several technical challenges.

First, to enable fast camera-side evaluation of the importance of different orientations, MadEye adopts a custom knowledge distillation~\cite{hinton2015distilling} strategy with edge-grade, ultra-compressed NN models. To cope with their potentially limited predictive power, we task them with modeling query sensitivities only to the point of accurately \emph{ranking} orientations in terms of impact on workload accuracy -- precise results are left to backend servers. Even with this relaxed framing, MadEye must employ several optimizations to achieve sufficient rank accuracy. Most notably, MadEye trains edge models using a common abstraction -- detection for objects of interest -- that reflects the minimum information needed to capture sensitivities and biases for popular tasks. Task-specific semantics need not be baked into edge models, and instead can be incorporated by post processing the generated results.

Edge models are \emph{continually} trained on MadEye’s backend using both the latest and historical workload results, with the goal of mitigating data skew towards recently-selected orientations  (given uncertainties in what will be selected next). Importantly, to balance resource costs and accuracy, each edge model covers only a single query but all orientations. The intuition is that, while model results can exhibit substantial divergence~\cite{boggart,balakrishnan2021towards,du2020fairness,khosla2012undoing}, feature-level variance between orientations for the same scene is considerably narrower, often smaller than that in typical pre-training datasets~\cite{ms-coco-eccv14}. Accordingly, MadEye freezes pre-trained feature extraction layers across queries, caching those weights on cameras, thereby lowering retraining and (downlink) model update overheads.


Second, we develop a novel, on-camera search strategy to explore orientations with the goal of capturing the best one (accuracy-wise) at each timestep. Three key empirical observations guide our search: (1) despite rapid temporal shifts, transitions between best orientations move slowly in the spatial dimension, (2) the best orientations are typically spatially clustered, and (3) neighboring orientations (with overlapping regions) exhibit highly correlated trends in efficacy.

Building on these observations, MadEye explores a flexible shape of contiguous orientations at each timestep, and considers shifting only towards neighboring orientations whose efficacy can be robustly predicted. Decisions to keep/remove orientations are governed by both response rates (and the corresponding time budgets) and \emph{relative} comparisons of recent edge model results. For the former, MadEye uses an efficient heuristic to determine path feasibility in the time budget (a variant of the NP-Hard Traveling Salesman Problem~\cite{held1970traveling}). For the latter, MadEye gracefully trades off exploration (i.e., shape size) for network usage (i.e., sending more orientations for backend inference) to bound the effects of edge model errors and maximize accuracy for the required response rate.



To evaluate MadEye, we developed the first (to our knowledge) dataset that supports tuning rotation and zoom at each time instant by splicing out scenes of interest from publicly available 360\degree~videos. Using this dataset, we evaluated MadEye on a variety of network conditions and workloads that incorporate multiple vision DNNs and query tasks: classification, counting (per-frame and aggregate), and detection. Across these settings, MadEye boosts accuracy by 2.9-25.7\% compared to an oracle fixed-orientation strategy without inflating resource usage; these wins are within 1.8-13.9\% of the oracle dynamic strategy. Framed differently, MadEye achieves those accuracy boosts with 2-3.7$\times$ lower resource footprints than the best strategy of using (multiple) fixed-orientation cameras. Moreover, MadEye outperforms recent PTZ tracking algorithms~\cite{panoptes,auto-tracking} (by 2.0-3.8$\times$) and multi-armed bandit solutions~\cite{mab_for_wireless_selection} (by 5.8$\times$). We will release MadEye and our datasets. This work does not raise any ethical issues.

\section{Background and Motivation}
\label{s:motivation}

We start with an overview of live video analytics deployments (\S\ref{ss:background}). We then show measurements highlighting the importance of dynamically adapting camera orientations to workloads and scenes (\S\ref{ss:potential}), and the challenges associated with realizing those benefits in practice (\S\ref{ss:challenges}).


\subsection{Overview of Live Video Analytics}
\label{ss:background}

In a live video analytics deployment, one or more cameras continually stream their video frames to servers for processing. Servers can range from distant (but powerful) cloud machines~\cite{videostorm-nsdi17,nexus} to nearby (but weaker) edge boxes~\cite{gemel,ekya,AzureStackEdge}, and are tasked with running queries on the incoming frames to support different applications. Queries most often involve running deep neural network (DNN) inference on individual frames, with the goals of locating and characterizing various objects in the scene, e.g., an intersection. Moreover, the queries for different applications can vary in terms of the tasks they perform, the objects they consider, the DNNs they use (different architectures and weights), and the response rates they require. For instance, footfall tracking for business analytics will count people passing through an area, with response rates at 1 fps or less~\cite{smart-mall}. In contrast, smart driving or sports analytics applications will detect the specific locations of cars or people, with response rates upwards of 30 fps~\cite{sports_va}. 

In this paper, we focus on the following four query tasks (and their corresponding accuracy metrics) that have been prevalent in recent literate~\cite{blazeit,noscope-vldb17,privid,reducto,dds} and real-world deployments~\cite{gemel,optasia-socc16}. We note that these query types also serve as the building blocks for complex applications and other tasks, e.g., tracking queries rely on object detections.
\squishlist
\item \textbf{Binary classification}: asks if any objects of interest are present in a frame. Accuracy across the video is measured as the fraction of frames with the correct binary decision.

\item \textbf{Counting}: counts the number of objects of interest in each frame. Accuracy for each frame is measured as the percent difference between the returned and ground truth counts.

\item \textbf{Detection}: finds the precise bounding box coordinates for objects of interest in a frame. Accuracy per frame is measured using mAP~\cite{Everingham:2010:PVO:1747084.1747104}, which evaluates the overlap between each returned box and its ground truth counterpart.

\item \textbf{Aggregate counting}: counts the \emph{unique} objects of interest that appear in a scene. Accuracy per video is the percent difference between the returned and ground truth counts.

\squishend

Over time, an analytics deployment will face diverse workloads to run on the video feeds it manages, each varying in query composition and size~\cite{gemel,boggart}. Yet, the overarching goals persist: subject to resource constraints, deliver low-latency results (at the desired response rate) with maximal accuracy.

\subsection{Opportunities with Tuning Camera Orientations}
\label{ss:potential}

Existing optimizations for video analytics (\S\ref{s:related}) assume that a stationary camera's orientation (rotation and zoom), and thus what it ingests from the target scene, is fixed and incapable of being adapted. To quantify the significance of this restriction, we run experiments on our 50-video dataset and workloads that incorporate 4 model architectures, the 4 tasks from \S\ref{ss:background}, and people/cars; \S\ref{ss:methodology} details both. Each video supports tuning of rotations (150\degree~horizontally by 30\degree, 75\degree~vertically by 15\degree) and zoom (1-3$\times$); we consider other granularities in \S\ref{ss:deepdive}.


For each video, we obtained per-frame (15 fps here) results for each workload by running its queries on all 75 orientations. We then define accuracy relative to the \emph{best} orientation for each frame, i.e., the orientation that maximized per-frame 
accuracy for the workload. For instance, for counting, an orientation's accuracy at any time is its object of interest count divided by the max count across all orientations at that time. Using this methodology, we compare three schemes: (1) \emph{one time fixed} which selects the best orientation at time=0 and keeps it throughout the video, (2) \emph{best fixed} which uses oracle knowledge to pick the best single orientation that maximizes average workload accuracy for the video, and (3) \emph{best dynamic} which selects the best orientation per frame in the video.


\begin{figure}[t]
    \centering
    \includegraphics[width=0.8\columnwidth]{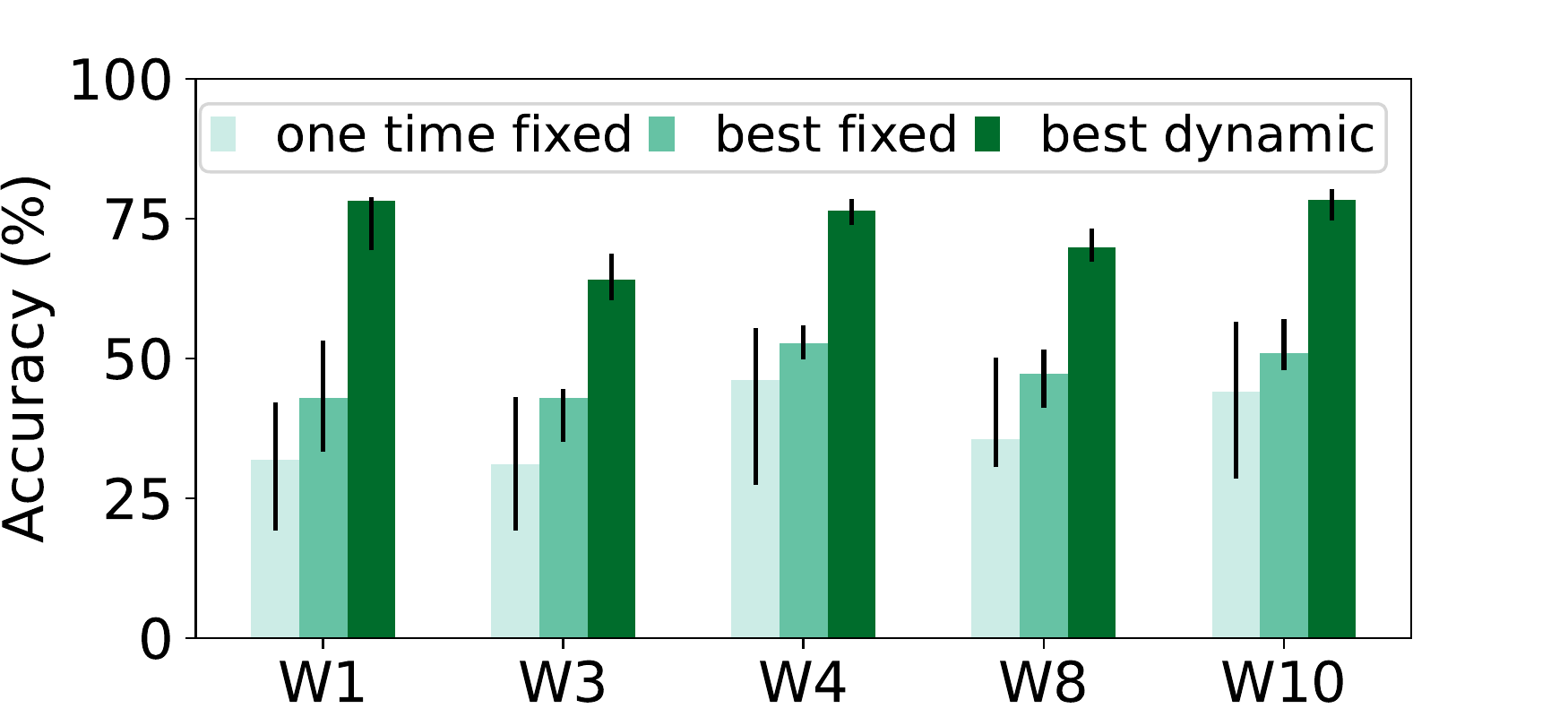}
    \tightcaption{Accuracy for 5 representative workloads when using varying degrees of orientation adaptation. Bars list results for the median video, with error bars spanning 25-75th percentiles.} 
    \vspace{3pt}
        \label{fig:workload_opportunity}
\end{figure}

\begin{figure}[t]
    \centering
    \includegraphics[width=0.8\columnwidth]{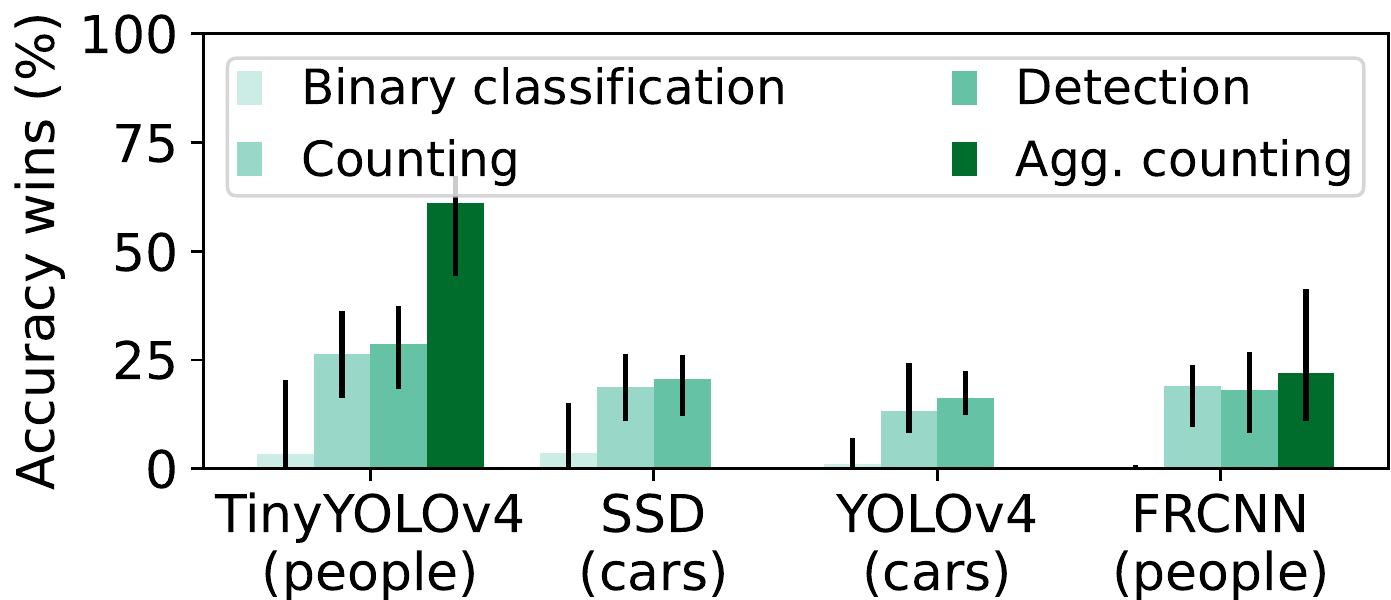}
\tightcaption{Accuracy wins from adapting orientations (compared to \emph{best fixed}) grow as query specificity grows. Bars list median videos, with error bars for 25-75th percentiles. We exclude agg. counting+cars due to limits of multi-object trackers (\S\ref{ss:methodology}).}
\vspace{6pt}
\label{fig:query_opportunity}
\end{figure}


As shown in Figure~\ref{fig:workload_opportunity}, adapting camera orientations brings substantial accuracy improvements without inflating resource usage, i.e., the same number of frames are transmitted and processed: median boosts with \emph{best dynamic} are 30.4-46.3\% over \emph{one time fixed} and 21.3-35.3\% over the \emph{best fixed} scheme that is an upper bound for any fixed-orientation approach. Figure~\ref{fig:query_opportunity} breaks down these results by query task. Notably, the importance of adapting orientations grows as query types become more specific. For instance, for YOLOv4 and cars, median accuracy improvements over \emph{best fixed} are 1.2\%, 13.4\%, and 16.4\% for binary classification, counting, and detection, respectively. The reason is that coarser queries mask certain differences across orientations, e.g., if many objects of interest are present in the scene, any orientation that catches a single object will deliver max accuracy for binary classification; counting, on the other hand, will favor the orientation with the most objects.


\para{Primer on PTZ cameras.} Pan-tilt-zoom (PTZ) cameras present an intuitive mechanism to realize such adaptation. PTZ cameras come in two forms, traditional~\cite{wisecam, ptzoptics-cameras} and electronic (ePTZ)~\cite{what-is-eptz,understainding-eptz-ptz}, both of which support software tuning of pan (horizontal rotation), tilt (vertical rotation), and zoom. The key difference between the two variants is in their tuning mechanisms. Traditional PTZ cameras embed physical motors to rotate at well over 360\degree-per-second and optically zoom (i.e., without reducing resolutions). In contrast, ePTZ cameras capture wide field-of-views and employ near-instantaneous digital rotation and zoom to focus on specific parts of the scene. ePTZ cameras change orientations faster and are cheaper, but also cover smaller rotation areas (150\degree~vs. 360\degree) and degrade image quality by using digital zoom. PTZ cameras rival traditional ones in on-board compute resources, with recent offerings housing edge-grade GPUs~\cite{NVIDIAJetsonNano}.

\subsection{Challenges}
\label{ss:challenges}

Despite the potential benefits of adapting camera orientations using PTZ cameras, three fundamental challenges complicate this approach in practice. We describe them in turn.

\begin{figure}[t]
    \centering
    \includegraphics[width=0.8\columnwidth]{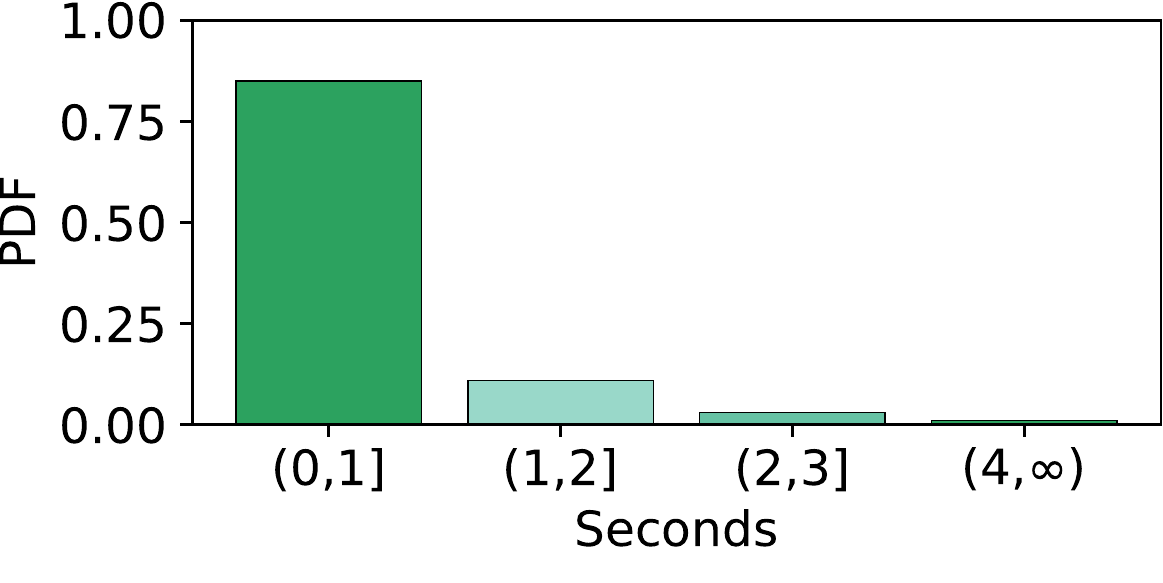}
    \vspace{-2pt}
    \tightcaption{Shifts in the best orientation are frequent. Results list a PDF (binned by 1 sec) of time between switches in best orientation across all videos and workloads.}
    \label{fig:c1_dynamism}
\end{figure}


\para{C1: rapid changes in best orientation over time.} As shown in Figure~\ref{fig:c1_dynamism}, due to the dynamic nature of video content, switches in best orientation are frequent: 85\% of switches occur in $\leq$1 sec since the last switch.


\begin{figure}[t]
    \centering
    \includegraphics[width=0.8\columnwidth]{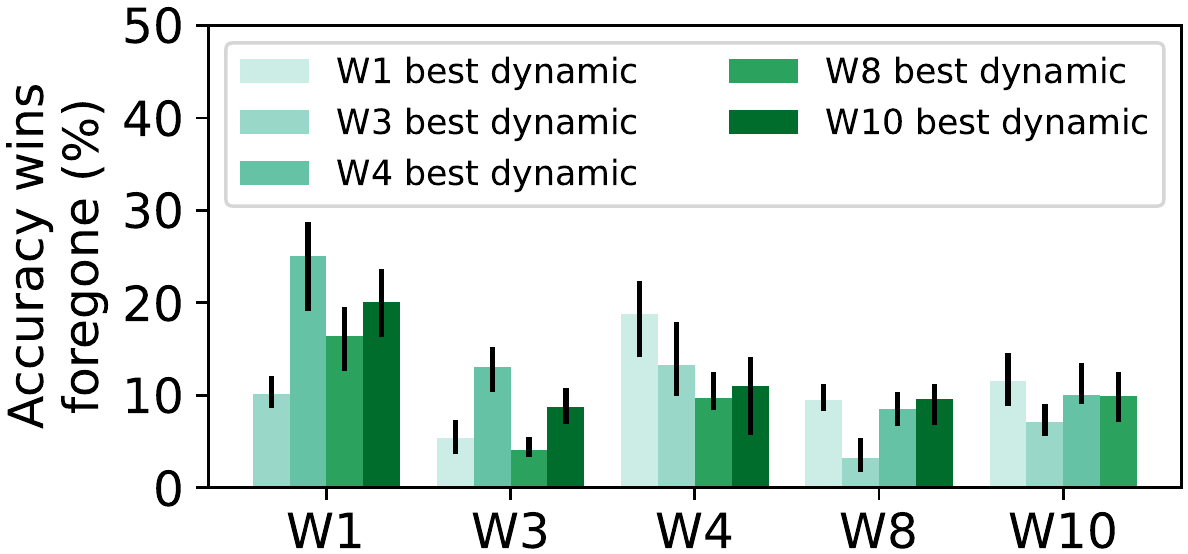}
    \tightcaption{Workloads exhibit different sensitivity to orientations. Results apply the best orientations for workload $X$ (legend) to workload $Y$ (x axis), and plot the accuracy wins (over best fixed for $Y$) that are lost from not using the best orientations for workload $Y$. Bars list medians; error bars for 25-75th percentiles.}
    \vspace{5pt}
    \label{fig:sensitivities}
\end{figure}

\para{C2: diverse workload sensitivities to zoom and rotation at each time.} At any point in time, the best orientation can vary across individual queries and workloads. Figure~\ref{fig:sensitivities} illustrates this, showing that adapting orientations to maximize accuracy for one workload can result in foregoing 3.2-25.1\% of the potential (median) accuracy wins for other workloads. 

\begin{figure}[t]
    \centering
    \includegraphics[width=0.8\columnwidth]{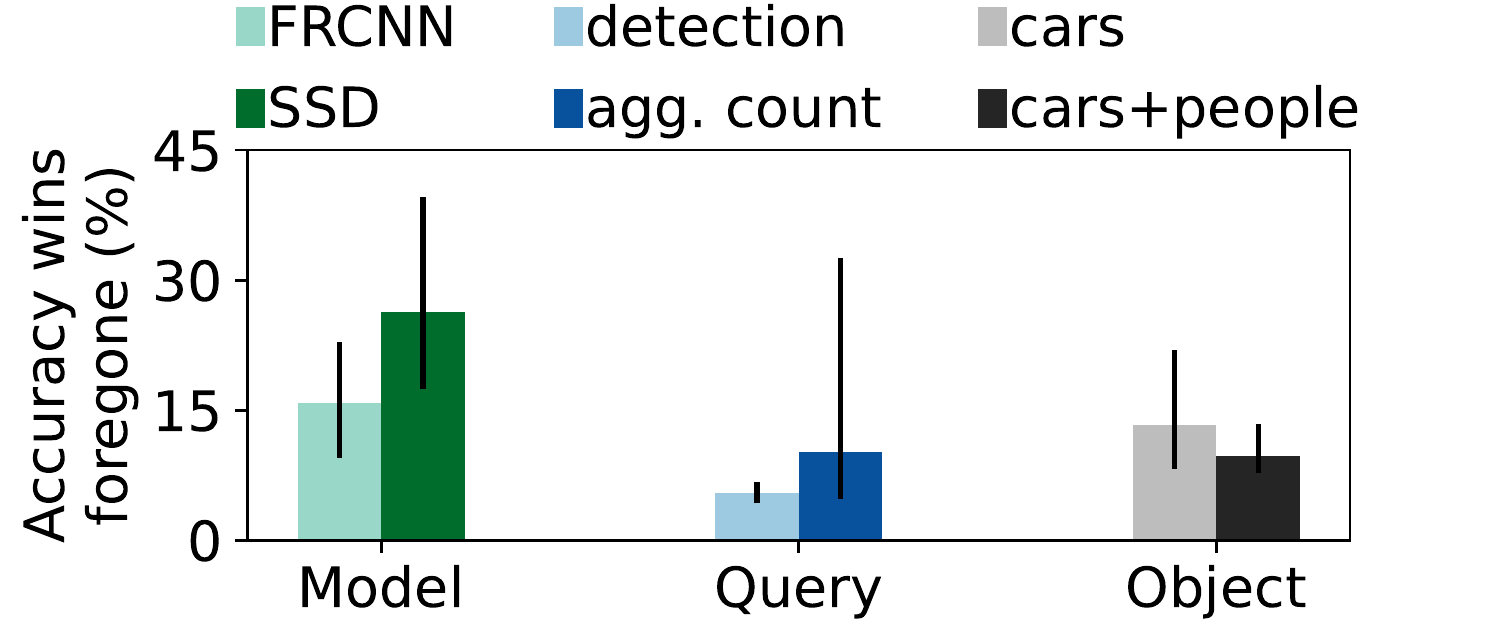} 
    \tightcaption{Applying the best orientations for a base query of \{YOLOv4, counting, people\} to a query $Y$ that modifies a single element in the base query; we compare the accuracy wins (over best fixed) to those when using the best orientations for $Y$. Bars list medians; error bars for 25-75th percentiles.}
    \vspace{4pt}
    \label{fig:query_sensitivities}
\end{figure}

Figure~\ref{fig:query_sensitivities} highlights this at a query level, showing that different models, objects, and tasks can all influence orientation selections. Model discrepancies influence what can be discerned in the scene during inference and under what orientations. For instance, with people counting, selecting best orientations for a query using YOLOv4 will miss out on 26.3\% median accuracy wins for the same task using SSD (even when trained on the same dataset). In contrast, tasks dictate the specificity needed in the collected results, e.g., optimizing for counting people (with YOLOv4) rather than aggregate people counting with the same model foregoes 10.2\% of potential wins. Lastly, objects of interest govern the importance of regions based on object densities, as well as the features used for and difficulty in detecting relevant objects (smaller objects are typically tougher to discern~\cite{redmon2018yolov3}). Thus, unsurprisingly, optimizing for a YOLOv4 people counting query would forego 13.3\% of wins if the query considered cars instead.

\begin{figure*}[t]
    \centering
    \includegraphics[width=0.87\textwidth,height=0.87\textheight,keepaspectratio]{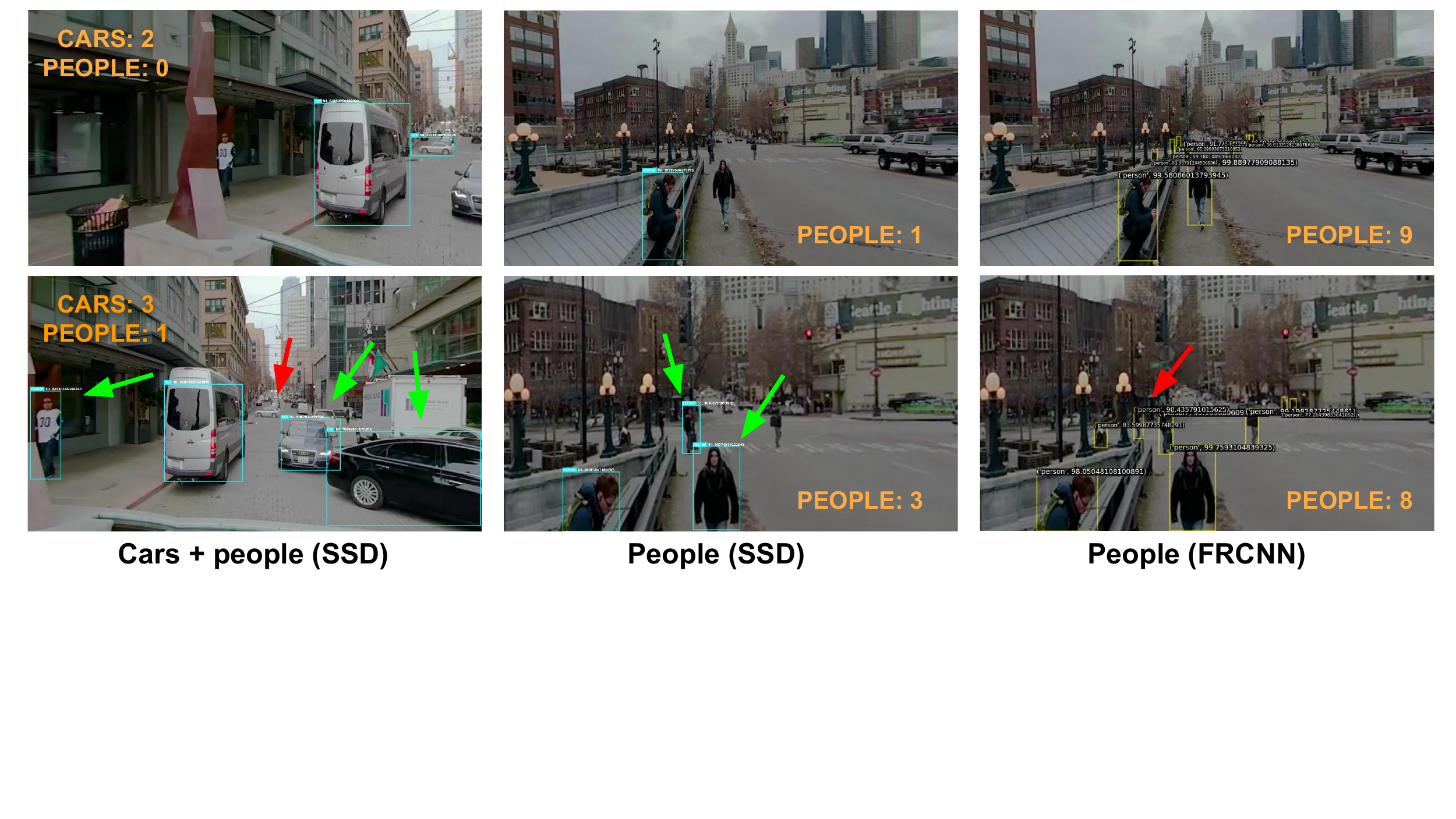}
     \vspace{-7.5em}
    \tightcaption{Screenshots showing the (diverse) impact of rotation and zoom for different queries. Each column shows two images from the same time instant that use either different rotation or zoom. On the bottom row, green arrows show newly captured objects, while red arrows show objects that are newly missed after the orientation change. Left: rotation brings a new object into the scene, helps detect 2 previously-visible objects, but loses a previously-detected object. Middle: zooming in helps detect new people. Right: after switching models, the same zoom from the middle column actually reduces the number of detected people.}
    
        \label{fig:screenshots}
   \end{figure*}

Figure~\ref{fig:screenshots} provides example screenshots to illustrate the benefits and harm of changing orientations. Importantly, tuning orientations does not simply bring new objects into field of view, and instead plays a large role in a model's ability to detect objects that were already visible.

\begin{figure}[t]
    \centering
    \includegraphics[width=0.8\columnwidth]{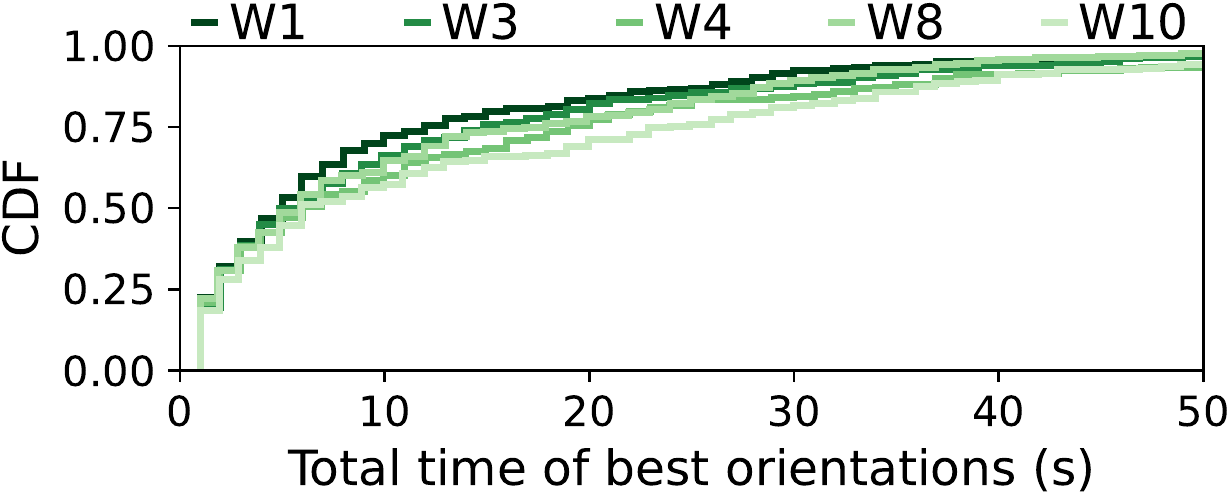}
    \tightcaption{Most orientations are best for short total times in each video. Results consider all orientation-video pairs per workload.}
    \vspace{3pt}
    \label{fig:best_duration}
\end{figure}

\para{C3: massive (but sparse) search space.} The orientation space exhibits substantial sparsity in the spatial and temporal dimensions. For the former, among the 75 orientations at any time, only 1 (or several, with ties) is best, with steady dropoff in accuracy to the others, e.g., median dips of 4.8\% and 20.7\% from the best to 2nd and 5th best. For the latter, most orientations are best for short total times in each video, with median durations of 5-6 sec across workloads (Figure~\ref{fig:best_duration}).

\section{Design}
\label{ss:design}

\begin{figure*}[t]
    \centering
    \includegraphics[width=0.82\textwidth,height=0.82\textheight,keepaspectratio]{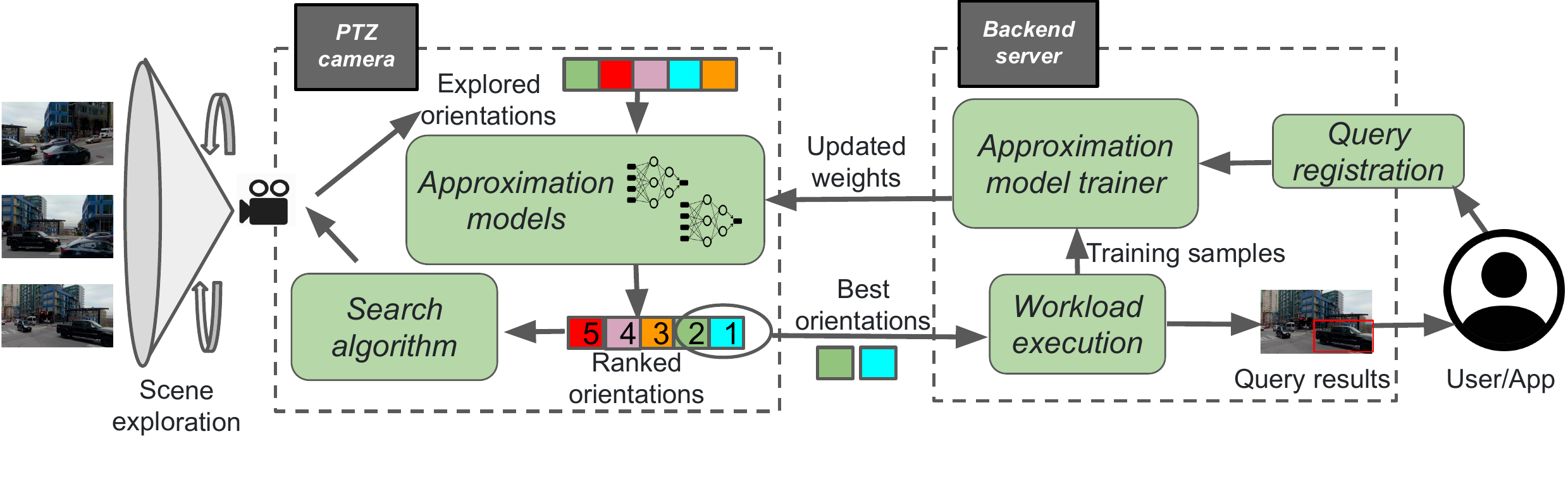}
     \vspace{-1.3em}
    \tightcaption{Overview of MadEye's end-to-end workflow.}
    \label{fig:overview}
\end{figure*}

Figure~\ref{fig:overview} shows the end-to-end operation of MadEye. The main insight behind MadEye is to leverage fast PTZ rotation speeds to explore many orientations in each timestep (i.e., between when results are needed for an fps), and then select, based on their \emph{current} content, the one(s) that maximize workload accuracy under resource constraints. The idea is to limit the ``guess work'' compared to prior search algorithms that rely only on past orientation efficacy (\S\ref{ss:sota}).

As in other video analytics systems~\cite{gemel,chameleon,reducto,dds}, users register queries with a backend agent (on an edge or cloud server), specifying a target scene, as well as a model to use, object(s) of interest, and a task. To operate under camera compute constraints, MadEye then trains edge-compatible (i.e., highly compressed) models (\S\ref{ss:approximate}), not to replace the original (more accurate) query models (as in typical knowledge distillation~\cite{hinton2015distilling}), but instead to \emph{approximately} extract information of importance in a frame for each query. In other words, approximation models are explicitly designed to estimate the inherent sensitivities of each query (C2 from \S\ref{ss:challenges}).

To cope with the large space of orientations and rapid shifts in best orientations (C1 and C3 from \S\ref{ss:challenges}), MadEye employs an efficient on-camera search strategy (\S\ref{ss:search}) that explores as many potentially fruitful orientations as possible while avoiding fps violations for results. The camera then runs approximation models on all captured orientations in each timestep and uses the results to (1) \emph{rank} the orientations in terms of their likelihood to maximize overall workload accuracy, and (2) determine the set of orientations to consider in the next time step. The highest ranked orientations that the network can support are sent to the backend for full workload inference; new results are used to continually adapt approximation models to the current scene (\S\ref{ss:server}).

\subsection{Designing Approximation Models}
\label{ss:approximate}



The primary objective of MadEye's approximation models is to quantify the \emph{relative} importance of orientations for the queries in a workload. However, this requires capturing the sensitivity of each query to different orientation and scene dynamics, subject to camera compute constraints. Given the potential complexity of workload queries, we eschew noisy (and limited) vision features based on local gradients~\cite{dalal2005histograms,lowe2004distinctive} in favor of knowledge distillation with compressed models~\cite{hinton2015distilling}. However, we alter this approach in several ways to favorably balance ranking accuracy and resource efficiency.

We design approximation models using a common abstraction that reflects the minimum amount of information needed to sufficiently rank orientations. The key idea is that the core elements of query sensitivity pertain to how models find and characterize objects, rather than how tasks post-process those results. Thus, MadEye's approximation models are structured purely as ultra-lightweight detectors for objects of interest; this strategy also avoids tricky development of compressed models per task. Concretely, we use the smallest variant of the edge EfficientDet family~\cite{efficientdet}, EfficientDet-D0 (3.9M parameters, $>$150 fps on a Jetson edge GPU). 
More complex detectors could be used, but cameras possess limited GPU memory~\cite{gemel,reducto}, and inference delays negatively influence the degree to which MadEye can explore orientations (\S\ref{ss:search}).


\para{Why a detector?} Two alternatives we considered for the approximation models are to directly estimate object counts in an image, and to directly output rank orderings across multiple images. However, we empirically observed high error rates with both. This is largely because such approaches can only relate the presence of features to objects via a global regression over an entire image (or multiple images), failing to leverage local regressions via bounding box predictions to boost precision. While image-level DNN object counters do exist~\cite{sindagi2017generating,yang2020embedding,zhang2019attentional,zhao2019cascaded}, they focus on large crowds of people. In contrast, there are often few objects of interest in an orientation at any time (\S\ref{ss:challenges}), making rank orderings extremely sensitive to small errors in count prediction.




MadEye uses one approximation model per query, rather than per workload or per object. Though more efficient, we avoid per-workload and per-object approximation models as we (like others~\cite{boggart}) find that different DNNs can exhibit wildly varying response profiles to even the same object classes due to object-independent factors like scale and resolution~\cite{huang2017speed}. Moreover, DNNs trained on very different datasets are known to inherit different algorithmic biases~\cite{balakrishnan2021towards,du2020fairness,khosla2012undoing,oksuz2020imbalance,steed2021image,wang2020towards}.


However, each approximation model is configured to support \emph{all} orientations for two reasons. First, the number of orientations is large (\S\ref{ss:challenges}), making per-orientation approximation models impractical with on-camera GPUs. Second, neighboring orientations exhibit substantial overlap, and since we only consider orientations for a given scene, divergence in background content, lighting, shadows, etc. are minimal. Indeed, we measured the perceptual distance~\cite{perceptual-similarity} of images (LPIPS) from different orientations in the same scene to be 0.30. For context, the same value for the popular MS-COCO and Pascal VOC datasets used to successfully pre-train many vision models (including EfficientDet) are 0.46 and 0.41.


\para{Estimating workload accuracies.} MadEye post-processes the generated bounding boxes from all approximation models to compute \emph{predicted workload accuracies} for orientation ranking. To do this, MadEye follows the per-task accuracy metrics from \S\ref{ss:background}, but computes per-orientation predicted accuracy in a relative manner compared to the other orientations under test. For instance, counting computes the ratio of object counts between each orientation and the max among the set of explored orientations at that timestep, while detection expands this to incorporate object area sizes (as per mAP score). Lastly, aggregate counting modulates count scores to favor less explored orientations (that may have unseen objects). 

\subsection{Continually Training Approximation Models}
\label{ss:server}

MadEye servers train a new approximation model for each new query, with the goals of being fast (since training blocks deployment) and accurate (in ranking orientation importance). Initial training uses a small set of 1000 historical images from the target scene that is then labeled (online) using the DNN in the registered query; label generation takes 7-90 sec depending on the DNN. However, to accelerate this process, MadEye begins with a version of EfficientDet that is pre-trained on Pascal VOC, and freezes both the backbone network and the BiFPN layers responsible for feature extraction and fusion. Only weights for the final 3 bounding box and class prediction layers are fine-tuned to mimic the target query's behavior. The rationale is that model features progressively move from general (e.g., textures, gradients) to task-specific (e.g., object prediction) as a function of layer depth~\cite{bau2017network,zeiler2014visualizing,NIPS2014_375c7134}. 
Initial fine-tuning lasts for 40 epochs ($\approx$25 mins).

Even after initial fine-tuning, approximation models may fail to generalize to changing scene dynamics~\cite{odin}, leading to degrading accuracy. To cope with such data drift, MadEye employs continual learning (every 400-500 ms) to update the model's weights using the latest query results on orientations sent to the server for full workload inference. While continual learning has been applied to edge video analytics~\cite{ekya,recl}, MadEye requires several alterations from prior efforts. The main challenge is that within each retraining window, samples are only available for the orientations that MadEye's camera-side component recently visited and deemed worthy of backend inference. Since orientations are typically best for short total times (\S\ref{ss:challenges}), there is often severe imbalance in the orientations covered by new training samples. For instance, with perfect rankings, the average 2-minute window sees only 9.3\% of orientations get sent to the backend. This can result in overfitting to certain orientations, and catastrophic forgetting~\cite{catfail} for others that may soon be ranked highly. 


%

To deal with this, MadEye retrieves the most recent historical training samples from each orientation and uses this to balance the dataset. As we will discuss in \S\ref{ss:search}, we find that orientation shifts are often spatially localized, with changes to distant orientations happening over longer timescales. Thus, MadEye pads the data samples for neighboring orientations (up to 3 away from the latest one) to match the count for the most popular orientation in the retraining window. The remaining orientations use an exponentially declining number of samples based on their distance from the latest orientation. 

\subsection{Exploring and Ranking Orientations}
\label{ss:search}

The primary goal of MadEye's on-camera component is to efficiently explore (a subset of) the large orientation space to capture the best orientation for each timestep. Realizing this is challenging for three reasons. First, MadEye only has visibility into the orientations that it has recently explored, but other orientations can change in content and importance at any time. Second, even among recently explored orientations, MadEye only has access to coarse results from approximation models (i.e., that accurately capture only relative importance) for most. Third, each timestep is not only dedicated to exploration, but also (1) running approximation models on explored orientations, (2) encoding and shipping select orientations to the server, and (3) running the workload on shipped images.

\begin{figure}[t]
    \centering
    \includegraphics[width=0.8\columnwidth]{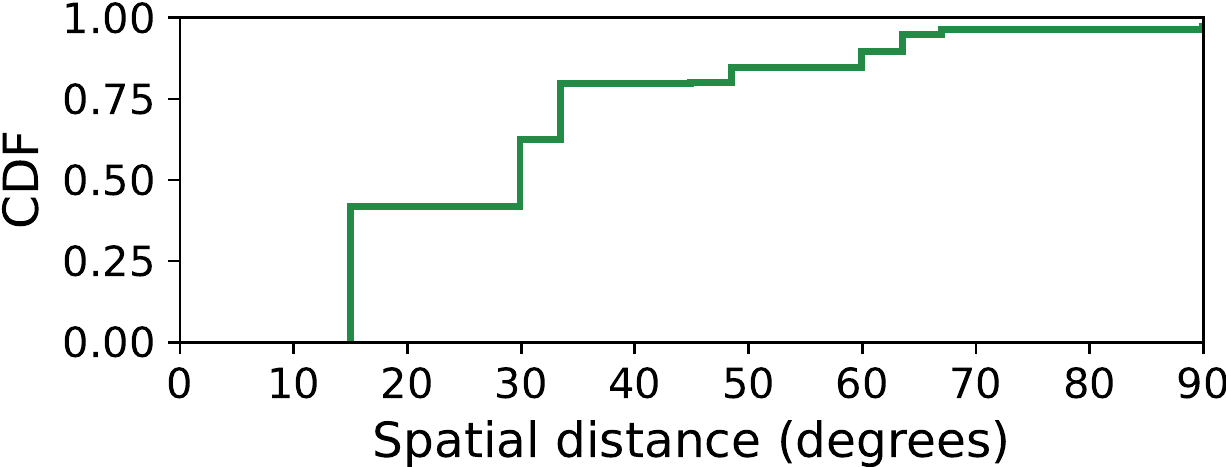}
    \tightcaption{Spatial distance between successive best orientations is small, with most transitions between neighboring orientations. Results aggregate across all videos and workloads for 15 fps.}
    \label{fig:spatial}
\end{figure}

\begin{figure}[t]
    \centering
    \includegraphics[width=0.8\columnwidth]{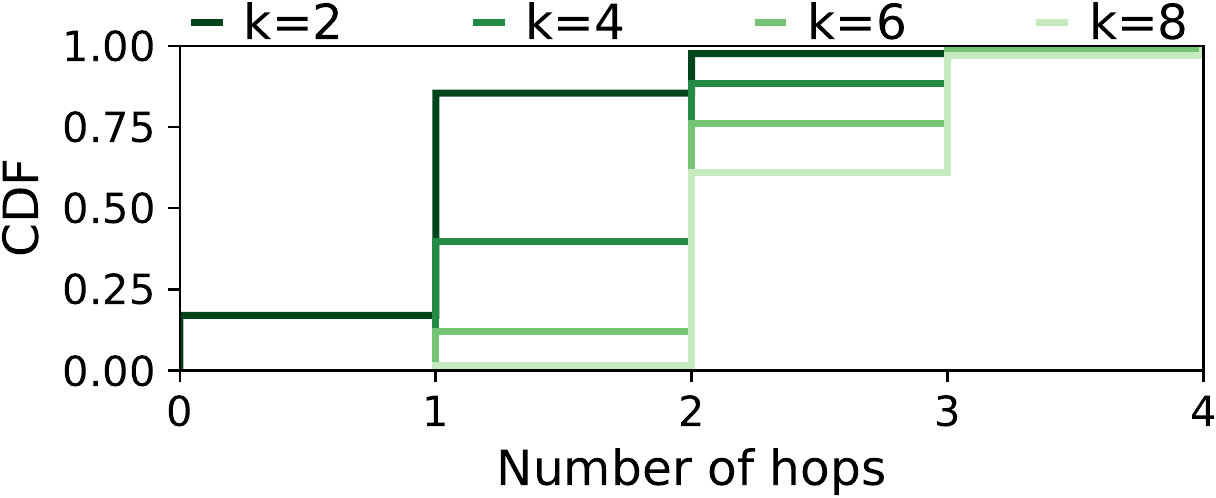}
    \tightcaption{Top ranked orientations are often spatially clustered. Results use 15 fps, are aggregated across all workloads and videos, and show the max distance between orientations in the top $k$ ranked orientations at each timestep.}
    \label{fig:spatial_cluster}
\end{figure}

Rather than relying on previous (and potentially stale) observations at each orientation (\S\ref{ss:sota}), MadEye opts for a more informed strategy guided by 3 empirical observations. 
\squishlist
\item Although best orientations change rapidly over time (\S\ref{ss:challenges}), those changes are far slower in the spatial dimension. 
Figure~\ref{fig:spatial} illustrates this, showing that the median and 90th percentile spatial distance between successive best orientations are 30\degree~and 63.5\degree,  
which pertains to shifts spanning only 1 or 2 orientations in our default grid (\S\ref{ss:methodology}). 

\item The best performing orientations (accuracy-wise) at any time are often spatially clustered (Figure~\ref{fig:spatial_cluster}). Concretely, across our dataset, the 75th percentile distance separating orientations in the top $k$ at each timestep is 1 and 2 orientations for $k$ values of 2 and 6.

\begin{figure}[t]
    \centering
    \includegraphics[width=0.85\columnwidth]{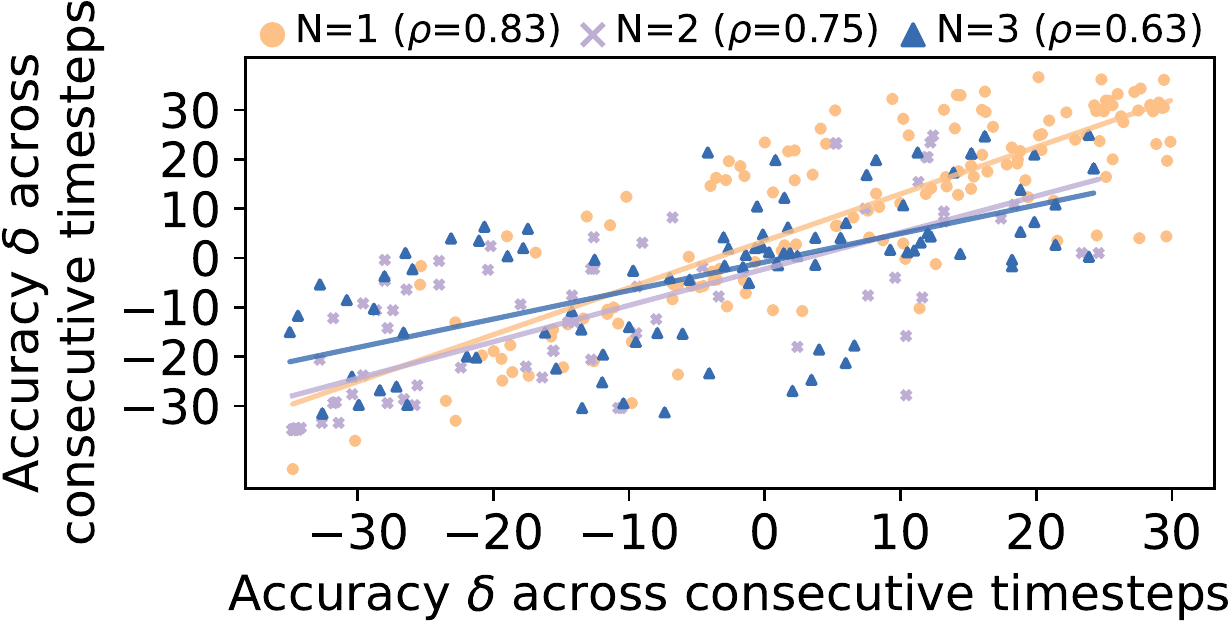}
    \tightcaption{Correlation in accuracy changes across orientations separated by $N$ hops. Results list Pearson Correlation Coefficients and cover 3 representative videos and workloads (15 fps).}
    \vspace{4pt}
    \label{fig:overlap}
\end{figure}

\item Accuracy for neighboring orientations often shift in tandem. Indeed, as shown in Figure~\ref{fig:overlap}, the correlation coefficient for accuracy changes in direct neighbors is 0.83; intuitively, this value shrinks to 0.75 when considering neighbors 2-hops away (that exhibit less content overlap).
\squishend
\noindent Taken together, these findings motivate a search strategy that considers a flexible shape of contiguous orientations at each timestep, and swaps out underperforming orientations in the previous shape only for neighboring ones whose trends we can robustly predict for the next timestep. 
We start with a description of the algorithm that does not account for zoom or resource constraints and later incorporate those elements. Common themes are: only relative comparisons of approximation model results are used, we leverage all outputs from those models (including bounding boxes), and search decisions are entirely local (i.e., on cameras) to remain rapid.

MadEye begins with a rectangular seed shape that reflects the largest coverable area in the time budget, thereby maximizing early exploration; we reset to this shape any time 0 objects of interest are found in a shape.
The corresponding orientations are captured and analyzed with approximation models to compute a predicted workload accuracy for each (\S\ref{ss:approximate}). After sending the top $k$ orientations to the server for workload inference, MadEye must use these prior results to determine the set of orientations to explore in the next timestep.

To do this, MadEye labels each orientation from the last timestep with a value that indicates the likelihood of being fruitful in the \emph{next} timestep. Concretely, we combine the exponentially weighted moving averages from recent (10) timesteps for (1) any computed predicted accuracy values, and (2) the deltas between those values. Weighted averages are used to remain robust to inconsistencies in DNN results across consecutive frames~\cite{boggart,aqua}, which is especially pronounced with MadEye's compressed approximation models.


Using those labels, MadEye must now determine which orientations to remove and add for the upcoming timestep. For this, MadEye sorts orientations into an ordered list based on their label values. Using pointers at the head $H$ (largest label) and tail $T$ (smallest label) of the list, MadEye iteratively compares orientations by asking: should we remove the orientation at $T$ in favor of adding a neighbor to $H$? 
Concretely, MadEye computes the ratio of label values for $H$/$T$. If (1) that ratio exceeds a threshold (indicating a substantial disparity in the potential of $H$ and $T$), (2) $H$ has neighbors not already in the shape, and (3) removing $T$ would not break contiguity, we remove the orientation at $T$ and increment the pointer. The process repeats by considering the addition of another neighbor for $H$, this time using a larger threshold to account for the additional uncertainty of adding more neighbors. $H$ is decremented when a neighbor cannot be added, and the process ends when even one neighbor for $H$ cannot be added.

For each iteration that results in a neighbor addition for $H$, MadEye selects among $H$'s neighbors by analyzing the bounding boxes that its approximation models generated in the last timestep. For each candidate neighbor, we compute the ratio of two values: normal distances to the center of $H$ and to the centroid of all bounding boxes in $H$. Values $<$1 indicate lower chances of $H$'s objects moving to the candidate in the next timestep. We repeat this process for all other orientations in the last shape that the candidate exhibits any non-zero overlap with. 
Candidate neighbor scores are computed as the weighted sum of these ratios (weights according to degree of overlap), and the candidate with the max score is selected.


\para{Reachability and path selection.}
The search algorithm thus far ignores whether a PTZ camera can sufficiently cover the selected shape in a given time budget. Formally, the shape of orientations can be represented as a fully-connected undirected graph with edge weights pertaining to the time taken to move between two adjacent orientations (given a rotation speed). Our goal is to determine whether the shape is coverable in a given time budget, and if so, what is the shortest path. The paths between orientations satisfy the triangle inequality property~\cite{tversky1982similarity}, so this can be modeled as a variant of the NP-Hard Traveling Salesman Problem (TSP)~\cite{bender1999performance}. Given our tight time budgets, MadEye employs the Minimum Spanning Tree (MST) heuristic~\cite{held1970traveling}, but optimizes it to minimize online delays. In particular, since our orientation grid is static, we precompute pairwise distances and the entire MST ahead of time. Online, for a given shape, we quickly extract and perform a preorder walk on the corresponding subgraph to get the shortest path. This reduces the heuristic to linear complexity (in orientations); each path computation takes 14 \micro$s$, and the resultant paths are within 92\% of optimal. Upon failure, MadEye greedily removes the orientation with the lowest potential (that does not break contiguity) and rechecks reachability.

\para{Balancing search size and network/compute delays.} MadEye pipelines its exploration through orientations with the running of approximation models on each one. However, network transmission to and workload inference on the backend do not overlap with orientation exploration. The reason is that transmissions are governed by global ranks across \emph{all} orientations explored in each timestep. Thus, in each timestep, we face a tradeoff between exploring more orientations and sending more orientations to the backend.

MadEye resolves this tension based on the expected difficulty for its approximation models to accurately rank the considered orientations, which in turn governs the risk associated with exploring more orientations (and sending fewer to get ground truth results). Intuitively, scenarios where the considered orientations are projected to contribute similar accuracies pose the biggest difficulty for approximation models (as the gaps between ranks shrinks). MadEye determines the right balance by first selecting a target number of frames to send according to the training accuracy for approximation models (provided by the backend) and the variance in predicted accuracy values in the last timestep, e.g., 85\% training accuracy and 25\% variance results in sending at least 2 frames. MadEye then computes a target shape size for exploration, accounting for network transmission delays (harmonic mean of past 5 transfers~\cite{mpc}), backend compute delays, camera rotation speeds, and approximation model inference delays. 




\para{Handling zoom.} After selecting the set of orientations to visit, the search algorithm must determine the zoom factor to use for each one. The challenge is that past accuracies are insufficient for determining zoom fidelity as MadEye cannot know what objects are being missed by not zooming in/out. Instead, we rely on bounding boxes from approximation models to determine the risk of zooming in. When an orientation is added to the shape, we start at the lowest zoom factor to gain visibility into its whole content. At each timestep, we compute the average distance between each bounding box and the centroid of all boxes; smaller distances indicate more clustering and less risk of zooming in. These values are compared with the area covered by each zoom factor to select one, and MadEye automatically zooms out after 3 seconds to avoid missing newly entering objects in the orientation.

\para{Transmitting images.} At the end of each timestep, MadEye must transmit select images to the server for workload inference. Unlike standard streaming, MadEye sends disjoint sets of images from each orientation's video stream. To keep bandwidth costs low, MadEye maintains a list of the last image shared for each orientation, and employs a functional encoder~\cite{salsify} that computes deltas relative to that image.

\section{Implementation}
\label{s:impl}

MadEye's core components are written in 9.1k lines of Python code, with all training and inference tasks across the backend and camera run in PyTorch. We use TensorRT~\cite{tensorRT} to accelerate inference on the backend, and a variant of Nexus~\cite{nexus} as a round-robin scheduler for approximation model inference on cameras. Orientations are first represented as rotational values, projected onto a 360\degree~space, and then converted using an in-house equirectangular-to-rectilinear image converter (written in C++) to match the APIs offered by recent PTZ cameras~\cite{ptz-web-dev-api}. For ground truth accuracy computations (\S\ref{ss:methodology}) that require a global (i.e., across all orientations) perspective on object locations and uniqueness, atop the ByteTrack multi-object tracker~\cite{bytetrack} that links objects across an orientation's video, we use \texttt{cv2} and \texttt{scikit-image} to extract image features (e.g., SIFT) that link objects across orientations.
\section{Evaluation}
\label{s:eval}

We evaluated MadEye across diverse workloads, network settings, and videos. Our key findings are:
\squishlist

\item MadEye increases median workload accuracies by 2.9-25.7\% compared to an oracle fixed-orientation strategy (while using the same amount of resources); wins are within 1.8-13.9\% of the oracle dynamic strategy.

\item Achieving MadEye's accuracy wins with 1 PTZ camera would require the best 4-6 fixed-orientation cameras, which comes with a 2-3.7$\times$ inflation in resource costs.

\item MadEye outperforms prior PTZ algorithms by 2.0-5.8$\times$, providing 31.1\%, 46.8\%, and 52.7\% higher accuracy than Panoptes~\cite{panoptes}, tracking~\cite{auto-tracking}, and multi-armed bandits~\cite{mab_for_wireless_selection}.

\item MadEye gracefully balances on-camera exploration and transmission of orientations to maximize accuracy even as resources shrink and response rates rise. 





\squishend

\subsection{Methodology}
\label{ss:methodology}

\para{Video dataset for PTZ analysis.} To the best of our knowledge, there does not exist a public video dataset for PTZ cameras that enables users to tune rotation and zoom knobs; instead, existing PTZ datasets reflect pre-determined knob decisions. Thus, to evaluate MadEye, we generate our own dataset. To construct our dataset, we begin with the abundance of 360\degree~datasets. Concretely, we use 50 360-degree videos from YouTube that incorporate scenes resembling those from prior work~\cite{reducto,dds,ekya}, e.g., traffic intersections, walkways, shopping centers. Each video lasts 5-10 minutes.

From each video, we carve out scenes of interest as regions spanning 150\degree~horizontally and 75\degree~vertically. We then subdivide those regions into grids of orientations to mimic recent PTZ offerings~\cite{understainding-eptz-ptz} (30\degree~and 15\degree~granularities for pan and tilt; we explore other grids in \S\ref{ss:deepdive}), and extract a full video per orientation. For zoom, since we operate on pre-captured videos, we employ digital zoom (1-3$\times$) by cropping images and scaling back the dimensions to match the original image. 

\para{Models and workloads.} We consider 4 popular architectures for vision tasks: SSD~\cite{ssd} and Faster RCNN \cite{faster-rcnn} with ResNet-50 backbones, YOLOv4 and Tiny-YOLOv4~\cite{yolov4} with CSPDarknet53 backbones. We consider two versions of each model trained on Pascal VOC and MS-COCO, but show results for the latter as the trends were similar. To construct queries, we follow the same methodology from recent work (based on production deployments)~\cite{gemel}. Each model can perform any of the four tasks from \S\ref{ss:background} with a focus on either people or cars. We enumerate all possible workloads sized between 2-20 queries and pick 10 randomly. The appendix details each workload. We run workloads on all videos and consider response rates from 1-30 fps.

\para{Hardware and networks.} On-camera computations run on an edge-grade Jetson Nano~\cite{NVIDIAJetsonNano} equipped with a 128-core Maxwell GPU, quad-core ARM CPU with 1.43 GHz clock speed, and 4 GB of memory. We consider default camera rotation speeds of 400\degree~per second; we study this parameter in \S\ref{ss:deepdive}. Workload inference and training of approximation models run on a server with an NVIDIA GTX 1080 GPU (8 GB RAM) and 18-core Intel Xeon 5220 CPU (2.2 GHz; 125 GB RAM). Camera and server components are connected with emulated Mahimahi networks~\cite{mahimahi} using fixed-capacity (24-60 Mbps; 5-20 ms) and real-world mobile traces.

\para{Metrics.} Our primary evaluation metric is average workload accuracy per video. For each frame, following the accuracy definitions from \S\ref{ss:background}, we compute per-orientation accuracy for each query relative to the orientation that delivers the max accuracy at that time. Per-query accuracies at each time are averaged to compute per-frame workload accuracies, which in turn are averaged to compute workload accuracy for a video. 

While computing these values for binary classification and counting are straightforward, detections and aggregate counting require slight alterations. For detections,  mAP scores depend on bounding box coordinates for specific objects and thus cannot be measured by comparing results directly across orientations. Thus, we consolidate the bounding boxes across orientations into a global view, and employ de-duplication~\cite{pan2010detecting} to eliminate redundant objects in overlapping regions. We then compute each orientation's mAP score relative to the global scene, and assign per-orientation accuracies as the ratio of its mAP score to the max one across orientations.

Aggregate counting queries are directly evaluated across the entire video (not per frame). Thus, we compute the ratio of unique objects across the orientations that a system selects compared to the total number of unique objects in the video. Note that ByteTrack (\S\ref{s:impl}) was unable to robustly support car tracking, so we exclude aggregate counting for cars.

\subsection{Overall Results}
\label{ss:main_results}

\begin{figure*}[t]
\begin{subfigure}[t]{0.33\textwidth}
  \centering
  \includegraphics[width=1\textwidth]{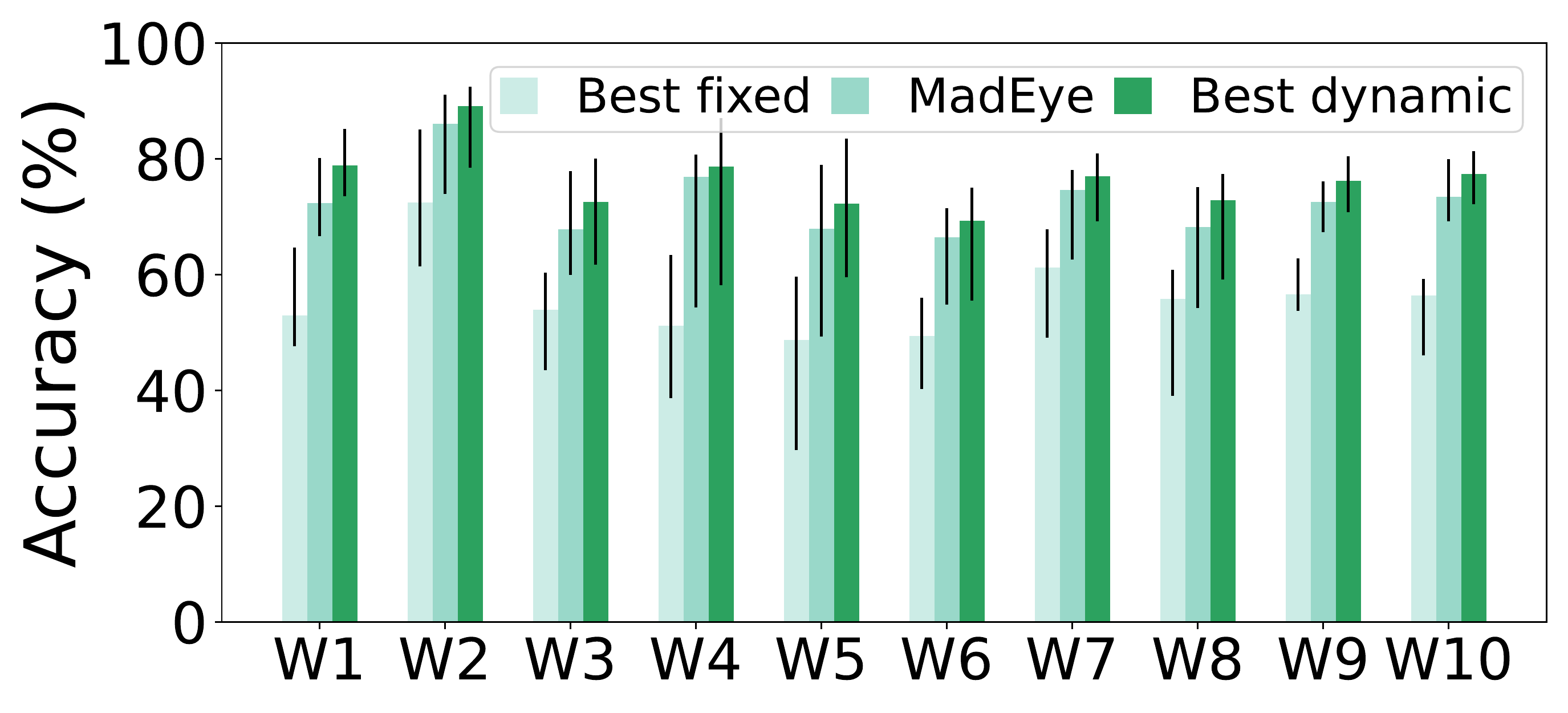}
 \vspace{-8pt}
  \tightcaption{1 fps}
\end{subfigure}
\begin{subfigure}[t]{0.33\textwidth}
  \centering
  \includegraphics[width=1\textwidth]{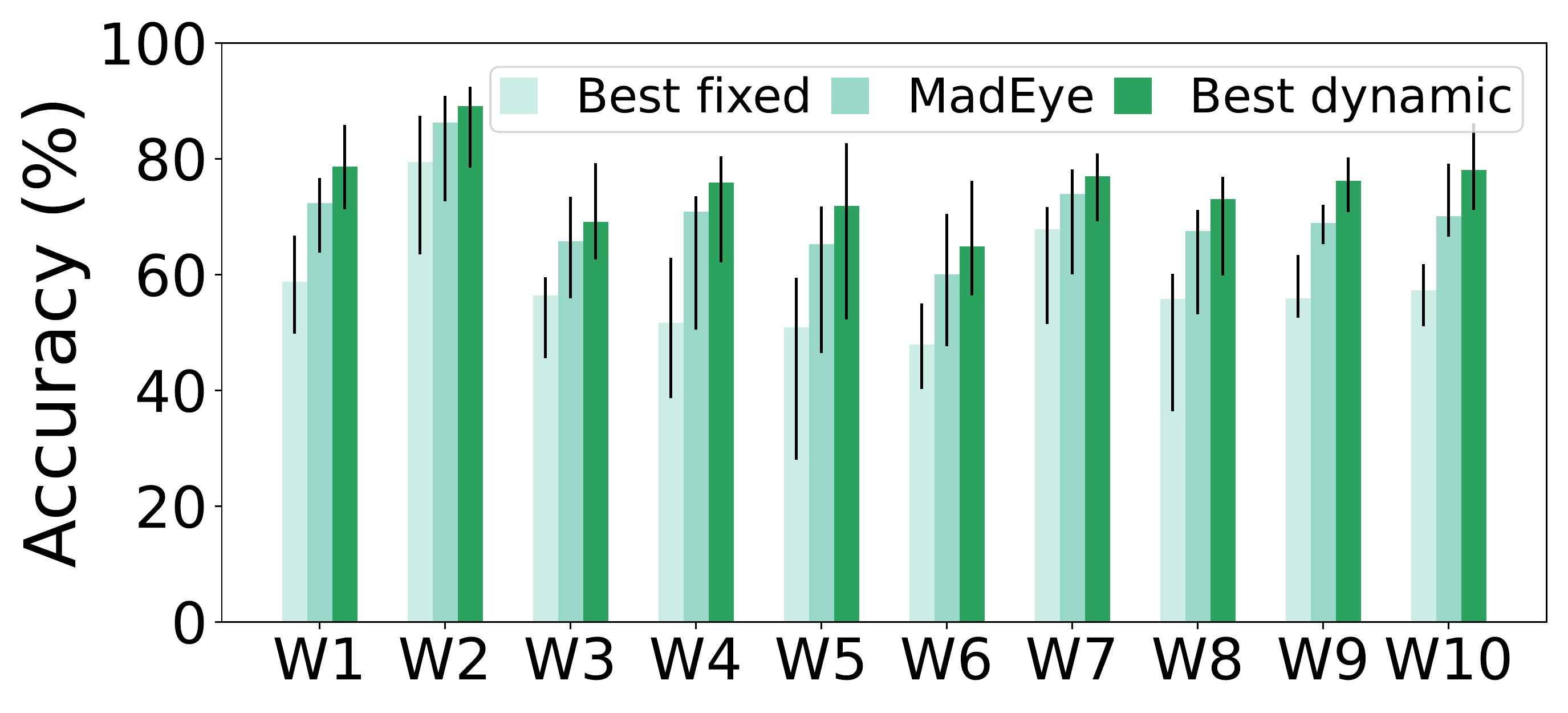}
 \vspace{-8pt}
  \tightcaption{15 fps}
\end{subfigure}
\begin{subfigure}[t]{0.33\textwidth}
  \centering
  \includegraphics[width=1\textwidth]{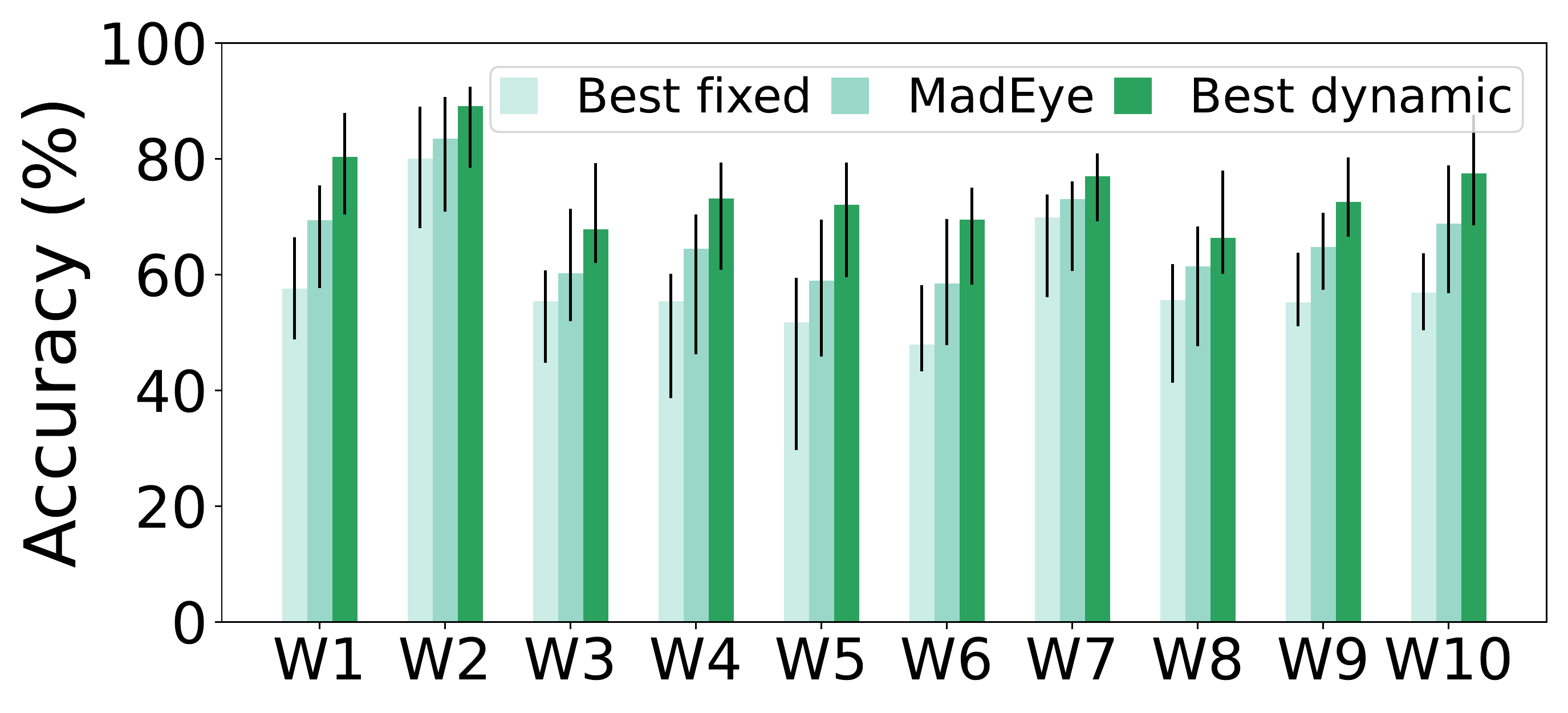}
 \vspace{-8pt}
  \tightcaption{30 fps}
\end{subfigure}
\vspace{3pt}
    \tightcaption{Comparing MadEye with the best possible fixed- and adaptive-orientation schemes across all videos and workloads with a \{24 Mbps, 20 ms\} network and varying fps. Bars list medians with errors bars spanning 25-75th percentiles.}
    \label{fig:fixed_fps}
\end{figure*}


We first compare MadEye with the two baselines from \S\ref{ss:potential}, \emph{best fixed} and \emph{best dynamic}, on different network and fps settings. Both baselines impractically rely on oracle knowledge of video content and workload accuracy, i.e., to pick the best orientation per video or per timestep, respectively, that maximizes accuracy for the target workload-video. Nonetheless, they serve as useful context for MadEye's performance. Note that MadEye automatically adapts the number of frames it explores and transmits based on network delays and response rates (\S\ref{ss:search}). For \emph{best fixed}, we leverage increasing network speeds by adding more fixed cameras (i.e., best, 2nd best, etc.), rather than simply capturing more (redundant) frames from 1 camera. \emph{Best dynamic} does not change for any query other than aggregate counting, for which we send the largest number of fruitful orientations that the network can support.
\begin{figure*}[t]
    \centering
\begin{subfigure}[t]{0.33\textwidth}
  \centering
  \includegraphics[width=\textwidth]{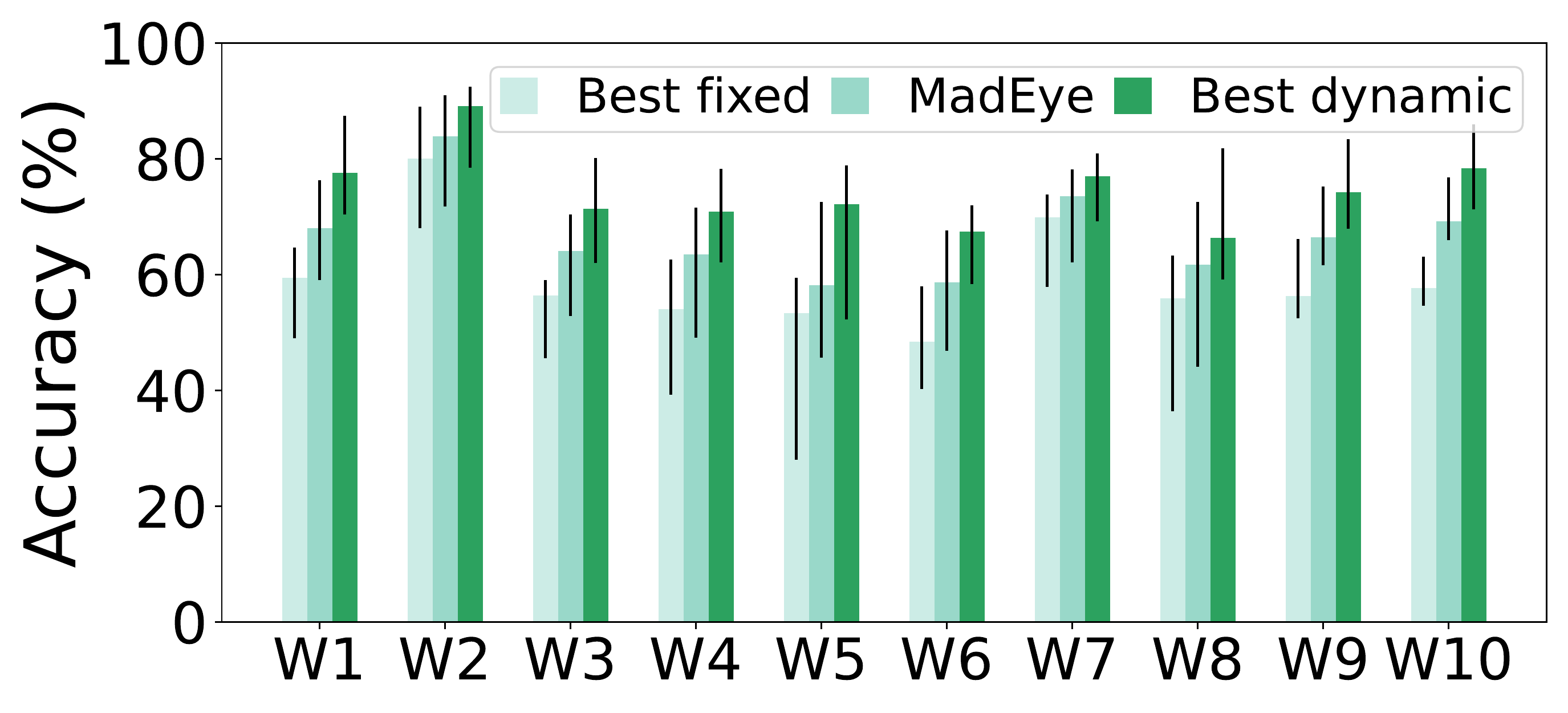}
 \vspace{-8pt}
  \tightcaption{Verizon LTE}
\end{subfigure}
\begin{subfigure}[t]{0.33\textwidth}
  \centering
  \includegraphics[width=\textwidth]{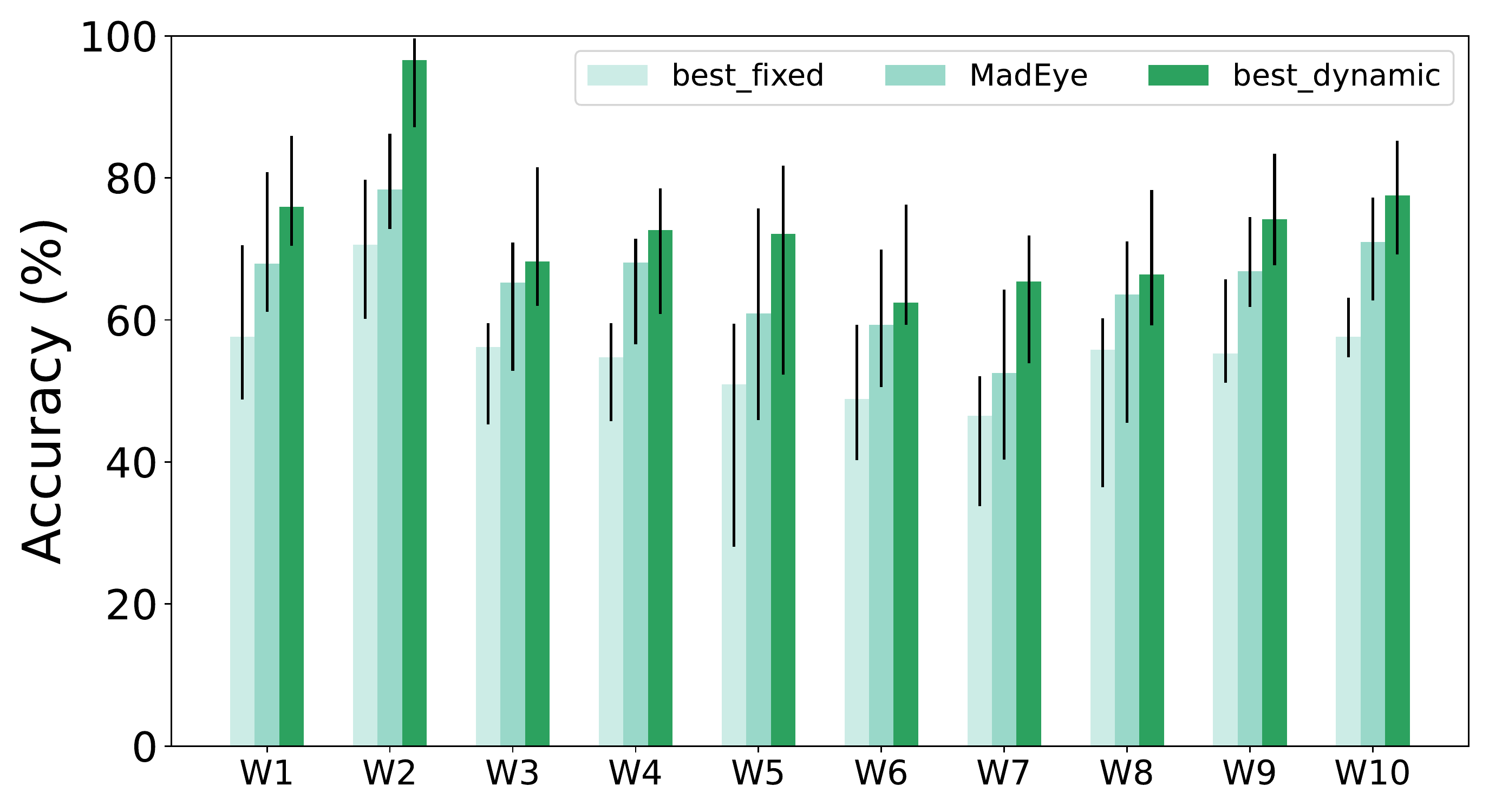}
 \vspace{-8pt}
  \tightcaption{24 Mbps; 20 ms}
\end{subfigure}
\begin{subfigure}[t]{0.33\textwidth}
  \centering
  \includegraphics[width=\textwidth]{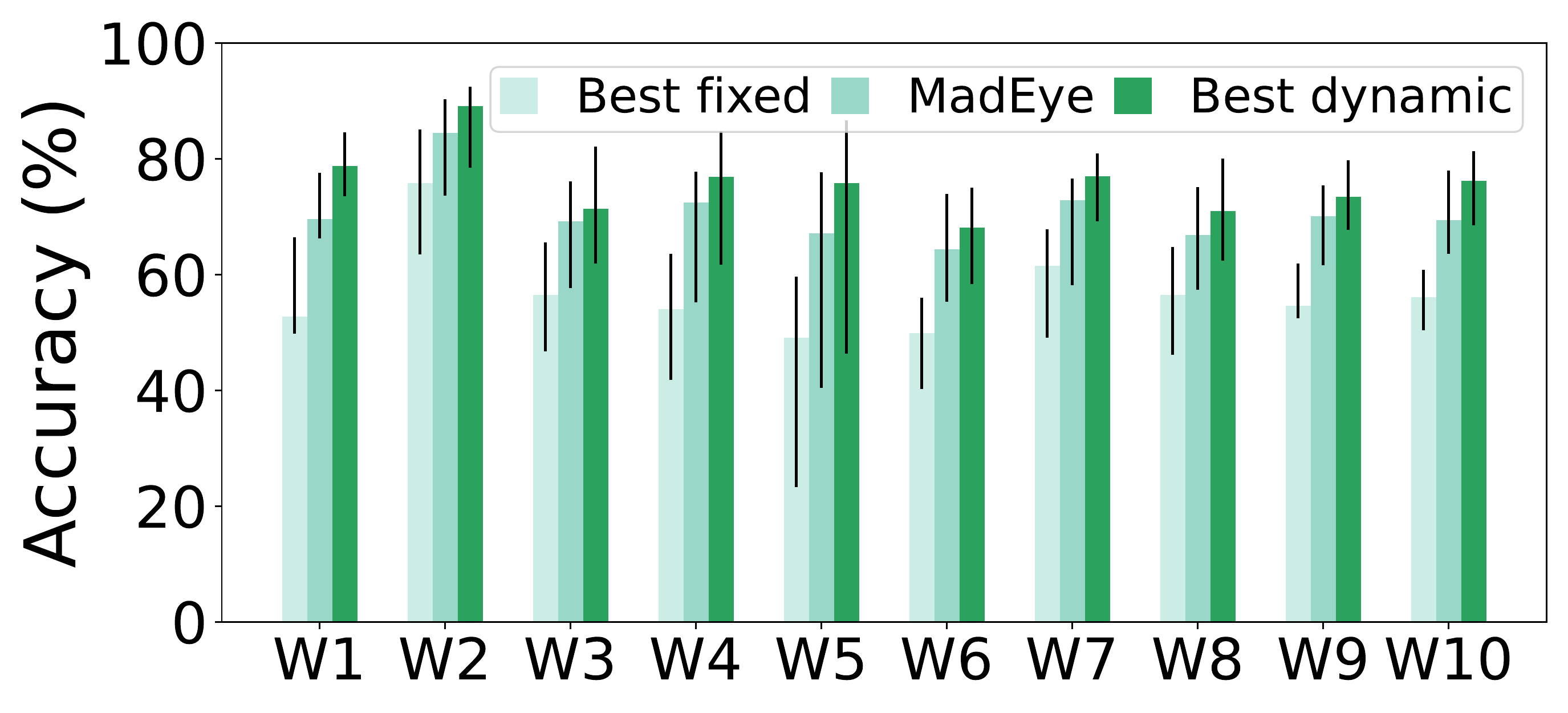}
 \vspace{-8pt}
  \tightcaption{60 Mbps; 5 ms}
\end{subfigure}
\vspace{3pt}
\tightcaption{Comparing MadEye with the best possible fixed- and adaptive-orientation schemes across all videos and workloads with fixed fps (15) and varying networks (improving from left to right). Bars list medians with errors bars spanning 25-75th percentiles.}
    \label{fig:fixed_network}
\end{figure*}

Our results are captured in Figures~\ref{fig:fixed_fps}-\ref{fig:fixed_network}. Across these settings, MadEye delivers median and 75th percentile accuracies that are 2.9-25.7\% and 1.6-20.7\% higher than \emph{best fixed}, and within 1.8-13.9\% and 1.3-12.5\% of \emph{best dynamic}. Digging deeper, our results show two key trends. First, as frame rates decrease (for a fixed network), MadEye's accuracies and wins over \emph{best fixed} grow, e.g., for a \{24 Mbps, 20 ms\} network, median wins improve from 5.8-13.3\% to 12.4-25.7\% as fps drops from 15 to 1. The reason is that lower fps yields larger timesteps (e.g., 1 sec for 1 fps, 66.7 ms for 15 fps), enabling more exploration and/or transmission. Second, as network speeds grow (for fixed fps), the same trends persist (since each network transfer is faster) but to a lesser extent, e.g., median 15 fps wins grow to 8.6-18.4\% for \{60 Mbps, 5 ms\}. 

\begin{figure}[t]
    \centering
    \begin{subfigure}[t]{0.23\textwidth}
  \centering
  \includegraphics[width=\textwidth]{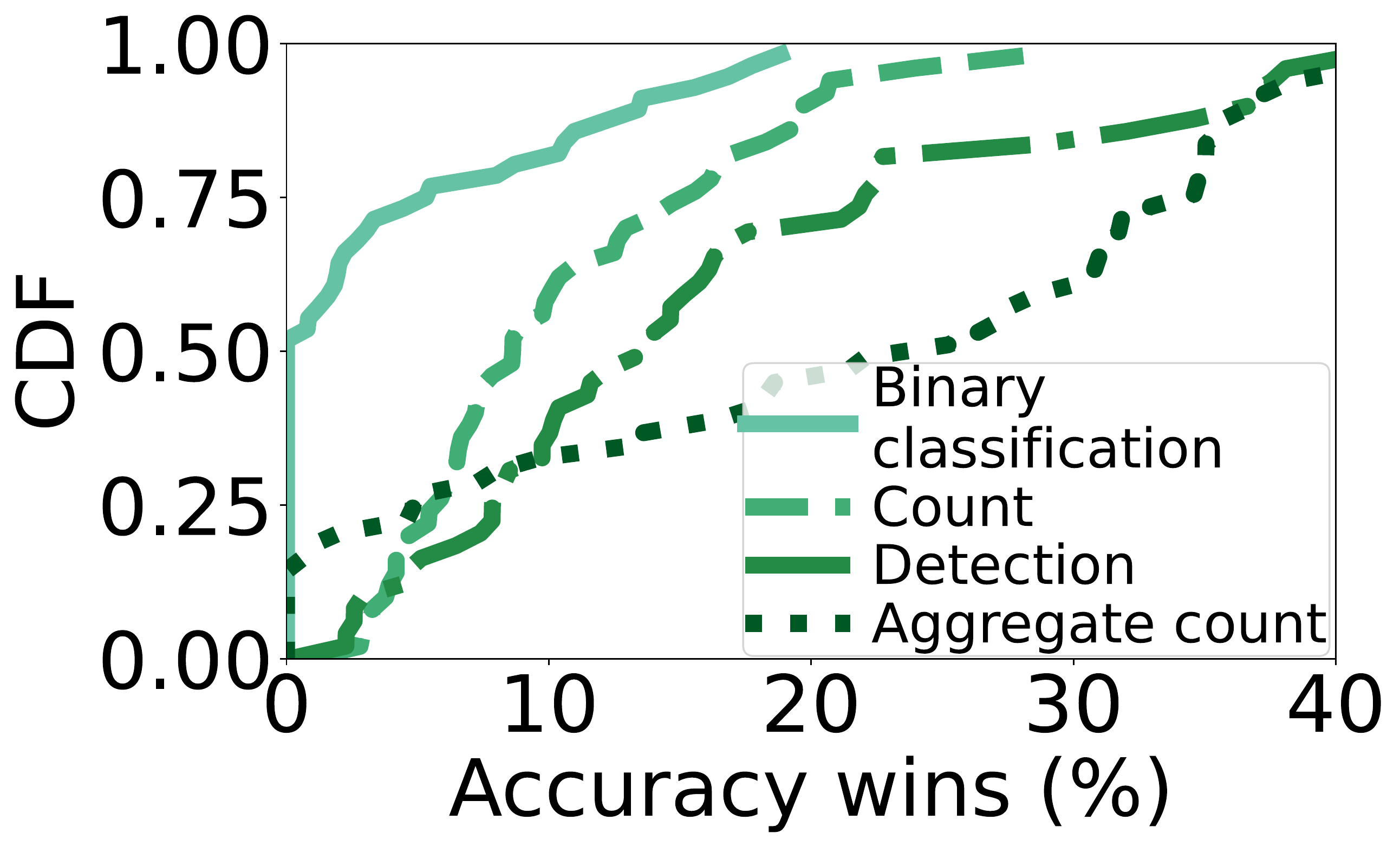}
 \vspace{-8pt}
  \tightcaption{People}
\end{subfigure}
        \begin{subfigure}[t]{0.23\textwidth}
  \centering
  \includegraphics[width=\textwidth]{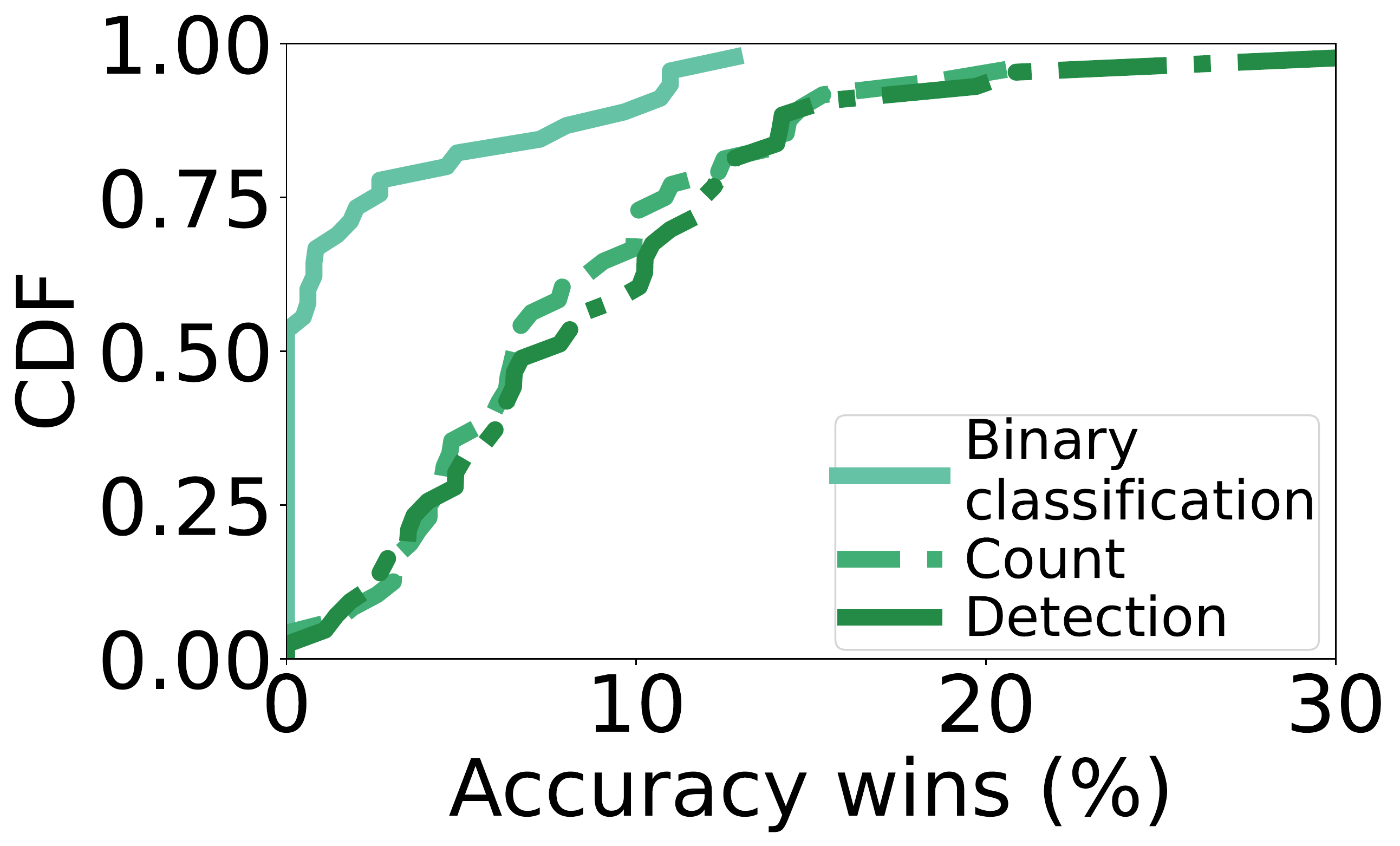}
 \vspace{-8pt}
  \tightcaption{Cars}
\end{subfigure}
\vspace{4pt}
    \tightcaption{MadEye's accuracy improvements (over best fixed) for different query tasks and objects. Results consider all videos and models, and use 15 fps and \{24 Mbps; 20 ms\}.}
    \vspace{4pt}
    \label{fig:query_breakdown}
\end{figure}

Figure~\ref{fig:query_breakdown} breaks down MadEye's wins over \emph{best fixed} by task and object. Following the rationale from \S\ref{ss:potential}, accuracy boosts with MadEye grow as task specificity grows: median wins grow from 8.6\% to 13.3\% to 22.1\% as we move from counting to detections to aggregate counting for people. We also observe consistently larger accuracy wins for people queries (rather than cars) due to their less structured motion patterns (more frequent and scattered orientation switches), e.g., for detections, wins for cars shrink to 6.7\%.






\begin{table}[t]
\footnotesize
\centering
\begin{tabular}{|l|l|l|} 
\hline
\textbf{MadEye Variant} & \textbf{Median Accuracy (\%)} & \textbf{\# Fixed Cameras} \\\hline
\textbf{MadEye-1} & 63.1 & 3.7 \\\hline 
\textbf{MadEye-2} & 66.3 &  5.5  \\\hline
\textbf{MadEye-3} & 66.8 &  6.1 \\\hline
\end{tabular}
\vspace{7pt}
\tightcaption{Number of optimally-configured fixed cameras needed to match the accuracy of MadEye. MadEye-$k$ refers to a version of MadEye that is restricted to sending the top $k$ frames to the server for workload inference. Results consider a \{24 Mbps; 20 ms\} network, 15 fps response rate, and all video-workload pairs.}
\label{t:efficiency}
\end{table}

Results thus far focus on accuracy improvements. However, a key goal with MadEye is to maximize accuracy for a given resource cost, i.e., network and backend inference overheads. Table~\ref{t:efficiency} lists the smallest number of optimally configured fixed cameras that would be required to match the accuracies that different versions of MadEye deliver, each of which sends a different number of frames per timestep. As shown, it would take 3.7 fixed cameras to realize the 63.1\% accuracy that MadEye-1 achieves, implying a 3.7$\times$ reduction in network and backend compute usage. MadEye-2 is matched by 5.5 fixed cameras; here, however, the resource reduction factor is 2.8$\times$ since MadEye also sends 2 frames per timestep.

\subsection{Comparisons with State-of-the-Art}
\label{ss:sota}

\begin{figure}[t]
    \centering
    \includegraphics[width=0.8\columnwidth]{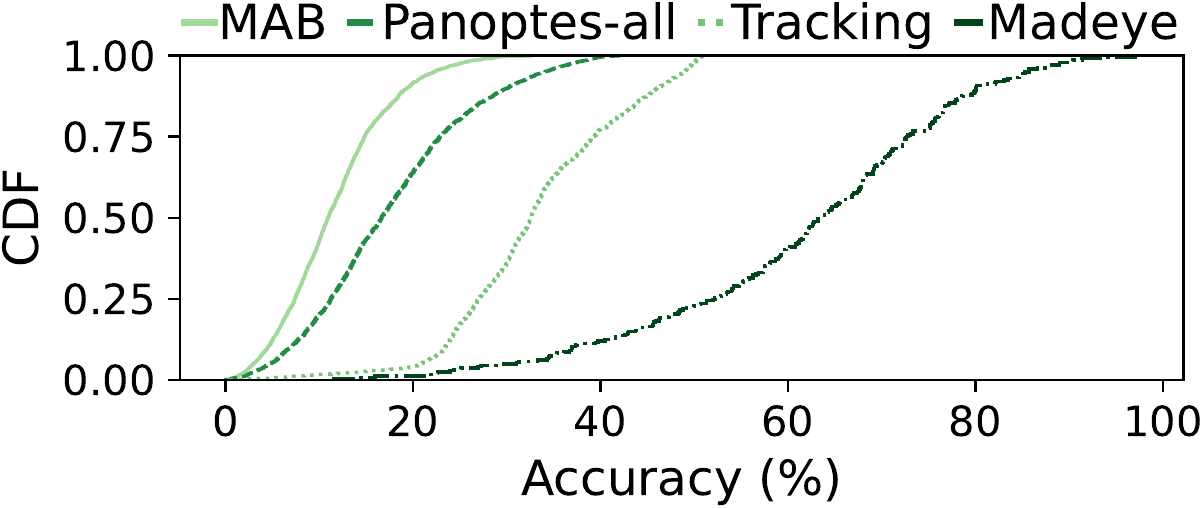}
    \tightcaption{MadEye vs. 3 camera tuning strategies. Results are for all workloads and videos, 15 fps, and \{24 Mbps; 20 ms\}.}
    \vspace{4pt}
    \label{fig:sota}
\end{figure}

We compare MadEye with 3 alternate approaches for adaptive camera orientations. Figure~\ref{fig:sota} shows results for a \{24 Mbps; 20 ms\} network and 15 fps; trends hold for all other scenarios.

First, we consider Panoptes~\cite{panoptes}, a recent PTZ system that configures orientations for workloads of applications, each explicitly concerned with specific orientation(s). For orientations of relevance, Panoptes generates a static round-robin schedule that is weighted according to how many queries an orientation is of interest to and how much motion has been detected historically in that orientation; higher weights indicate staying in an orientation for longer. Panoptes then switches between orientations according to this schedule with one exception: if motion gradients in the direction of any overlapping orientation of interest exceed a threshold, Panoptes switches there for several sec before resuming the round robin. Panoptes does not specify a zoom strategy, so we consider the best zoom (accuracy-wise) for any orientation it visits.

We consider two versions of Panoptes, \emph{Panoptes-all} and \emph{Panoptes-few}, in which each workload query is interested in all orientations or only its best fixed orientation, respectively. Max accuracy in both cases is defined relative to the best orientation among only the set of considered ones. As shown in Figure~\ref{fig:sota}, MadEye outperforms Panoptes-all by 3.8$\times$, with 46.8\% higher accuracy at the median. The reason is that Panoptes cycles through orientations based on a pre-determined schedule and motion gradients \emph{in the current orientation}, neither of which are sufficient indicators of importance of other orientations at the current time, e.g., orientations are suboptimal most of the time (\S\ref{ss:challenges}). In contrast, MadEye considers many orientations per timestep, ranking them based on current content. The wins persist compared to Panoptes-few (not shown due to the different accuracy metric), but are less pronounced (median of 40.5\%) as there are fewer unfruitful orientations for Panoptes to consider.

Next, we consider tracking algorithms that most PTZ cameras come equipped with today~\cite{auto-tracking}. This algorithm starts in a home region (best fixed in our experiment), selects the largest object it finds, and tracks that object continually across orientations aiming to keep it as centered as possible. The algorithm resets to the home region upon losing the tracked object. We consider a favorable variant in which all orientations explored in a timestep are shared with the backend, which uses the one with the highest accuracy. As shown, MadEye delivers 2.0$\times$ higher workload accuracies (31.1\% more at the median) compared to this tracking scheme. The main reason again is that the presence of a large object is a poor indicator of accuracy importance as it fails to capture more general scene properties and the sensitivity of the queries under test. In contrast, MadEye directly estimates query sensitivity to make workload-aware, informed orientation selections.

Finally, we consider the common UCB1 multi-armed bandit (MAB) algorithm~\cite{mab_for_wireless_selection}. Each orientation is considered a lever with a weight set to the average observed accuracy across all past visits (we seed this with historical data). The algorithm continually selects an orientation to visit as the one with the highest sum of weighted average and upper confidence bound (which favors less-visited orientations). As with tracking, we send all visited orientations to the backend, which selects the best one per timestep. MadEye delivers 52.7\% higher median accuracies than this scheme, i.e., a 5.8$\times$ win. Unlike the schemes above, MAB does factor in workload accuracies in selecting orientations. However, its adaptation considers only historical efficacy (not current content), and scene dynamics have shifted by the time it updates its patterns. 

\begin{table}[t]
\footnotesize
\centering
\begin{tabular}{|l|l|l|} 
\hline
\textbf{System} & \textbf{Resource reduction} & \textbf{Median accuracy} \\\hline
\textbf{Chameleon~\cite{chameleon}} & 2.4$\times$ & 46.3\% \\ \hline 
\textbf{Chameleon + MadEye} & 2.4$\times$ & 56.1\% \\ \hline
\end{tabular}
\vspace{6pt}
\tightcaption{MadEye preserves resource savings of recent systems, while improving accuracy. Results use 15 fps, \{24 Mbps; 20 ms\}.}
\label{t:chameleon}
\end{table}

\para{Compatibility with other optimizations.} By focusing on previously un-tuned knobs (rotation and zoom) to boost accuracy, MadEye is largely compatible with prior efforts that optimize resource overheads. To illustrate this, we consider a variant of Chameleon~\cite{chameleon} that dynamically tunes pipeline knobs (resolution and frame rate) to lower network and backend inference resource costs without harming accuracy; we brute force selections per frame focused on the best fixed orientation. We then run MadEye atop the fps and resolution selections that Chameleon makes, sending the same amount of network data. As shown in Table~\ref{t:chameleon}, Chameleon lowers resource costs by 2.4$\times$ compared to the naive scheme that sends all frames at the highest resolution; MadEye preserves these efficiency wins, while increasing accuracy by 9.8\%.


\subsection{Deep Dive Results}
\label{ss:deepdive}


\para{Rotation speeds.} We evaluated the impact of camera rotation speed on MadEye's performance by considering values of \{200, 400, 500, infinite\}\degree~per second, a fixed network (\{24 Mbps; 20 ms\}), and 15 fps. Intuitively, accuracy grows as rotation speeds increase, e.g., jumping from 54.2\% to 64.9\% as rotation speed grows from 200 to 500\degree~per second. The reason is that faster rotations enable the exploration of additional orientations or, in rarer instances, additional transmissions. Importantly, benefits plateau since most queries (other than aggregate counting) are fully satisfied accuracy-wise as long as MadEye finds the best orientation at each timestep.




\para{Grid granularity.} To understand the effect of grid granularity (with other settings fixed), we focus on the pan dimension (since it is wider) and consider steps of \{15, 30, 45, 60\}\degree. Overall, MadEye's accuracy benefits shrink as grids become more fine-grained (with more orientations), e.g., median accuracies drop from 67.5\% to 51.8\% when pan steps drop from 45 to 15. This is because, although exploration in a time budget is governed by rotation speeds rather than grid granularity, the same distance (in \degree) of exploration will warrant more approximation model inference on more orientations, thereby shrinking each timestep's exploration budget.

\para{Overheads.} On MadEye's backend, the primary overheads are in initializing approximation models and continually sharing model updates with the camera. Across our workloads, we find median bootstrapping delays to be 27 mins (including labeling and initial fine-tuning). Downlink streaming consumes 3.2 Mbps for the median experiment. Recall that both overheads are mitigated by MadEye's fine-tuning strategy (\S\ref{ss:server}). On cameras, the main overheads are in selecting orientations to explore and running approximation models; for the median workload-video pair, per-timestep delays for each task were 17\micro$s$ and 6.7 ms for 15 fps and \{24 Mbps; 20 ms\}. The former benefits from pre-computed reachability analysis (\S\ref{ss:search}).

\begin{figure}[!t]
  \includegraphics[width=0.8\columnwidth]{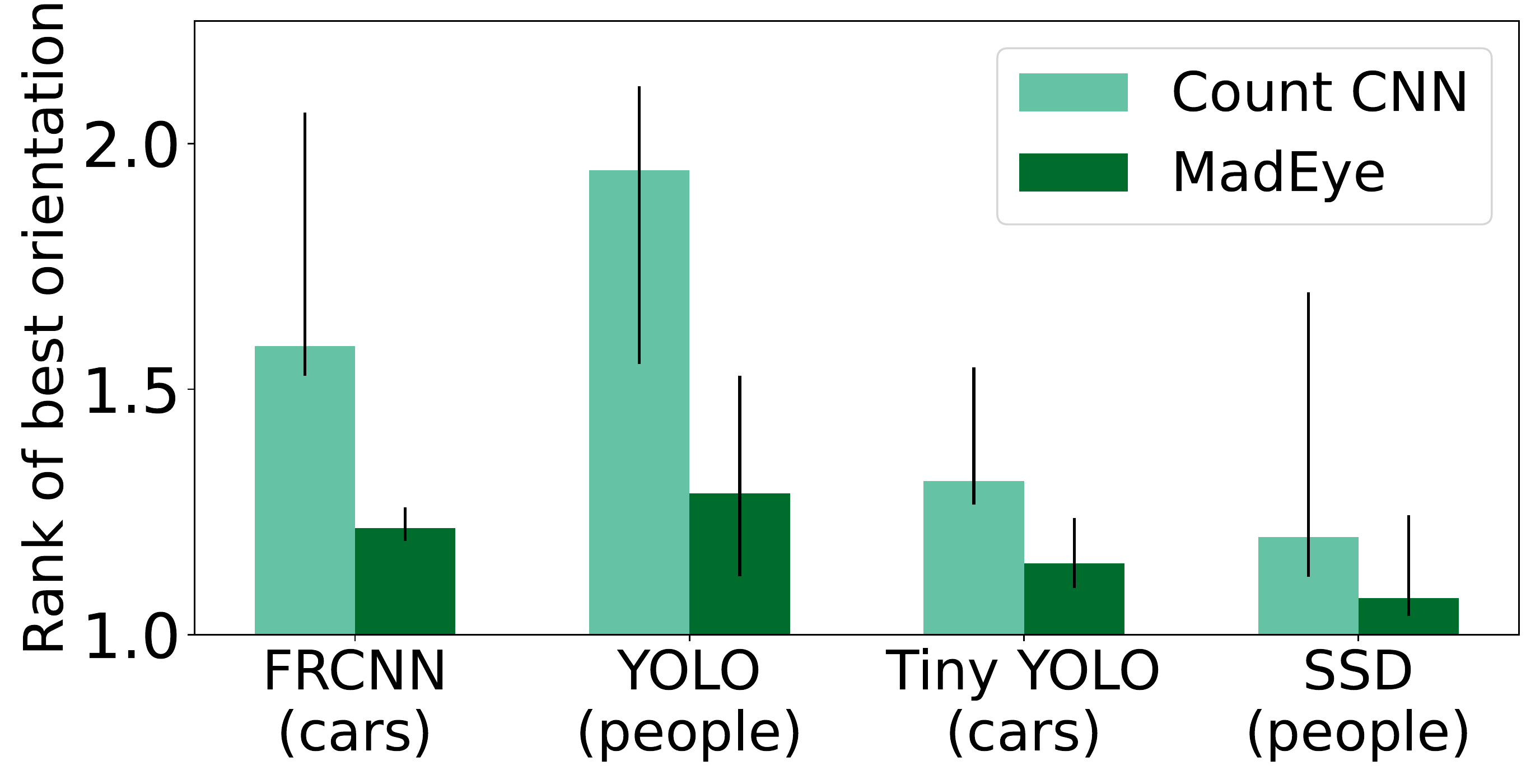}
 \tightcaption{Comparing different approximation model designs: MadEye's lightweight detection models and compressed counting models (Count CNN). Results use all videos, \{24 Mbps; 20 ms\}, 15 fps, and list median rank assigned to the best explored orientation at each timestep (error bars for 25-75th percentiles).}
  \vspace{6pt}
 \label{fig:ranks}
 \end{figure}



\para{Microbenchmarks.} MadEye's performance is governed by two main tasks: (1) ranking orientations with approximation models, and (2) selecting orientations to explore to find the best one(s) per timestep. For the former, Figure~\ref{fig:ranks} show that MadEye's approximation models assign median ranks of 1.1-1.3 to the best explored orientation at each timestep, significantly outperforming the variant that relies on counting directly on images. For the latter, for the median workload-video pair on \{24 Mbps; 20 ms\} and 15 fps, MadEye explores best orientation 89.3\% of the time, with 6.8\% of errors coming from our conservative zoom strategy (\S\ref{ss:search}).


\section{Related Work}
\label{s:related}

\para{Adapting video analytics knobs.}
VideoStorm~\cite{VideoStormNSDI2017} selects an input knob configuration (e.g., frame rate, resolution) per workload to lower resource costs and facilitate job scheduling on backend servers. Chameleon~\cite{chameleon} extends such configuration tuning to be adaptive in order to cope with ever-changing scene dynamics while keep resource costs low. As shown in \S\ref{ss:sota}, by focusing on tuning camera orientations (and not backend pipeline knobs), MadEye provides complementary benefits to these efforts, boosting accuracies while preserving the resource efficiency wins they bring. Other efforts focus on camera-side knobs as MadEye does. For example, CamTuner~\cite{camtuner} uses SARSA Reinforcement Learning to boost accuracy by automatically tuning capture knobs that cameras do not usually auto-adjust, e.g., brightness, contrast, and sharpness. AccMPEG~\cite{accmpeg} predicts the effects of macroblock encoding settings on server-side DNNs, and tunes encoding to maximize accuracy. MadEye shares the same goal as these efforts -- tune camera knobs to boost workload accuracy -- but focuses on complementary knobs, i.e., camera orientations.

\para{Frame filtering and result reuse.}
Many prior efforts exploit temporal redundancies in video data by filtering out frames for network transfer and processing, and reusing results accordingly~\cite{glimpse-sensys15, reducto, filterforward, vigil, potluck, cachier, deepcache, marlin, euphrates, foggycache,aqua,FreezingHotCloud19,ClipperNSDI17}. Spatula~\cite{spatula} extends this to multi-camera settings, selecting among cameras in a network.
These optimizations are logically similar to MadEye, which also aims to maximize accuracy per network usage. However, the techniques are largely complementary: filtering decisions could be made among explored orientations to maximize new content in transfers.

\para{Computation and network optimizations.} Several efforts seek to lower compute footprints either by identifying lightweight model variants~\cite{InFaaSATC21,mcdnn,hinton2015distilling,courbariaux2016binarized, jacob2018quantization,zhu2016trained,liu2017learning, blalock2020state}, sharing model layers during inference~\cite{gemel, MainstreamATC2018}, or using smarter job scheduling strategies~\cite{nexus, VideoStormNSDI2017}. Other systems target lower network overheads by intelligently compressing transmitted frames in a manner that is recoverable on the server or does not negatively impact accuracy~\cite{dds,cloudseg,dnn-blackbox}. MadEye is entirely complementary to both directions in that it solely focuses on judiciously selecting images (i.e., orientations) to process at any time for an application-provided model (which can be compressed); MadEye is agnostic to the way that selected frames are transmitted or processed on the backed. 

\para{Drone coordination.} Numerous efforts aim to adapt drone flight plans (and thus the content on-board cameras see) to maximize analytics accuracy or scene coverage~\cite{beecluster,visage,farmbeats,skyquery}. However, these systems focus on identifying events of interest (e.g., wildfires, objects) in a geographically dispersed area for a preset application. In contrast, MadEye focuses on tuning camera orientations for a single scene to cope with workload nuances and maximize accuracy.







\section{Conclusion}
\label{s:concl}

This paper presents MadEye, a system that continually tunes PTZ camera orientations to maximize accuracy for a given analytics workload and resource setting. Key to MadEye are a rapid algorithm that searches through the large space of orientations at each time, and a new, approximate transfer learning strategy that efficiently selects the most fruitful (accuracy-wise) orientations from those explored. Across many videos, workloads, and resource conditions, MadEye increases accuracy by 2.9-25.7\% for the same resource usage, or achieves the same accuracy with 2.0-3.7$\times$ lower resource costs.

\phantomsection
\label{EndOfPaper}
\balance
\Urlmuskip=0mu plus 1mu\relax
\bibliographystyle{ACM-Reference-Format}
\bibliography{references}


\begin{thebibliography}{115}


\ifx \showCODEN    \undefined \def \showCODEN     #1{\unskip}     \fi
\ifx \showDOI      \undefined \def \showDOI       #1{#1}\fi
\ifx \showISBNx    \undefined \def \showISBNx     #1{\unskip}     \fi
\ifx \showISBNxiii \undefined \def \showISBNxiii  #1{\unskip}     \fi
\ifx \showISSN     \undefined \def \showISSN      #1{\unskip}     \fi
\ifx \showLCCN     \undefined \def \showLCCN      #1{\unskip}     \fi
\ifx \shownote     \undefined \def \shownote      #1{#1}          \fi
\ifx \showarticletitle \undefined \def \showarticletitle #1{#1}   \fi
\ifx \showURL      \undefined \def \showURL       {\relax}        \fi
\providecommand\bibfield[2]{#2}
\providecommand\bibinfo[2]{#2}
\providecommand\natexlab[1]{#1}
\providecommand\showeprint[2][]{arXiv:#2}

\bibitem[\protect\citeauthoryear{??}{tra}{2020}]%
        {traffic-analysis}
 \bibinfo{year}{2020}\natexlab{}.
\newblock \bibinfo{title}{Video Analytics Traffic Study Creates Baseline for
  Change}.
\newblock
  \bibinfo{howpublished}{\url{https://www.govtech.com/analytics/Video-Analytics-Traffic-Study-Creates-Baseline-for-Change.html}}.
\newblock


\bibitem[\protect\citeauthoryear{??}{ten}{2021}]%
        {tensorRT}
 \bibinfo{year}{2021}\natexlab{}.
\newblock \bibinfo{title}{NVIDIA TensorRT}.
\newblock
\newblock
\urldef\tempurl%
\url{https://developer.nvidia.com/tensorrt}
\showURL{%
\tempurl}


\bibitem[\protect\citeauthoryear{??}{pyt}{2021}]%
        {pytorchyolov3}
 \bibinfo{year}{2021}\natexlab{}.
\newblock \bibinfo{title}{PyTorch-YOLOv3}.
\newblock
  \bibinfo{howpublished}{\url{https://github.com/eriklindernoren/PyTorch-YOLOv3}}.
\newblock


\bibitem[\protect\citeauthoryear{??}{eky}{2022}]%
        {ekya}
 \bibinfo{year}{2022}\natexlab{}.
\newblock \showarticletitle{Ekya: Continuous Learning of Video Analytics Models
  on Edge Compute Servers}. In \bibinfo{booktitle}{\emph{19th USENIX Symposium
  on Networked Systems Design and Implementation (NSDI 22)}}.
  \bibinfo{publisher}{USENIX Association}, \bibinfo{address}{Renton, WA},
  \bibinfo{pages}{119--135}.
\newblock
\showISBNx{978-1-939133-27-4}
\urldef\tempurl%
\url{https://www.usenix.org/conference/nsdi22/presentation/bhardwaj}
\showURL{%
\tempurl}


\bibitem[\protect\citeauthoryear{??}{for}{2022}]%
        {fortune_va_market_report}
 \bibinfo{year}{2022}\natexlab{}.
\newblock \bibinfo{title}{Video Analytics Market}.
\newblock
  \bibinfo{howpublished}{\url{https://www.fortunebusinessinsights.com/industry-reports/video-analytics-market-101114}}.
\newblock


\bibitem[\protect\citeauthoryear{Agarwal and Netravali}{Agarwal and
  Netravali}{[n.d.]}]%
        {boggart}
\bibfield{author}{\bibinfo{person}{Neil Agarwal} {and} \bibinfo{person}{Ravi
  Netravali}.} \bibinfo{year}{[n.d.]}\natexlab{}.
\newblock \showarticletitle{Boggart: Towards General-Purpose Acceleration of
  Retrospective Video Analytics}.
\newblock


\bibitem[\protect\citeauthoryear{Ananthanarayanan, Bahl, Cox, Crown, Nogbahi,
  and Shu}{Ananthanarayanan et~al\mbox{.}}{2019}]%
        {VideoKillerApp2}
\bibfield{author}{\bibinfo{person}{Ganesh Ananthanarayanan},
  \bibinfo{person}{Victor Bahl}, \bibinfo{person}{Landon Cox},
  \bibinfo{person}{Alex Crown}, \bibinfo{person}{Shadi Nogbahi}, {and}
  \bibinfo{person}{Yuanchao Shu}.} \bibinfo{year}{2019}\natexlab{}.
\newblock \showarticletitle{Video Analytics - Killer App for Edge Computing}.
  In \bibinfo{booktitle}{\emph{Proceedings of the 17th Annual International
  Conference on Mobile Systems, Applications, and Services}} (Seoul, Republic
  of Korea) \emph{(\bibinfo{series}{MobiSys '19})}.
  \bibinfo{publisher}{Association for Computing Machinery},
  \bibinfo{address}{New York, NY, USA}, \bibinfo{pages}{695--696}.
\newblock
\showISBNx{9781450366618}
\urldef\tempurl%
\url{https://doi.org/10.1145/3307334.3328589}
\showDOI{\tempurl}


\bibitem[\protect\citeauthoryear{Anderson}{Anderson}{[n.d.]}]%
        {smart-mall}
\bibfield{author}{\bibinfo{person}{Larry Anderson}.}
  \bibinfo{year}{[n.d.]}\natexlab{}.
\newblock \bibinfo{title}{Video Analytics Applications In Retail - Beyond
  Security}.
\newblock
  \bibinfo{howpublished}{\url{https://www.securityinformed.com/insights/co-2603-ga-co-2214-ga-co-1880-ga.16620.html/}}.
\newblock


\bibitem[\protect\citeauthoryear{Apicharttrisorn, Ran, Chen, Krishnamurthy, and
  Roy-Chowdhury}{Apicharttrisorn et~al\mbox{.}}{2019}]%
        {marlin}
\bibfield{author}{\bibinfo{person}{Kittipat Apicharttrisorn},
  \bibinfo{person}{Xukan Ran}, \bibinfo{person}{Jiasi Chen},
  \bibinfo{person}{Srikanth~V. Krishnamurthy}, {and} \bibinfo{person}{Amit~K.
  Roy-Chowdhury}.} \bibinfo{year}{2019}\natexlab{}.
\newblock \showarticletitle{Frugal Following: Power Thrifty Object Detection
  and Tracking for Mobile Augmented Reality}. In
  \bibinfo{booktitle}{\emph{Proceedings of the 17th Conference on Embedded
  Networked Sensor Systems}} (New York, New York)
  \emph{(\bibinfo{series}{SenSys '19})}. \bibinfo{publisher}{Association for
  Computing Machinery}, \bibinfo{address}{New York, NY, USA},
  \bibinfo{pages}{96–109}.
\newblock
\showISBNx{9781450369503}
\urldef\tempurl%
\url{https://doi.org/10.1145/3356250.3360044}
\showDOI{\tempurl}


\bibitem[\protect\citeauthoryear{Balakrishnan, Xiong, Xia, and
  Perona}{Balakrishnan et~al\mbox{.}}{2021}]%
        {balakrishnan2021towards}
\bibfield{author}{\bibinfo{person}{Guha Balakrishnan}, \bibinfo{person}{Yuanjun
  Xiong}, \bibinfo{person}{Wei Xia}, {and} \bibinfo{person}{Pietro Perona}.}
  \bibinfo{year}{2021}\natexlab{}.
\newblock \showarticletitle{Towards causal benchmarking of biasin face analysis
  algorithms}.
\newblock In \bibinfo{booktitle}{\emph{Deep Learning-Based Face Analytics}}.
  \bibinfo{publisher}{Springer}, \bibinfo{pages}{327--359}.
\newblock


\bibitem[\protect\citeauthoryear{Bastani, He, Jiang, Bastani, and
  Madden}{Bastani et~al\mbox{.}}{2021}]%
        {skyquery}
\bibfield{author}{\bibinfo{person}{Favyen Bastani}, \bibinfo{person}{Songtao
  He}, \bibinfo{person}{Ziwen Jiang}, \bibinfo{person}{Osbert Bastani}, {and}
  \bibinfo{person}{Sam Madden}.} \bibinfo{year}{2021}\natexlab{}.
\newblock \showarticletitle{SkyQuery: An Aerial Drone Video Sensing Platform}.
  In \bibinfo{booktitle}{\emph{Proceedings of the 2021 ACM SIGPLAN
  International Symposium on New Ideas, New Paradigms, and Reflections on
  Programming and Software}} (Chicago, IL, USA) \emph{(\bibinfo{series}{Onward!
  2021})}. \bibinfo{publisher}{Association for Computing Machinery},
  \bibinfo{address}{New York, NY, USA}, \bibinfo{pages}{56–67}.
\newblock
\showISBNx{9781450391108}
\urldef\tempurl%
\url{https://doi.org/10.1145/3486607.3486750}
\showDOI{\tempurl}


\bibitem[\protect\citeauthoryear{Bau, Zhou, Khosla, Oliva, and Torralba}{Bau
  et~al\mbox{.}}{2017}]%
        {bau2017network}
\bibfield{author}{\bibinfo{person}{David Bau}, \bibinfo{person}{Bolei Zhou},
  \bibinfo{person}{Aditya Khosla}, \bibinfo{person}{Aude Oliva}, {and}
  \bibinfo{person}{Antonio Torralba}.} \bibinfo{year}{2017}\natexlab{}.
\newblock \showarticletitle{Network dissection: Quantifying interpretability of
  deep visual representations}. In \bibinfo{booktitle}{\emph{Proceedings of the
  IEEE conference on computer vision and pattern recognition}}.
  \bibinfo{pages}{6541--6549}.
\newblock


\bibitem[\protect\citeauthoryear{Beaufort}{Beaufort}{[n.d.]}]%
        {ptz-web-dev-api}
\bibfield{author}{\bibinfo{person}{Francois Beaufort}.}
  \bibinfo{year}{[n.d.]}\natexlab{}.
\newblock \bibinfo{title}{{Control camera pan, tilt, and zoom -- Pan, tilt, and
  zoom features on cameras are finally controllable on the web.}}
\newblock \bibinfo{howpublished}{\url{https://web.dev/camera-pan-tilt-zoom/}}.
\newblock


\bibitem[\protect\citeauthoryear{Bender and Chekuri}{Bender and
  Chekuri}{1999}]%
        {bender1999performance}
\bibfield{author}{\bibinfo{person}{Michael~A Bender} {and}
  \bibinfo{person}{Chandra Chekuri}.} \bibinfo{year}{1999}\natexlab{}.
\newblock \showarticletitle{Performance guarantees for the TSP with a
  parameterized triangle inequality}. In \bibinfo{booktitle}{\emph{Algorithms
  and Data Structures: 6th International Workshop, WADS’99 Vancouver, Canada,
  August 11--14, 1999 Proceedings 6}}. Springer, \bibinfo{pages}{80--85}.
\newblock


\bibitem[\protect\citeauthoryear{Blalock, Ortiz, Frankle, and Guttag}{Blalock
  et~al\mbox{.}}{2020}]%
        {blalock2020state}
\bibfield{author}{\bibinfo{person}{Davis Blalock}, \bibinfo{person}{Jose
  Javier~Gonzalez Ortiz}, \bibinfo{person}{Jonathan Frankle}, {and}
  \bibinfo{person}{John Guttag}.} \bibinfo{year}{2020}\natexlab{}.
\newblock \showarticletitle{What is the state of neural network pruning?}
\newblock \bibinfo{journal}{\emph{arXiv preprint arXiv:2003.03033}}
  (\bibinfo{year}{2020}).
\newblock


\bibitem[\protect\citeauthoryear{{Business Research Insights}}{{Business
  Research Insights}}{[n.d.]}]%
        {global_ptz_report}
\bibfield{author}{\bibinfo{person}{{Business Research Insights}}.}
  \bibinfo{year}{[n.d.]}\natexlab{}.
\newblock \bibinfo{title}{Global PTZ Camera Market Research Report 2020}.
\newblock
  \bibinfo{howpublished}{\url{https://www.businessresearchinsights.com/market-reports/ptz-cameras-market-100130}}.
\newblock


\bibitem[\protect\citeauthoryear{Cai, Saberian, and Vasconcelos}{Cai
  et~al\mbox{.}}{2015}]%
        {pedestrian-detection-iccv15}
\bibfield{author}{\bibinfo{person}{Zhaowei Cai}, \bibinfo{person}{Mohammad
  Saberian}, {and} \bibinfo{person}{Nuno Vasconcelos}.}
  \bibinfo{year}{2015}\natexlab{}.
\newblock \showarticletitle{Learning Complexity-Aware Cascades for Deep
  Pedestrian Detection}. In \bibinfo{booktitle}{\emph{Proceedings of the 2015
  IEEE International Conference on Computer Vision (ICCV)}}
  \emph{(\bibinfo{series}{ICCV '15})}. \bibinfo{publisher}{IEEE Computer
  Society}, \bibinfo{address}{Washington, DC, USA},
  \bibinfo{pages}{3361--3369}.
\newblock
\showISBNx{978-1-4673-8391-2}
\urldef\tempurl%
\url{https://doi.org/10.1109/ICCV.2015.384}
\showDOI{\tempurl}


\bibitem[\protect\citeauthoryear{Canel, Kim, Zhou, Li, Lim, Andersen, Kaminsky,
  and Dulloor}{Canel et~al\mbox{.}}{2019}]%
        {filterforward}
\bibfield{author}{\bibinfo{person}{Christopher Canel}, \bibinfo{person}{Thomas
  Kim}, \bibinfo{person}{Giulio Zhou}, \bibinfo{person}{Conglong Li},
  \bibinfo{person}{Hyeontaek Lim}, \bibinfo{person}{David~G. Andersen},
  \bibinfo{person}{Michael Kaminsky}, {and} \bibinfo{person}{Subramanya~R.
  Dulloor}.} \bibinfo{year}{2019}\natexlab{}.
\newblock \showarticletitle{Scaling Video Analytics on Constrained Edge Nodes}.
  In \bibinfo{booktitle}{\emph{2nd SysML Conference}}.
\newblock


\bibitem[\protect\citeauthoryear{Cangialosi, Agarwal, Arun, Jiang, Narayana,
  Sarwate, and Netravali}{Cangialosi et~al\mbox{.}}{2022}]%
        {privid}
\bibfield{author}{\bibinfo{person}{Frank Cangialosi}, \bibinfo{person}{Neil
  Agarwal}, \bibinfo{person}{Venkat Arun}, \bibinfo{person}{Junchen Jiang},
  \bibinfo{person}{Srinivas Narayana}, \bibinfo{person}{Anand Sarwate}, {and}
  \bibinfo{person}{Ravi Netravali}.} \bibinfo{year}{2022}\natexlab{}.
\newblock \showarticletitle{Privid: Practical, Privacy-Preserving Video
  Analytics Queries}. In \bibinfo{booktitle}{\emph{Proceedings of the 19th
  USENIX Conference on Networked Systems Design and Implementation}} (Renton,
  WA, USA) \emph{(\bibinfo{series}{NSDI'22})}. \bibinfo{publisher}{USENIX
  Association}, \bibinfo{address}{Berkeley, CA, USA}.
\newblock


\bibitem[\protect\citeauthoryear{Cassel}{Cassel}{[n.d.]}]%
        {are-we-ready-for-ai-powered-security-cameras}
\bibfield{author}{\bibinfo{person}{David Cassel}.}
  \bibinfo{year}{[n.d.]}\natexlab{}.
\newblock \bibinfo{title}{Are We Ready for AI-Powered Security Cameras?}
\newblock
  \bibinfo{howpublished}{\url{https://thenewstack.io/are-we-ready-for-ai-powered-security-cameras/}}.
\newblock


\bibitem[\protect\citeauthoryear{Chen, Ravindranath, Deng, Bahl, and
  Balakrishnan}{Chen et~al\mbox{.}}{2015}]%
        {glimpse-sensys15}
\bibfield{author}{\bibinfo{person}{Tiffany Yu-Han Chen}, \bibinfo{person}{Lenin
  Ravindranath}, \bibinfo{person}{Shuo Deng}, \bibinfo{person}{Paramvir Bahl},
  {and} \bibinfo{person}{Hari Balakrishnan}.} \bibinfo{year}{2015}\natexlab{}.
\newblock \showarticletitle{Glimpse: Continuous, Real-Time Object Recognition
  on Mobile Devices}. In \bibinfo{booktitle}{\emph{Proceedings of the 13th ACM
  Conference on Embedded Networked Sensor Systems}}. \bibinfo{pages}{155--168}.
\newblock


\bibitem[\protect\citeauthoryear{Collins}{Collins}{[n.d.]}]%
        {soccer_tracking}
\bibfield{author}{\bibinfo{person}{Mary Collins}.}
  \bibinfo{year}{[n.d.]}\natexlab{}.
\newblock \bibinfo{title}{{The Hudl Algorithm: Turning Video into Player
  Tracking Data}}.
\newblock
  \bibinfo{howpublished}{{\url{https://www.maryecollins.com/hudl-tracking}}}.
\newblock


\bibitem[\protect\citeauthoryear{Courbariaux, Hubara, Soudry, El-Yaniv, and
  Bengio}{Courbariaux et~al\mbox{.}}{2016}]%
        {courbariaux2016binarized}
\bibfield{author}{\bibinfo{person}{Matthieu Courbariaux}, \bibinfo{person}{Itay
  Hubara}, \bibinfo{person}{Daniel Soudry}, \bibinfo{person}{Ran El-Yaniv},
  {and} \bibinfo{person}{Yoshua Bengio}.} \bibinfo{year}{2016}\natexlab{}.
\newblock \showarticletitle{Binarized neural networks: Training deep neural
  networks with weights and activations constrained to+ 1 or-1}.
\newblock \bibinfo{journal}{\emph{arXiv preprint arXiv:1602.02830}}
  (\bibinfo{year}{2016}).
\newblock


\bibitem[\protect\citeauthoryear{Crankshaw, Wang, Zhou, Franklin, Gonzalez, and
  Stoica}{Crankshaw et~al\mbox{.}}{2017}]%
        {ClipperNSDI17}
\bibfield{author}{\bibinfo{person}{Daniel Crankshaw}, \bibinfo{person}{Xin
  Wang}, \bibinfo{person}{Guilio Zhou}, \bibinfo{person}{Michael~J. Franklin},
  \bibinfo{person}{Joseph~E. Gonzalez}, {and} \bibinfo{person}{Ion Stoica}.}
  \bibinfo{year}{2017}\natexlab{}.
\newblock \showarticletitle{Clipper: A {Low-Latency} Online Prediction Serving
  System}. In \bibinfo{booktitle}{\emph{14th USENIX Symposium on Networked
  Systems Design and Implementation (NSDI 17)}}. \bibinfo{publisher}{USENIX
  Association}, \bibinfo{address}{Boston, MA}, \bibinfo{pages}{613--627}.
\newblock
\showISBNx{978-1-931971-37-9}
\urldef\tempurl%
\url{https://www.usenix.org/conference/nsdi17/technical-sessions/presentation/crankshaw}
\showURL{%
\tempurl}


\bibitem[\protect\citeauthoryear{Dalal and Triggs}{Dalal and Triggs}{2005}]%
        {dalal2005histograms}
\bibfield{author}{\bibinfo{person}{Navneet Dalal} {and} \bibinfo{person}{Bill
  Triggs}.} \bibinfo{year}{2005}\natexlab{}.
\newblock \showarticletitle{Histograms of oriented gradients for human
  detection}. In \bibinfo{booktitle}{\emph{2005 IEEE computer society
  conference on computer vision and pattern recognition (CVPR'05)}},
  Vol.~\bibinfo{volume}{1}. Ieee, \bibinfo{pages}{886--893}.
\newblock


\bibitem[\protect\citeauthoryear{Datondji, Dupuis, Subirats, and
  Vasseur}{Datondji et~al\mbox{.}}{2016}]%
        {traff2}
\bibfield{author}{\bibinfo{person}{Sokemi Rene~Emmanuel Datondji},
  \bibinfo{person}{Yohan Dupuis}, \bibinfo{person}{Peggy Subirats}, {and}
  \bibinfo{person}{Pascal Vasseur}.} \bibinfo{year}{2016}\natexlab{}.
\newblock \showarticletitle{A Survey of Vision-Based Traffic Monitoring of Road
  Intersections}.
\newblock \bibinfo{journal}{\emph{Trans. Intell. Transport. Sys.}}
  \bibinfo{volume}{17}, \bibinfo{number}{10} (\bibinfo{date}{Oct.}
  \bibinfo{year}{2016}), \bibinfo{pages}{2681–2698}.
\newblock
\showISSN{1524-9050}
\urldef\tempurl%
\url{https://doi.org/10.1109/TITS.2016.2530146}
\showDOI{\tempurl}


\bibitem[\protect\citeauthoryear{Drolia, Guo, Tan, Gandhi, and
  Narasimhan}{Drolia et~al\mbox{.}}{2017}]%
        {cachier}
\bibfield{author}{\bibinfo{person}{Utsav Drolia}, \bibinfo{person}{Katherine
  Guo}, \bibinfo{person}{Jiaqi Tan}, \bibinfo{person}{Rajeev Gandhi}, {and}
  \bibinfo{person}{Priya Narasimhan}.} \bibinfo{year}{2017}\natexlab{}.
\newblock \showarticletitle{Cachier: Edge-Caching for Recognition
  Applications}. In \bibinfo{booktitle}{\emph{2017 IEEE 37th International
  Conference on Distributed Computing Systems (ICDCS)}}.
  \bibinfo{pages}{276--286}.
\newblock
\urldef\tempurl%
\url{https://doi.org/10.1109/ICDCS.2017.94}
\showDOI{\tempurl}


\bibitem[\protect\citeauthoryear{Du, Pervaiz, Yuan, Chowdhery, Zhang, Hoffmann,
  and Jiang}{Du et~al\mbox{.}}{2020a}]%
        {dds}
\bibfield{author}{\bibinfo{person}{Kuntai Du}, \bibinfo{person}{Ahsan Pervaiz},
  \bibinfo{person}{Xin Yuan}, \bibinfo{person}{Aakanksha Chowdhery},
  \bibinfo{person}{Qizheng Zhang}, \bibinfo{person}{Henry Hoffmann}, {and}
  \bibinfo{person}{Junchen Jiang}.} \bibinfo{year}{2020}\natexlab{a}.
\newblock \showarticletitle{Server-Driven Video Streaming for Deep Learning
  Inference}. In \bibinfo{booktitle}{\emph{Proceedings of the Annual Conference
  of the ACM Special Interest Group on Data Communication on the Applications,
  Technologies, Architectures, and Protocols for Computer Communication}}
  (Virtual Event, USA) \emph{(\bibinfo{series}{SIGCOMM '20})}.
  \bibinfo{publisher}{Association for Computing Machinery},
  \bibinfo{address}{New York, NY, USA}, \bibinfo{pages}{557–570}.
\newblock
\showISBNx{9781450379557}
\urldef\tempurl%
\url{https://doi.org/10.1145/3387514.3405887}
\showDOI{\tempurl}


\bibitem[\protect\citeauthoryear{Du, Zhang, Arapin, Wang, Xia, and Jiang}{Du
  et~al\mbox{.}}{2022}]%
        {accmpeg}
\bibfield{author}{\bibinfo{person}{Kuntai Du}, \bibinfo{person}{Qizheng Zhang},
  \bibinfo{person}{Anton Arapin}, \bibinfo{person}{Haodong Wang},
  \bibinfo{person}{Zhengxu Xia}, {and} \bibinfo{person}{Junchen Jiang}.}
  \bibinfo{year}{2022}\natexlab{}.
\newblock \showarticletitle{AccMPEG: Optimizing Video Encoding for Accurate
  Video Analytics}. In \bibinfo{booktitle}{\emph{Proceedings of Machine
  Learning and Systems}}, \bibfield{editor}{\bibinfo{person}{D.~Marculescu},
  \bibinfo{person}{Y.~Chi}, {and} \bibinfo{person}{C.~Wu}} (Eds.),
  Vol.~\bibinfo{volume}{4}. \bibinfo{pages}{450--466}.
\newblock
\urldef\tempurl%
\url{https://proceedings.mlsys.org/paper/2022/file/98f13708210194c475687be6106a3b84-Paper.pdf}
\showURL{%
\tempurl}


\bibitem[\protect\citeauthoryear{Du, Yang, Zou, and Hu}{Du
  et~al\mbox{.}}{2020b}]%
        {du2020fairness}
\bibfield{author}{\bibinfo{person}{Mengnan Du}, \bibinfo{person}{Fan Yang},
  \bibinfo{person}{Na Zou}, {and} \bibinfo{person}{Xia Hu}.}
  \bibinfo{year}{2020}\natexlab{b}.
\newblock \showarticletitle{Fairness in deep learning: A computational
  perspective}.
\newblock \bibinfo{journal}{\emph{IEEE Intelligent Systems}}
  \bibinfo{volume}{36}, \bibinfo{number}{4} (\bibinfo{year}{2020}),
  \bibinfo{pages}{25--34}.
\newblock


\bibitem[\protect\citeauthoryear{E, He, Li, and Liu}{E et~al\mbox{.}}{2023}]%
        {wisecam}
\bibfield{author}{\bibinfo{person}{Jinlong E}, \bibinfo{person}{Lin He},
  \bibinfo{person}{Zhenhua Li}, {and} \bibinfo{person}{Yunhao Liu}.}
  \bibinfo{year}{2023}\natexlab{}.
\newblock \showarticletitle{WiseCam: Wisely Tuning Wireless Pan-Tilt Cameras
  for Cost-Effective Moving Object Tracking}. In \bibinfo{booktitle}{\emph{IEEE
  INFOCOM 2023-IEEE Conference on Computer Communications}}. IEEE.
\newblock


\bibitem[\protect\citeauthoryear{Emmons, Fouladi, Ananthanarayanan,
  Venkataraman, Savarese, and Winstein}{Emmons et~al\mbox{.}}{2019}]%
        {dnn-blackbox}
\bibfield{author}{\bibinfo{person}{John Emmons}, \bibinfo{person}{Sadjad
  Fouladi}, \bibinfo{person}{Ganesh Ananthanarayanan},
  \bibinfo{person}{Shivaram Venkataraman}, \bibinfo{person}{Silvio Savarese},
  {and} \bibinfo{person}{Keith Winstein}.} \bibinfo{year}{2019}\natexlab{}.
\newblock \showarticletitle{Cracking Open the DNN Black-Box: Video Analytics
  with DNNs across the Camera-Cloud Boundary}. In
  \bibinfo{booktitle}{\emph{Proceedings of the 2019 Workshop on Hot Topics in
  Video Analytics and Intelligent Edges}} (Los Cabos, Mexico)
  \emph{(\bibinfo{series}{HotEdgeVideo'19})}. \bibinfo{publisher}{Association
  for Computing Machinery}, \bibinfo{address}{New York, NY, USA},
  \bibinfo{pages}{27–32}.
\newblock
\showISBNx{9781450369282}
\urldef\tempurl%
\url{https://doi.org/10.1145/3349614.3356023}
\showDOI{\tempurl}


\bibitem[\protect\citeauthoryear{Everingham, Gool, Williams, Winn, and
  Zisserman}{Everingham et~al\mbox{.}}{2010}]%
        {Everingham:2010:PVO:1747084.1747104}
\bibfield{author}{\bibinfo{person}{Mark Everingham}, \bibinfo{person}{Luc
  Gool}, \bibinfo{person}{Christopher~K. Williams}, \bibinfo{person}{John
  Winn}, {and} \bibinfo{person}{Andrew Zisserman}.}
  \bibinfo{year}{2010}\natexlab{}.
\newblock \showarticletitle{The Pascal Visual Object Classes (VOC) Challenge}.
\newblock \bibinfo{journal}{\emph{Int. J. Comput. Vision}}
  \bibinfo{volume}{88}, \bibinfo{number}{2} (\bibinfo{date}{June}
  \bibinfo{year}{2010}), \bibinfo{pages}{303--338}.
\newblock
\showISSN{0920-5691}
\urldef\tempurl%
\url{https://doi.org/10.1007/s11263-009-0275-4}
\showDOI{\tempurl}


\bibitem[\protect\citeauthoryear{Fouladi, Emmons, Orbay, Wu, Wahby, and
  Winstein}{Fouladi et~al\mbox{.}}{2018}]%
        {salsify}
\bibfield{author}{\bibinfo{person}{Sadjad Fouladi}, \bibinfo{person}{John
  Emmons}, \bibinfo{person}{Emre Orbay}, \bibinfo{person}{Catherine Wu},
  \bibinfo{person}{Riad~S. Wahby}, {and} \bibinfo{person}{Keith Winstein}.}
  \bibinfo{year}{2018}\natexlab{}.
\newblock \showarticletitle{Salsify: Low-Latency Network Video through Tighter
  Integration between a Video Codec and a Transport Protocol}. In
  \bibinfo{booktitle}{\emph{Proceedings of the 15th USENIX Conference on
  Networked Systems Design and Implementation}} (Renton, WA, USA)
  \emph{(\bibinfo{series}{NSDI'18})}. \bibinfo{publisher}{USENIX Association},
  \bibinfo{address}{USA}, \bibinfo{pages}{267–282}.
\newblock
\showISBNx{9781931971430}


\bibitem[\protect\citeauthoryear{Ghodgaonkar, Chakraborty, Banna, Allcroft,
  Metwaly, Bordwell, Kimura, Zhao, Goel, Tung, et~al\mbox{.}}{Ghodgaonkar
  et~al\mbox{.}}{2020}]%
        {analyzing-social-distancing}
\bibfield{author}{\bibinfo{person}{Isha Ghodgaonkar},
  \bibinfo{person}{Subhankar Chakraborty}, \bibinfo{person}{Vishnu Banna},
  \bibinfo{person}{Shane Allcroft}, \bibinfo{person}{Mohammed Metwaly},
  \bibinfo{person}{Fischer Bordwell}, \bibinfo{person}{Kohsuke Kimura},
  \bibinfo{person}{Xinxin Zhao}, \bibinfo{person}{Abhinav Goel},
  \bibinfo{person}{Caleb Tung}, {et~al\mbox{.}}}
  \bibinfo{year}{2020}\natexlab{}.
\newblock \showarticletitle{Analyzing Worldwide Social Distancing through
  Large-Scale Computer Vision}.
\newblock \bibinfo{journal}{\emph{arXiv preprint arXiv:2008.12363}}
  (\bibinfo{year}{2020}).
\newblock


\bibitem[\protect\citeauthoryear{{Grand View Research}}{{Grand View
  Research}}{[n.d.]}]%
        {sports}
\bibfield{author}{\bibinfo{person}{{Grand View Research}}.}
  \bibinfo{year}{[n.d.]}\natexlab{}.
\newblock \bibinfo{title}{{Global Sports Analytics Market Size Report,
  2021-2028}}.
\newblock
  \bibinfo{howpublished}{\url{https://www.grandviewresearch.com/industry-analysis/sports-analytics-market}}.
\newblock


\bibitem[\protect\citeauthoryear{Guo, Hu, Li, and Hu}{Guo
  et~al\mbox{.}}{2018}]%
        {foggycache}
\bibfield{author}{\bibinfo{person}{Peizhen Guo}, \bibinfo{person}{Bo Hu},
  \bibinfo{person}{Rui Li}, {and} \bibinfo{person}{Wenjun Hu}.}
  \bibinfo{year}{2018}\natexlab{}.
\newblock \showarticletitle{FoggyCache: Cross-Device Approximate Computation
  Reuse}. In \bibinfo{booktitle}{\emph{Proceedings of the 24th Annual
  International Conference on Mobile Computing and Networking}} (New Delhi,
  India) \emph{(\bibinfo{series}{MobiCom '18})}.
  \bibinfo{publisher}{Association for Computing Machinery},
  \bibinfo{address}{New York, NY, USA}, \bibinfo{pages}{19–34}.
\newblock
\showISBNx{9781450359030}
\urldef\tempurl%
\url{https://doi.org/10.1145/3241539.3241557}
\showDOI{\tempurl}


\bibitem[\protect\citeauthoryear{Guo and Hu}{Guo and Hu}{2018}]%
        {potluck}
\bibfield{author}{\bibinfo{person}{Peizhen Guo} {and} \bibinfo{person}{Wenjun
  Hu}.} \bibinfo{year}{2018}\natexlab{}.
\newblock \showarticletitle{Potluck: Cross-Application Approximate
  Deduplication for Computation-Intensive Mobile Applications}.
\newblock \bibinfo{journal}{\emph{SIGPLAN Not.}} \bibinfo{volume}{53},
  \bibinfo{number}{2} (\bibinfo{date}{mar} \bibinfo{year}{2018}),
  \bibinfo{pages}{271–284}.
\newblock
\showISSN{0362-1340}
\urldef\tempurl%
\url{https://doi.org/10.1145/3296957.3173185}
\showDOI{\tempurl}


\bibitem[\protect\citeauthoryear{Han, Shen, Philipose, Agarwal, Wolman, and
  Krishnamurthy}{Han et~al\mbox{.}}{2016}]%
        {mcdnn}
\bibfield{author}{\bibinfo{person}{Seungyeop Han}, \bibinfo{person}{Haichen
  Shen}, \bibinfo{person}{Matthai Philipose}, \bibinfo{person}{Sharad Agarwal},
  \bibinfo{person}{Alec Wolman}, {and} \bibinfo{person}{Arvind Krishnamurthy}.}
  \bibinfo{year}{2016}\natexlab{}.
\newblock \showarticletitle{MCDNN: An Approximation-Based Execution Framework
  for Deep Stream Processing Under Resource Constraints}. In
  \bibinfo{booktitle}{\emph{Proceedings of the 14th Annual International
  Conference on Mobile Systems, Applications, and Services}} (Singapore,
  Singapore) \emph{(\bibinfo{series}{MobiSys '16})}.
  \bibinfo{publisher}{Association for Computing Machinery},
  \bibinfo{address}{New York, NY, USA}, \bibinfo{pages}{123–136}.
\newblock
\showISBNx{9781450342698}
\urldef\tempurl%
\url{https://doi.org/10.1145/2906388.2906396}
\showDOI{\tempurl}


\bibitem[\protect\citeauthoryear{He, Gkioxari, Doll{\'{a}}r, and Girshick}{He
  et~al\mbox{.}}{2017}]%
        {maskrcnn}
\bibfield{author}{\bibinfo{person}{Kaiming He}, \bibinfo{person}{Georgia
  Gkioxari}, \bibinfo{person}{Piotr Doll{\'{a}}r}, {and}
  \bibinfo{person}{Ross~B. Girshick}.} \bibinfo{year}{2017}\natexlab{}.
\newblock \showarticletitle{Mask {R-CNN}}.
\newblock \bibinfo{journal}{\emph{CoRR}}  \bibinfo{volume}{abs/1703.06870}
  (\bibinfo{year}{2017}).
\newblock
\showeprint[arxiv]{1703.06870}
\urldef\tempurl%
\url{http://arxiv.org/abs/1703.06870}
\showURL{%
\tempurl}


\bibitem[\protect\citeauthoryear{He, Bastani, Balasingam, Gopalakrishna, Jiang,
  Alizadeh, Balakrishnan, Cafarella, Kraska, and Madden}{He
  et~al\mbox{.}}{2020}]%
        {beecluster}
\bibfield{author}{\bibinfo{person}{Songtao He}, \bibinfo{person}{Favyen
  Bastani}, \bibinfo{person}{Arjun Balasingam}, \bibinfo{person}{Karthik
  Gopalakrishna}, \bibinfo{person}{Ziwen Jiang}, \bibinfo{person}{Mohammad
  Alizadeh}, \bibinfo{person}{Hari Balakrishnan}, \bibinfo{person}{Michael
  Cafarella}, \bibinfo{person}{Tim Kraska}, {and} \bibinfo{person}{Sam
  Madden}.} \bibinfo{year}{2020}\natexlab{}.
\newblock \showarticletitle{BeeCluster: Drone Orchestration via Predictive
  Optimization}. In \bibinfo{booktitle}{\emph{Proceedings of the 18th
  International Conference on Mobile Systems, Applications, and Services}}
  (Toronto, Ontario, Canada) \emph{(\bibinfo{series}{MobiSys '20})}.
  \bibinfo{publisher}{Association for Computing Machinery},
  \bibinfo{address}{New York, NY, USA}, \bibinfo{pages}{299–311}.
\newblock
\showISBNx{9781450379540}
\urldef\tempurl%
\url{https://doi.org/10.1145/3386901.3388912}
\showDOI{\tempurl}


\bibitem[\protect\citeauthoryear{Held and Karp}{Held and Karp}{1970}]%
        {held1970traveling}
\bibfield{author}{\bibinfo{person}{Michael Held} {and}
  \bibinfo{person}{Richard~M Karp}.} \bibinfo{year}{1970}\natexlab{}.
\newblock \showarticletitle{The traveling-salesman problem and minimum spanning
  trees}.
\newblock \bibinfo{journal}{\emph{Operations Research}} \bibinfo{volume}{18},
  \bibinfo{number}{6} (\bibinfo{year}{1970}), \bibinfo{pages}{1138--1162}.
\newblock


\bibitem[\protect\citeauthoryear{Hinton, Vinyals, and Dean}{Hinton
  et~al\mbox{.}}{2015}]%
        {hinton2015distilling}
\bibfield{author}{\bibinfo{person}{Geoffrey Hinton}, \bibinfo{person}{Oriol
  Vinyals}, {and} \bibinfo{person}{Jeff Dean}.}
  \bibinfo{year}{2015}\natexlab{}.
\newblock \showarticletitle{Distilling the knowledge in a neural network}.
\newblock \bibinfo{journal}{\emph{arXiv preprint arXiv:1503.02531}}
  (\bibinfo{year}{2015}).
\newblock


\bibitem[\protect\citeauthoryear{{Honey Optics}}{{Honey Optics}}{[n.d.]}]%
        {ptz_camera_cost}
\bibfield{author}{\bibinfo{person}{{Honey Optics}}.}
  \bibinfo{year}{[n.d.]}\natexlab{}.
\newblock \bibinfo{title}{{ How much do PTZ cameras cost?}}
\newblock
  \bibinfo{howpublished}{\url{https://honeyoptics.com/how-much-do-ptz-cameras-cost/}}.
\newblock


\bibitem[\protect\citeauthoryear{Huang, Rathod, Sun, Zhu, Korattikara, Fathi,
  Fischer, Wojna, Song, Guadarrama, et~al\mbox{.}}{Huang et~al\mbox{.}}{2017}]%
        {huang2017speed}
\bibfield{author}{\bibinfo{person}{Jonathan Huang}, \bibinfo{person}{Vivek
  Rathod}, \bibinfo{person}{Chen Sun}, \bibinfo{person}{Menglong Zhu},
  \bibinfo{person}{Anoop Korattikara}, \bibinfo{person}{Alireza Fathi},
  \bibinfo{person}{Ian Fischer}, \bibinfo{person}{Zbigniew Wojna},
  \bibinfo{person}{Yang Song}, \bibinfo{person}{Sergio Guadarrama},
  {et~al\mbox{.}}} \bibinfo{year}{2017}\natexlab{}.
\newblock \showarticletitle{Speed/accuracy trade-offs for modern convolutional
  object detectors}. In \bibinfo{booktitle}{\emph{Proceedings of the IEEE
  conference on computer vision and pattern recognition}}.
  \bibinfo{pages}{7310--7311}.
\newblock


\bibitem[\protect\citeauthoryear{{HuddleCamHD}}{{HuddleCamHD}}{[n.d.]}]%
        {understainding-eptz-ptz}
\bibfield{author}{\bibinfo{person}{{HuddleCamHD}}.}
  \bibinfo{year}{[n.d.]}\natexlab{}.
\newblock \bibinfo{title}{{Understanding the difference between EPTZ and PTZ
  }}.
\newblock \bibinfo{howpublished}{\url{https://huddlecamhd.com/eptz-and-ptz/}}.
\newblock


\bibitem[\protect\citeauthoryear{Jacob, Kligys, Chen, Zhu, Tang, Howard, Adam,
  and Kalenichenko}{Jacob et~al\mbox{.}}{2018}]%
        {jacob2018quantization}
\bibfield{author}{\bibinfo{person}{Benoit Jacob}, \bibinfo{person}{Skirmantas
  Kligys}, \bibinfo{person}{Bo Chen}, \bibinfo{person}{Menglong Zhu},
  \bibinfo{person}{Matthew Tang}, \bibinfo{person}{Andrew Howard},
  \bibinfo{person}{Hartwig Adam}, {and} \bibinfo{person}{Dmitry Kalenichenko}.}
  \bibinfo{year}{2018}\natexlab{}.
\newblock \showarticletitle{Quantization and training of neural networks for
  efficient integer-arithmetic-only inference}. In
  \bibinfo{booktitle}{\emph{Proceedings of the IEEE Conference on Computer
  Vision and Pattern Recognition}}. \bibinfo{pages}{2704--2713}.
\newblock


\bibitem[\protect\citeauthoryear{Jain, Zhang, Zhou, Ananthanarayanan, Jiang,
  Shu, Bahl, and Gonzalez}{Jain et~al\mbox{.}}{2020}]%
        {spatula}
\bibfield{author}{\bibinfo{person}{Samvit Jain}, \bibinfo{person}{Xun Zhang},
  \bibinfo{person}{Yuhao Zhou}, \bibinfo{person}{Ganesh Ananthanarayanan},
  \bibinfo{person}{Junchen Jiang}, \bibinfo{person}{Yuanchao Shu},
  \bibinfo{person}{Victor Bahl}, {and} \bibinfo{person}{Joseph Gonzalez}.}
  \bibinfo{year}{2020}\natexlab{}.
\newblock \showarticletitle{Spatula: Efficient cross-camera video analytics on
  large camera networks}. In \bibinfo{booktitle}{\emph{ACM/IEEE Symposium on
  Edge Computing (SEC 2020)}}.
\newblock


\bibitem[\protect\citeauthoryear{Jha, Li, Noghabi, Ranganathan, Kumar, Nelson,
  Toelle, Sinha, Chandra, and Badam}{Jha et~al\mbox{.}}{2021}]%
        {visage}
\bibfield{author}{\bibinfo{person}{Sagar Jha}, \bibinfo{person}{Youjie Li},
  \bibinfo{person}{Shadi Noghabi}, \bibinfo{person}{Vaishnavi Ranganathan},
  \bibinfo{person}{Peeyush Kumar}, \bibinfo{person}{Andrew Nelson},
  \bibinfo{person}{Michael Toelle}, \bibinfo{person}{Sudipta Sinha},
  \bibinfo{person}{Ranveer Chandra}, {and} \bibinfo{person}{Anirudh Badam}.}
  \bibinfo{year}{2021}\natexlab{}.
\newblock \showarticletitle{Visage: Enabling Timely Analytics for Drone
  Imagery}. In \bibinfo{booktitle}{\emph{Proceedings of the 27th Annual
  International Conference on Mobile Computing and Networking}} (New Orleans,
  Louisiana) \emph{(\bibinfo{series}{MobiCom '21})}.
  \bibinfo{publisher}{Association for Computing Machinery},
  \bibinfo{address}{New York, NY, USA}, \bibinfo{pages}{789–803}.
\newblock
\showISBNx{9781450383424}
\urldef\tempurl%
\url{https://doi.org/10.1145/3447993.3483273}
\showDOI{\tempurl}


\bibitem[\protect\citeauthoryear{Jiang, Wong, Canel, Tang, Misra, Kaminsky,
  Kozuch, Pillai, Andersen, and Ganger}{Jiang et~al\mbox{.}}{2018c}]%
        {MainstreamATC2018}
\bibfield{author}{\bibinfo{person}{Angela~H. Jiang}, \bibinfo{person}{Daniel
  L.-K. Wong}, \bibinfo{person}{Christopher Canel}, \bibinfo{person}{Lilia
  Tang}, \bibinfo{person}{Ishan Misra}, \bibinfo{person}{Michael Kaminsky},
  \bibinfo{person}{Michael~A. Kozuch}, \bibinfo{person}{Padmanabhan Pillai},
  \bibinfo{person}{David~G. Andersen}, {and} \bibinfo{person}{Gregory~R.
  Ganger}.} \bibinfo{year}{2018}\natexlab{c}.
\newblock \showarticletitle{Mainstream: Dynamic Stem-Sharing for Multi-Tenant
  Video Processing}. In \bibinfo{booktitle}{\emph{2018 {USENIX} Annual
  Technical Conference ({USENIX} {ATC} 18)}}. \bibinfo{publisher}{{USENIX}
  Association}, \bibinfo{address}{Boston, MA}, \bibinfo{pages}{29--42}.
\newblock
\showISBNx{978-1-931971-44-7}
\urldef\tempurl%
\url{https://www.usenix.org/conference/atc18/presentation/jiang}
\showURL{%
\tempurl}


\bibitem[\protect\citeauthoryear{Jiang, Ananthanarayanan, Bodik, Sen, and
  Stoica}{Jiang et~al\mbox{.}}{2018a}]%
        {chameleon-sigcomm18}
\bibfield{author}{\bibinfo{person}{Junchen Jiang}, \bibinfo{person}{Ganesh
  Ananthanarayanan}, \bibinfo{person}{Peter Bodik}, \bibinfo{person}{Siddhartha
  Sen}, {and} \bibinfo{person}{Ion Stoica}.} \bibinfo{year}{2018}\natexlab{a}.
\newblock \showarticletitle{Chameleon: Scalable Adaptation of Video Analytics}.
  In \bibinfo{booktitle}{\emph{Proceedings of the 2018 Conference of the ACM
  Special Interest Group on Data Communication}} (Budapest, Hungary)
  \emph{(\bibinfo{series}{SIGCOMM '18})}. \bibinfo{publisher}{ACM},
  \bibinfo{address}{New York, NY, USA}, \bibinfo{pages}{253--266}.
\newblock
\showISBNx{978-1-4503-5567-4}
\urldef\tempurl%
\url{https://doi.org/10.1145/3230543.3230574}
\showDOI{\tempurl}


\bibitem[\protect\citeauthoryear{Jiang, Ananthanarayanan, Bodik, Sen, and
  Stoica}{Jiang et~al\mbox{.}}{2018b}]%
        {chameleon}
\bibfield{author}{\bibinfo{person}{Junchen Jiang}, \bibinfo{person}{Ganesh
  Ananthanarayanan}, \bibinfo{person}{Peter Bodik}, \bibinfo{person}{Siddhartha
  Sen}, {and} \bibinfo{person}{Ion Stoica}.} \bibinfo{year}{2018}\natexlab{b}.
\newblock \showarticletitle{Chameleon: Scalable Adaptation of Video Analytics}.
  In \bibinfo{booktitle}{\emph{Proceedings of the 2018 Conference of the ACM
  Special Interest Group on Data Communication}} (Budapest, Hungary)
  \emph{(\bibinfo{series}{SIGCOMM '18})}. \bibinfo{publisher}{Association for
  Computing Machinery}, \bibinfo{address}{New York, NY, USA},
  \bibinfo{pages}{253–266}.
\newblock
\showISBNx{9781450355674}
\urldef\tempurl%
\url{https://doi.org/10.1145/3230543.3230574}
\showDOI{\tempurl}


\bibitem[\protect\citeauthoryear{Kang, Bailis, and Zaharia}{Kang
  et~al\mbox{.}}{2019}]%
        {blazeit}
\bibfield{author}{\bibinfo{person}{Daniel Kang}, \bibinfo{person}{Peter
  Bailis}, {and} \bibinfo{person}{Matei Zaharia}.}
  \bibinfo{year}{2019}\natexlab{}.
\newblock \showarticletitle{BlazeIt: Optimizing Declarative Aggregation and
  Limit Queries for Neural Network-Based Video Analytics}.
\newblock \bibinfo{journal}{\emph{Proc. VLDB Endow.}} \bibinfo{volume}{13},
  \bibinfo{number}{4} (\bibinfo{date}{Dec.} \bibinfo{year}{2019}),
  \bibinfo{pages}{533–546}.
\newblock
\showISSN{2150-8097}
\urldef\tempurl%
\url{https://doi.org/10.14778/3372716.3372725}
\showDOI{\tempurl}


\bibitem[\protect\citeauthoryear{Kang, Emmons, Abuzaid, Bailis, and
  Zaharia}{Kang et~al\mbox{.}}{2017}]%
        {noscope-vldb17}
\bibfield{author}{\bibinfo{person}{Daniel Kang}, \bibinfo{person}{John Emmons},
  \bibinfo{person}{Firas Abuzaid}, \bibinfo{person}{Peter Bailis}, {and}
  \bibinfo{person}{Matei Zaharia}.} \bibinfo{year}{2017}\natexlab{}.
\newblock \showarticletitle{NoScope: Optimizing Neural Network Queries over
  Video at Scale}.
\newblock \bibinfo{journal}{\emph{Proc. VLDB Endow.}} \bibinfo{volume}{10},
  \bibinfo{number}{11} (\bibinfo{date}{Aug.} \bibinfo{year}{2017}),
  \bibinfo{pages}{1586--1597}.
\newblock
\showISSN{2150-8097}
\urldef\tempurl%
\url{https://doi.org/10.14778/3137628.3137664}
\showDOI{\tempurl}


\bibitem[\protect\citeauthoryear{Kemker, McClure, Abitino, Hayes, and
  Kanan}{Kemker et~al\mbox{.}}{2018}]%
        {catfail}
\bibfield{author}{\bibinfo{person}{Ronald Kemker}, \bibinfo{person}{Marc
  McClure}, \bibinfo{person}{Angelina Abitino}, \bibinfo{person}{Tyler Hayes},
  {and} \bibinfo{person}{Christopher Kanan}.} \bibinfo{year}{2018}\natexlab{}.
\newblock \showarticletitle{Measuring Catastrophic Forgetting in Neural
  Networks}.
\newblock \bibinfo{journal}{\emph{Proceedings of the AAAI Conference on
  Artificial Intelligence}} \bibinfo{volume}{32}, \bibinfo{number}{1}
  (\bibinfo{date}{Apr.} \bibinfo{year}{2018}).
\newblock
\urldef\tempurl%
\url{https://doi.org/10.1609/aaai.v32i1.11651}
\showDOI{\tempurl}


\bibitem[\protect\citeauthoryear{Khosla, Zhou, Malisiewicz, Efros, and
  Torralba}{Khosla et~al\mbox{.}}{2012}]%
        {khosla2012undoing}
\bibfield{author}{\bibinfo{person}{Aditya Khosla}, \bibinfo{person}{Tinghui
  Zhou}, \bibinfo{person}{Tomasz Malisiewicz}, \bibinfo{person}{Alexei~A
  Efros}, {and} \bibinfo{person}{Antonio Torralba}.}
  \bibinfo{year}{2012}\natexlab{}.
\newblock \showarticletitle{Undoing the damage of dataset bias}. In
  \bibinfo{booktitle}{\emph{Computer Vision--ECCV 2012: 12th European
  Conference on Computer Vision, Florence, Italy, October 7-13, 2012,
  Proceedings, Part I 12}}. Springer, \bibinfo{pages}{158--171}.
\newblock


\bibitem[\protect\citeauthoryear{Kshitija~Taywade}{Kshitija~Taywade}{2022}]%
        {mab-cournot-games}
\bibfield{author}{\bibinfo{person}{Judy~Goldsmith Kshitija~Taywade,
  Brent~Harrison}.} \bibinfo{year}{2022}\natexlab{}.
\newblock \showarticletitle{USING NON-STATIONARY BANDITS FOR LEARNING IN
  REPEATED COURNOT GAMES WITH NON-STATIONARY DEMAND}.
\newblock \bibinfo{journal}{\emph{arXiv preprint arXiv:2201.00486}}
  (\bibinfo{year}{2022}).
\newblock


\bibitem[\protect\citeauthoryear{Kumar, Balasubramanian, Venkataraman, and
  Akella}{Kumar et~al\mbox{.}}{2019}]%
        {FreezingHotCloud19}
\bibfield{author}{\bibinfo{person}{Adarsh Kumar}, \bibinfo{person}{Arjun
  Balasubramanian}, \bibinfo{person}{Shivaram Venkataraman}, {and}
  \bibinfo{person}{Aditya Akella}.} \bibinfo{year}{2019}\natexlab{}.
\newblock \showarticletitle{Accelerating Deep Learning Inference via Freezing}.
  In \bibinfo{booktitle}{\emph{11th USENIX Workshop on Hot Topics in Cloud
  Computing (HotCloud 19)}}. \bibinfo{publisher}{USENIX Association},
  \bibinfo{address}{Renton, WA}.
\newblock
\urldef\tempurl%
\url{https://www.usenix.org/conference/hotcloud19/presentation/kumar}
\showURL{%
\tempurl}


\bibitem[\protect\citeauthoryear{{Li}, {Lin}, {Shen}, {Brandt}, and {Hua}}{{Li}
  et~al\mbox{.}}{2015}]%
        {cnn-face-cvpr15}
\bibfield{author}{\bibinfo{person}{H. {Li}}, \bibinfo{person}{Z. {Lin}},
  \bibinfo{person}{X. {Shen}}, \bibinfo{person}{J. {Brandt}}, {and}
  \bibinfo{person}{G. {Hua}}.} \bibinfo{year}{2015}\natexlab{}.
\newblock \showarticletitle{A convolutional neural network cascade for face
  detection}. In \bibinfo{booktitle}{\emph{2015 IEEE Conference on Computer
  Vision and Pattern Recognition (CVPR)}}. \bibinfo{pages}{5325--5334}.
\newblock


\bibitem[\protect\citeauthoryear{Li, Padmanabhan, Zhao, Wang, Xu, and
  Netravali}{Li et~al\mbox{.}}{2020}]%
        {reducto}
\bibfield{author}{\bibinfo{person}{Yuanqi Li}, \bibinfo{person}{Arthi
  Padmanabhan}, \bibinfo{person}{Pengzhan Zhao}, \bibinfo{person}{Yufei Wang},
  \bibinfo{person}{Guoqing~Harry Xu}, {and} \bibinfo{person}{Ravi Netravali}.}
  \bibinfo{year}{2020}\natexlab{}.
\newblock \showarticletitle{Reducto: On-Camera Filtering for Resource-Efficient
  Real-Time Video Analytics}. In \bibinfo{booktitle}{\emph{Proceedings of the
  Annual Conference of the ACM Special Interest Group on Data Communication on
  the Applications, Technologies, Architectures, and Protocols for Computer
  Communication}} (Virtual Event, USA) \emph{(\bibinfo{series}{SIGCOMM '20})}.
  \bibinfo{publisher}{Association for Computing Machinery},
  \bibinfo{address}{New York, NY, USA}, \bibinfo{pages}{359–376}.
\newblock
\showISBNx{9781450379557}
\urldef\tempurl%
\url{https://doi.org/10.1145/3387514.3405874}
\showDOI{\tempurl}


\bibitem[\protect\citeauthoryear{Li, Shu, Ananthanarayanan, Shangguan,
  Jamieson, and Bahl}{Li et~al\mbox{.}}{2021}]%
        {spider}
\bibfield{author}{\bibinfo{person}{Zhuqi Li}, \bibinfo{person}{Yuanchao Shu},
  \bibinfo{person}{Ganesh Ananthanarayanan}, \bibinfo{person}{Longfei
  Shangguan}, \bibinfo{person}{Kyle Jamieson}, {and} \bibinfo{person}{Victor
  Bahl}.} \bibinfo{year}{2021}\natexlab{}.
\newblock \showarticletitle{Spider: A Multi-Hop Millimeter-Wave Network for
  Live Video Analytics}. In \bibinfo{booktitle}{\emph{ACM/IEEE Symposium on
  Edge Computing}}. ACM/IEEE.
\newblock
\urldef\tempurl%
\url{https://www.microsoft.com/en-us/research/publication/spider-a-multi-hop-millimeter-wave-network-for-live-video-analytics/}
\showURL{%
\tempurl}


\bibitem[\protect\citeauthoryear{Lin, Belongie, Hays, Perona, Ramanan,
  Doll{\'a}r, and Zitnick}{Lin et~al\mbox{.}}{2014}]%
        {ms-coco-eccv14}
\bibfield{author}{\bibinfo{person}{Michael Lin, Tsung-Yiand~Maire},
  \bibinfo{person}{Serge Belongie}, \bibinfo{person}{James Hays},
  \bibinfo{person}{Pietro Perona}, \bibinfo{person}{Deva Ramanan},
  \bibinfo{person}{Piotr Doll{\'a}r}, {and} \bibinfo{person}{C.~Lawrence
  Zitnick}.} \bibinfo{year}{2014}\natexlab{}.
\newblock \showarticletitle{{Microsoft COCO}: Common Objects in Context}. In
  \bibinfo{booktitle}{\emph{Computer Vision -- ECCV 2014}}.
  \bibinfo{pages}{740--755}.
\newblock


\bibitem[\protect\citeauthoryear{{Lin}, {Doll{\'a}r}, {Girshick}, {He},
  {Hariharan}, and {Belongie}}{{Lin} et~al\mbox{.}}{2017}]%
        {pyramid-network-cvpr17}
\bibfield{author}{\bibinfo{person}{T. {Lin}}, \bibinfo{person}{P.
  {Doll{\'a}r}}, \bibinfo{person}{R. {Girshick}}, \bibinfo{person}{K. {He}},
  \bibinfo{person}{B. {Hariharan}}, {and} \bibinfo{person}{S. {Belongie}}.}
  \bibinfo{year}{2017}\natexlab{}.
\newblock \showarticletitle{Feature Pyramid Networks for Object Detection}. In
  \bibinfo{booktitle}{\emph{2017 IEEE Conference on Computer Vision and Pattern
  Recognition (CVPR)}}. \bibinfo{pages}{936--944}.
\newblock
\showISSN{1063-6919}
\urldef\tempurl%
\url{https://doi.org/10.1109/CVPR.2017.106}
\showDOI{\tempurl}


\bibitem[\protect\citeauthoryear{Liu, Anguelov, Erhan, Szegedy, Reed, Fu, and
  Berg}{Liu et~al\mbox{.}}{2016}]%
        {ssd}
\bibfield{author}{\bibinfo{person}{Wei Liu}, \bibinfo{person}{Dragomir
  Anguelov}, \bibinfo{person}{Dumitru Erhan}, \bibinfo{person}{Christian
  Szegedy}, \bibinfo{person}{Scott Reed}, \bibinfo{person}{Cheng-Yang Fu},
  {and} \bibinfo{person}{Alexander~C. Berg}.} \bibinfo{year}{2016}\natexlab{}.
\newblock \showarticletitle{SSD: Single Shot MultiBox Detector}. In
  \bibinfo{booktitle}{\emph{Computer Vision -- ECCV 2016}}.
  \bibinfo{publisher}{Springer International Publishing},
  \bibinfo{pages}{21--37}.
\newblock


\bibitem[\protect\citeauthoryear{Liu, Li, Shen, Huang, Yan, and Zhang}{Liu
  et~al\mbox{.}}{2017}]%
        {liu2017learning}
\bibfield{author}{\bibinfo{person}{Zhuang Liu}, \bibinfo{person}{Jianguo Li},
  \bibinfo{person}{Zhiqiang Shen}, \bibinfo{person}{Gao Huang},
  \bibinfo{person}{Shoumeng Yan}, {and} \bibinfo{person}{Changshui Zhang}.}
  \bibinfo{year}{2017}\natexlab{}.
\newblock \showarticletitle{Learning efficient convolutional networks through
  network slimming}. In \bibinfo{booktitle}{\emph{Proceedings of the IEEE
  International Conference on Computer Vision}}. \bibinfo{pages}{2736--2744}.
\newblock


\bibitem[\protect\citeauthoryear{Lowe}{Lowe}{2004}]%
        {lowe2004distinctive}
\bibfield{author}{\bibinfo{person}{David~G Lowe}.}
  \bibinfo{year}{2004}\natexlab{}.
\newblock \showarticletitle{Distinctive image features from scale-invariant
  keypoints}.
\newblock \bibinfo{journal}{\emph{International journal of computer vision}}
  \bibinfo{volume}{60} (\bibinfo{year}{2004}).
\newblock


\bibitem[\protect\citeauthoryear{Lu, Chowdhery, and Kandula}{Lu
  et~al\mbox{.}}{2016}]%
        {optasia-socc16}
\bibfield{author}{\bibinfo{person}{Yao Lu}, \bibinfo{person}{Aakanksha
  Chowdhery}, {and} \bibinfo{person}{Srikanth Kandula}.}
  \bibinfo{year}{2016}\natexlab{}.
\newblock \showarticletitle{Optasia: A Relational Platform for Efficient
  Large-Scale Video Analytics}. In \bibinfo{booktitle}{\emph{Proceedings of the
  Seventh ACM Symposium on Cloud Computing}} (Santa Clara, CA, USA)
  \emph{(\bibinfo{series}{SoCC '16})}. \bibinfo{publisher}{ACM},
  \bibinfo{address}{New York, NY, USA}, \bibinfo{pages}{57--70}.
\newblock
\showISBNx{978-1-4503-4525-5}
\urldef\tempurl%
\url{https://doi.org/10.1145/2987550.2987564}
\showDOI{\tempurl}


\bibitem[\protect\citeauthoryear{{Mehrdad Khani, Ganesh Ananthanarayanan, Kevin
  Hsieh, Junchen Jiang, Ravi Netravali , Yuanchao Shu, Mohammad Alizadeh ,
  Victor Bahl}}{{Mehrdad Khani, Ganesh Ananthanarayanan, Kevin Hsieh, Junchen
  Jiang, Ravi Netravali , Yuanchao Shu, Mohammad Alizadeh , Victor
  Bahl}}{2023}]%
        {recl}
\bibfield{author}{\bibinfo{person}{{Mehrdad Khani, Ganesh Ananthanarayanan,
  Kevin Hsieh, Junchen Jiang, Ravi Netravali , Yuanchao Shu, Mohammad Alizadeh
  , Victor Bahl}}.} \bibinfo{year}{2023}\natexlab{}.
\newblock \showarticletitle{RECL: Responsive Resource-Efficient Continuous
  Learning for Video Analytics}. In \bibinfo{booktitle}{\emph{20th USENIX
  Symposium on Networked Systems Design and Implementation (NSDI 23)}}.
  \bibinfo{publisher}{USENIX Association}, \bibinfo{address}{Boston, MA}.
\newblock


\bibitem[\protect\citeauthoryear{{Microsoft Azure}}{{Microsoft Azure}}{2021}]%
        {AzureStackEdge}
\bibfield{author}{\bibinfo{person}{{Microsoft Azure}}.}
  \bibinfo{year}{2021}\natexlab{}.
\newblock \bibinfo{title}{Azure Stack Edge}.
\newblock
\newblock
\urldef\tempurl%
\url{https://azure.microsoft.com/en-us/services/databox/edge/}
\showURL{%
\tempurl}


\bibitem[\protect\citeauthoryear{Netravali, Sivaraman, Winstein, Das, Goyal,
  Mickens, and Balakrishnan}{Netravali et~al\mbox{.}}{2015}]%
        {mahimahi}
\bibfield{author}{\bibinfo{person}{R. Netravali}, \bibinfo{person}{A.
  Sivaraman}, \bibinfo{person}{K. Winstein}, \bibinfo{person}{S. Das},
  \bibinfo{person}{A. Goyal}, \bibinfo{person}{J. Mickens}, {and}
  \bibinfo{person}{H. Balakrishnan}.} \bibinfo{year}{2015}\natexlab{}.
\newblock \showarticletitle{Mahimahi: {A}ccurate {R}ecord-and-{R}eplay for
  {HTTP}} \emph{(\bibinfo{series}{Proceedings of ATC '15})}.
  \bibinfo{publisher}{USENIX}.
\newblock


\bibitem[\protect\citeauthoryear{{NVIDIA}}{{NVIDIA}}{2021}]%
        {NVIDIAJetsonNano}
\bibfield{author}{\bibinfo{person}{{NVIDIA}}.} \bibinfo{year}{2021}\natexlab{}.
\newblock \bibinfo{title}{NVIDIA JetsonNano}.
\newblock
\newblock
\urldef\tempurl%
\url{https://www.nvidia.com/en-us/autonomous-machines/embedded-systems/jetson-nano/product-development/}
\showURL{%
\tempurl}


\bibitem[\protect\citeauthoryear{Oksuz, Cam, Kalkan, and Akbas}{Oksuz
  et~al\mbox{.}}{2020}]%
        {oksuz2020imbalance}
\bibfield{author}{\bibinfo{person}{Kemal Oksuz}, \bibinfo{person}{Baris~Can
  Cam}, \bibinfo{person}{Sinan Kalkan}, {and} \bibinfo{person}{Emre Akbas}.}
  \bibinfo{year}{2020}\natexlab{}.
\newblock \showarticletitle{Imbalance problems in object detection: A review}.
\newblock \bibinfo{journal}{\emph{IEEE transactions on pattern analysis and
  machine intelligence}} \bibinfo{volume}{43}, \bibinfo{number}{10}
  (\bibinfo{year}{2020}).
\newblock


\bibitem[\protect\citeauthoryear{{Omar Besbes, Yonatan Gur, Assaf Zeevi}}{{Omar
  Besbes, Yonatan Gur, Assaf Zeevi}}{2014}]%
        {mab-variation-budget}
\bibfield{author}{\bibinfo{person}{{Omar Besbes, Yonatan Gur, Assaf Zeevi}}.}
  \bibinfo{year}{2014}\natexlab{}.
\newblock \showarticletitle{{ Stochastic Multi-Armed-Bandit Problem with
  Non-stationary Rewards }}. In \bibinfo{booktitle}{\emph{NeurIPS}}.
\newblock


\bibitem[\protect\citeauthoryear{Padmanabhan, Agarwal, Iyer, Ananthanarayanan,
  Shu, Karianakis, Xu, and Netravali}{Padmanabhan et~al\mbox{.}}{2022}]%
        {gemel}
\bibfield{author}{\bibinfo{person}{Arthi Padmanabhan}, \bibinfo{person}{Neil
  Agarwal}, \bibinfo{person}{Anand Iyer}, \bibinfo{person}{Ganesh
  Ananthanarayanan}, \bibinfo{person}{Yuanchao Shu}, \bibinfo{person}{Nikolaos
  Karianakis}, \bibinfo{person}{Guoqing~Harry Xu}, {and} \bibinfo{person}{Ravi
  Netravali}.} \bibinfo{year}{2022}\natexlab{}.
\newblock \bibinfo{title}{GEMEL: Model Merging for Memory-Efficient, Real-Time
  Video Analytics at the Edge}.
\newblock
\newblock
\urldef\tempurl%
\url{https://doi.org/10.48550/ARXIV.2201.07705}
\showDOI{\tempurl}


\bibitem[\protect\citeauthoryear{Pan and Lyu}{Pan and Lyu}{2010}]%
        {pan2010detecting}
\bibfield{author}{\bibinfo{person}{Xunyu Pan} {and} \bibinfo{person}{Siwei
  Lyu}.} \bibinfo{year}{2010}\natexlab{}.
\newblock \showarticletitle{Detecting image region duplication using SIFT
  features}. In \bibinfo{booktitle}{\emph{2010 IEEE International Conference on
  Acoustics, Speech and Signal Processing}}. IEEE, \bibinfo{pages}{1706--1709}.
\newblock


\bibitem[\protect\citeauthoryear{Paul, Drolia, Hu, and Chakradhar}{Paul
  et~al\mbox{.}}{2021}]%
        {aqua}
\bibfield{author}{\bibinfo{person}{Sibendu Paul}, \bibinfo{person}{Utsav
  Drolia}, \bibinfo{person}{Y.~Charlie Hu}, {and} \bibinfo{person}{Srimat~T.
  Chakradhar}.} \bibinfo{year}{2021}\natexlab{}.
\newblock \bibinfo{title}{AQuA: Analytical Quality Assessment for Optimizing
  Video Analytics Systems}.
\newblock
\newblock
\urldef\tempurl%
\url{https://doi.org/10.48550/ARXIV.2101.09752}
\showDOI{\tempurl}


\bibitem[\protect\citeauthoryear{Paul, Rao, Coviello, Sankaradas, Po, Hu, and
  Chakradhar}{Paul et~al\mbox{.}}{2023}]%
        {camtuner}
\bibfield{author}{\bibinfo{person}{Sibendu Paul}, \bibinfo{person}{Kunal Rao},
  \bibinfo{person}{Giuseppe Coviello}, \bibinfo{person}{Murugan Sankaradas},
  \bibinfo{person}{Oliver Po}, \bibinfo{person}{Y.~Charlie Hu}, {and}
  \bibinfo{person}{Srimat Chakradhar}.} \bibinfo{year}{2023}\natexlab{}.
\newblock \showarticletitle{Enhancing Video Analytics Accuracy via Real-Time
  Automated Camera Parameter Tuning}. In \bibinfo{booktitle}{\emph{Proceedings
  of the 20th ACM Conference on Embedded Networked Sensor Systems}} (Boston,
  Massachusetts) \emph{(\bibinfo{series}{SenSys '22})}.
  \bibinfo{publisher}{Association for Computing Machinery},
  \bibinfo{address}{New York, NY, USA}, \bibinfo{pages}{291–304}.
\newblock
\showISBNx{9781450398862}
\urldef\tempurl%
\url{https://doi.org/10.1145/3560905.3568527}
\showDOI{\tempurl}


\bibitem[\protect\citeauthoryear{{PTZ Optics}}{{PTZ Optics}}{[n.d.]a}]%
        {ptzoptics-cameras}
\bibfield{author}{\bibinfo{person}{{PTZ Optics}}.}
  \bibinfo{year}{[n.d.]}\natexlab{a}.
\newblock \bibinfo{title}{{PTZ Optics PTZ cameras}}.
\newblock \bibinfo{howpublished}{\url{https://ptzoptics.com/products/}}.
\newblock


\bibitem[\protect\citeauthoryear{{PTZ Optics}}{{PTZ Optics}}{[n.d.]b}]%
        {what-is-eptz}
\bibfield{author}{\bibinfo{person}{{PTZ Optics}}.}
  \bibinfo{year}{[n.d.]}\natexlab{b}.
\newblock \bibinfo{title}{{What is ePTZ and how does it compare with true
  PTZ?}}
\newblock \bibinfo{howpublished}{\url{https://ptzoptics.com/what-is-eptz/}}.
\newblock


\bibitem[\protect\citeauthoryear{Redmon and Farhadi}{Redmon and
  Farhadi}{2018}]%
        {redmon2018yolov3}
\bibfield{author}{\bibinfo{person}{Joseph Redmon} {and} \bibinfo{person}{Ali
  Farhadi}.} \bibinfo{year}{2018}\natexlab{}.
\newblock \showarticletitle{Yolov3: An incremental improvement}.
\newblock \bibinfo{journal}{\emph{arXiv preprint arXiv:1804.02767}}
  (\bibinfo{year}{2018}).
\newblock


\bibitem[\protect\citeauthoryear{{Richard Zhang, Phillip Isola, Alexei A.
  Efros, Eli Shechtman, Oliver Wang}}{{Richard Zhang, Phillip Isola, Alexei A.
  Efros, Eli Shechtman, Oliver Wang}}{2018}]%
        {perceptual-similarity}
\bibfield{author}{\bibinfo{person}{{Richard Zhang, Phillip Isola, Alexei A.
  Efros, Eli Shechtman, Oliver Wang}}.} \bibinfo{year}{2018}\natexlab{}.
\newblock \showarticletitle{{The Unreasonable Effectiveness of Deep Features as
  a Perceptual Metric}}. In \bibinfo{booktitle}{\emph{CVPR}}.
\newblock


\bibitem[\protect\citeauthoryear{Rijas}{Rijas}{[n.d.]}]%
        {powering-the-edge-with-ai-in-an-iot-world}
\bibfield{author}{\bibinfo{person}{Mohammed Rijas}.}
  \bibinfo{year}{[n.d.]}\natexlab{}.
\newblock \bibinfo{title}{Powering The Edge With AI In An IoT World}.
\newblock
  \bibinfo{howpublished}{\url{https://www.forbes.com/sites/forbestechcouncil/2020/04/06/powering-the-edge-with-ai-in-an-iot-world/}}.
\newblock


\bibitem[\protect\citeauthoryear{Rizzoli}{Rizzoli}{2022}]%
        {sports_va}
\bibfield{author}{\bibinfo{person}{Alberto Rizzoli}.}
  \bibinfo{year}{2022}\natexlab{}.
\newblock \bibinfo{title}{{7 Game-Changing AI Applications in the Sports
  Industry}}.
\newblock
  \bibinfo{howpublished}{\url{https://www.v7labs.com/blog/ai-in-sports}}.
\newblock


\bibitem[\protect\citeauthoryear{Romero, Li, Yadwadkar, and Kozyrakis}{Romero
  et~al\mbox{.}}{2021}]%
        {InFaaSATC21}
\bibfield{author}{\bibinfo{person}{Francisco Romero}, \bibinfo{person}{Qian
  Li}, \bibinfo{person}{Neeraja~J. Yadwadkar}, {and} \bibinfo{person}{Christos
  Kozyrakis}.} \bibinfo{year}{2021}\natexlab{}.
\newblock \showarticletitle{{INFaaS}: Automated Model-less Inference Serving}.
  In \bibinfo{booktitle}{\emph{2021 USENIX Annual Technical Conference (USENIX
  ATC 21)}}. \bibinfo{publisher}{USENIX Association},
  \bibinfo{pages}{397--411}.
\newblock
\showISBNx{978-1-939133-23-6}
\urldef\tempurl%
\url{https://www.usenix.org/conference/atc21/presentation/romero}
\showURL{%
\tempurl}


\bibitem[\protect\citeauthoryear{{SCW}}{{SCW}}{[n.d.]}]%
        {auto-tracking}
\bibfield{author}{\bibinfo{person}{{SCW}}.} \bibinfo{year}{[n.d.]}\natexlab{}.
\newblock \bibinfo{title}{{PTZ Auto-Tracking Explained}}.
\newblock
  \bibinfo{howpublished}{\url{https://www.getscw.com/knowledge-base/auto-tracking-explained}}.
\newblock


\bibitem[\protect\citeauthoryear{{SecurityBros}}{{SecurityBros}}{[n.d.]}]%
        {cheap_ptz_comparison}
\bibfield{author}{\bibinfo{person}{{SecurityBros}}.}
  \bibinfo{year}{[n.d.]}\natexlab{}.
\newblock \bibinfo{title}{{ Which Cheap Outdoor WiFi PTZ IP Camera is Best?
  Boavision vs Inqmega}}.
\newblock
  \bibinfo{howpublished}{\url{https://securitybros.com/which-cheap-outdoor-wifi-ptz-ip-camera-is-best-boavision-vs-inqmega/}}.
\newblock


\bibitem[\protect\citeauthoryear{{Shaoqing Ren, Kaiming He, Ross Girshick, Jian
  Sun}}{{Shaoqing Ren, Kaiming He, Ross Girshick, Jian Sun}}{2015}]%
        {faster-rcnn}
\bibfield{author}{\bibinfo{person}{{Shaoqing Ren, Kaiming He, Ross Girshick,
  Jian Sun}}.} \bibinfo{year}{2015}\natexlab{}.
\newblock \showarticletitle{{Faster R-CNN: Towards Real-Time Object Detection
  with Region Proposal Networks}}. In \bibinfo{booktitle}{\emph{NeurIPS}}.
\newblock


\bibitem[\protect\citeauthoryear{Shen, Chen, Jin, Zhao, Kong, Philipose,
  Krishnamurthy, and Sundaram}{Shen et~al\mbox{.}}{2019}]%
        {nexus}
\bibfield{author}{\bibinfo{person}{Haichen Shen}, \bibinfo{person}{Lequn Chen},
  \bibinfo{person}{Yuchen Jin}, \bibinfo{person}{Liangyu Zhao},
  \bibinfo{person}{Bingyu Kong}, \bibinfo{person}{Matthai Philipose},
  \bibinfo{person}{Arvind Krishnamurthy}, {and} \bibinfo{person}{Ravi
  Sundaram}.} \bibinfo{year}{2019}\natexlab{}.
\newblock \showarticletitle{Nexus: A GPU Cluster Engine for Accelerating
  DNN-Based Video Analysis}. In \bibinfo{booktitle}{\emph{Proceedings of the
  27th ACM Symposium on Operating Systems Principles}} (Huntsville, Ontario,
  Canada) \emph{(\bibinfo{series}{SOSP '19})}. \bibinfo{publisher}{Association
  for Computing Machinery}, \bibinfo{address}{New York, NY, USA},
  \bibinfo{pages}{322--337}.
\newblock
\showISBNx{9781450368735}
\urldef\tempurl%
\url{https://doi.org/10.1145/3341301.3359658}
\showDOI{\tempurl}


\bibitem[\protect\citeauthoryear{Shi, Cao, Zhang, Li, and Xu}{Shi
  et~al\mbox{.}}{2016}]%
        {shi2016edge}
\bibfield{author}{\bibinfo{person}{Weisong Shi}, \bibinfo{person}{Jie Cao},
  \bibinfo{person}{Quan Zhang}, \bibinfo{person}{Youhuizi Li}, {and}
  \bibinfo{person}{Lanyu Xu}.} \bibinfo{year}{2016}\natexlab{}.
\newblock \showarticletitle{Edge computing: Vision and challenges}.
\newblock \bibinfo{journal}{\emph{IEEE internet of things journal}}
  \bibinfo{volume}{3}, \bibinfo{number}{5} (\bibinfo{year}{2016}),
  \bibinfo{pages}{637--646}.
\newblock


\bibitem[\protect\citeauthoryear{{Shubham Jain, Viet Nguyen, Marco Gruteser,
  Paramvir Bahl}}{{Shubham Jain, Viet Nguyen, Marco Gruteser, Paramvir
  Bahl}}{2017}]%
        {panoptes}
\bibfield{author}{\bibinfo{person}{{Shubham Jain, Viet Nguyen, Marco Gruteser,
  Paramvir Bahl}}.} \bibinfo{year}{2017}\natexlab{}.
\newblock \showarticletitle{{Panoptes: Servicing Multiple Applications
  Simultaneously using Steerable Cameras}}. In
  \bibinfo{booktitle}{\emph{IPSN}}.
\newblock


\bibitem[\protect\citeauthoryear{Sindagi and Patel}{Sindagi and Patel}{2017}]%
        {sindagi2017generating}
\bibfield{author}{\bibinfo{person}{Vishwanath~A Sindagi} {and}
  \bibinfo{person}{Vishal~M Patel}.} \bibinfo{year}{2017}\natexlab{}.
\newblock \showarticletitle{Generating high-quality crowd density maps using
  contextual pyramid cnns}. In \bibinfo{booktitle}{\emph{Proceedings of the
  IEEE international conference on computer vision}}.
  \bibinfo{pages}{1861--1870}.
\newblock


\bibitem[\protect\citeauthoryear{Steed and Caliskan}{Steed and
  Caliskan}{2021}]%
        {steed2021image}
\bibfield{author}{\bibinfo{person}{Ryan Steed} {and} \bibinfo{person}{Aylin
  Caliskan}.} \bibinfo{year}{2021}\natexlab{}.
\newblock \showarticletitle{Image representations learned with unsupervised
  pre-training contain human-like biases}. In
  \bibinfo{booktitle}{\emph{Proceedings of the 2021 ACM conference on fairness,
  accountability, and transparency}}. \bibinfo{pages}{701--713}.
\newblock


\bibitem[\protect\citeauthoryear{Suprem, Arulraj, Pu, and Ferreira}{Suprem
  et~al\mbox{.}}{2020}]%
        {odin}
\bibfield{author}{\bibinfo{person}{Abhijit Suprem}, \bibinfo{person}{Joy
  Arulraj}, \bibinfo{person}{Calton Pu}, {and} \bibinfo{person}{Joao
  Ferreira}.} \bibinfo{year}{2020}\natexlab{}.
\newblock \showarticletitle{ODIN: Automated Drift Detection and Recovery in
  Video Analytics}.
\newblock \bibinfo{journal}{\emph{Proc. VLDB Endow.}} \bibinfo{volume}{13},
  \bibinfo{number}{12} (\bibinfo{date}{July} \bibinfo{year}{2020}),
  \bibinfo{pages}{2453–2465}.
\newblock
\showISSN{2150-8097}
\urldef\tempurl%
\url{https://doi.org/10.14778/3407790.3407837}
\showDOI{\tempurl}


\bibitem[\protect\citeauthoryear{Tan, Pang, and Le}{Tan et~al\mbox{.}}{2019}]%
        {efficientdet}
\bibfield{author}{\bibinfo{person}{Mingxing Tan}, \bibinfo{person}{Ruoming
  Pang}, {and} \bibinfo{person}{Quoc~V. Le}.} \bibinfo{year}{2019}\natexlab{}.
\newblock \showarticletitle{EfficientDet: Scalable and Efficient Object
  Detection}.
\newblock \bibinfo{journal}{\emph{CoRR}}  \bibinfo{volume}{abs/1911.09070}
  (\bibinfo{year}{2019}).
\newblock
\showeprint[arXiv]{1911.09070}
\urldef\tempurl%
\url{http://arxiv.org/abs/1911.09070}
\showURL{%
\tempurl}


\bibitem[\protect\citeauthoryear{Tversky and Gati}{Tversky and Gati}{1982}]%
        {tversky1982similarity}
\bibfield{author}{\bibinfo{person}{Amos Tversky} {and} \bibinfo{person}{Itamar
  Gati}.} \bibinfo{year}{1982}\natexlab{}.
\newblock \showarticletitle{Similarity, separability, and the triangle
  inequality.}
\newblock \bibinfo{journal}{\emph{Psychological review}} \bibinfo{volume}{89},
  \bibinfo{number}{2} (\bibinfo{year}{1982}), \bibinfo{pages}{123}.
\newblock


\bibitem[\protect\citeauthoryear{Vasisht, Kapetanovic, Won, Jin, Chandra,
  Kapoor, Sinha, Sudarshan, and Stratman}{Vasisht et~al\mbox{.}}{2017}]%
        {farmbeats}
\bibfield{author}{\bibinfo{person}{Deepak Vasisht}, \bibinfo{person}{Zerina
  Kapetanovic}, \bibinfo{person}{Jong-ho Won}, \bibinfo{person}{Xinxin Jin},
  \bibinfo{person}{Ranveer Chandra}, \bibinfo{person}{Ashish Kapoor},
  \bibinfo{person}{Sudipta~N. Sinha}, \bibinfo{person}{Madhusudhan Sudarshan},
  {and} \bibinfo{person}{Sean Stratman}.} \bibinfo{year}{2017}\natexlab{}.
\newblock \showarticletitle{Farmbeats: An IoT Platform for Data-Driven
  Agriculture}. In \bibinfo{booktitle}{\emph{Proceedings of the 14th USENIX
  Conference on Networked Systems Design and Implementation}} (Boston, MA, USA)
  \emph{(\bibinfo{series}{NSDI'17})}. \bibinfo{publisher}{USENIX Association},
  \bibinfo{address}{USA}, \bibinfo{pages}{515–528}.
\newblock
\showISBNx{9781931971379}


\bibitem[\protect\citeauthoryear{Vidal}{Vidal}{2021}]%
        {mab_for_wireless_selection}
\bibfield{author}{\bibinfo{person}{Lluís Martínez; Margarita
  Cabrera-Bean;~Josep Vidal}.} \bibinfo{year}{2021}\natexlab{}.
\newblock \showarticletitle{{A Multi-Armed Bandit Model for Non-Stationary
  Wireless Network Selection}}. In \bibinfo{booktitle}{\emph{2021 IEEE Globecom
  Workshops}}. Ieee.
\newblock


\bibitem[\protect\citeauthoryear{{Vidit Saxena, Joakim Jaldén, Joseph E.
  Gonzalez, Mats Bengtsson, Hugo Tullberg, Ion Stoica}}{{Vidit Saxena, Joakim
  Jaldén, Joseph E. Gonzalez, Mats Bengtsson, Hugo Tullberg, Ion
  Stoica}}{2019}]%
        {mab-link-adaption}
\bibfield{author}{\bibinfo{person}{{Vidit Saxena, Joakim Jaldén, Joseph E.
  Gonzalez, Mats Bengtsson, Hugo Tullberg, Ion Stoica}}.}
  \bibinfo{year}{2019}\natexlab{}.
\newblock \showarticletitle{{ Contextual Multi-Armed Bandits for Link
  Adaptation in Cellular Networks}}. In \bibinfo{booktitle}{\emph{NetAI}}.
\newblock


\bibitem[\protect\citeauthoryear{Wang, Bochkovskiy, and Liao}{Wang
  et~al\mbox{.}}{2021}]%
        {yolov4}
\bibfield{author}{\bibinfo{person}{Chien-Yao Wang}, \bibinfo{person}{Alexey
  Bochkovskiy}, {and} \bibinfo{person}{Hong-Yuan~Mark Liao}.}
  \bibinfo{year}{2021}\natexlab{}.
\newblock \bibinfo{title}{Scaled-YOLOv4: Scaling Cross Stage Partial Network}.
\newblock
\newblock
\showeprint[arxiv]{2011.08036}~[cs.CV]


\bibitem[\protect\citeauthoryear{Wang, Feng, George, Iyengar, Pillai, and
  Satyanarayanan}{Wang et~al\mbox{.}}{2019a}]%
        {edgenative}
\bibfield{author}{\bibinfo{person}{Junjue Wang}, \bibinfo{person}{Ziqiang
  Feng}, \bibinfo{person}{Shilpa George}, \bibinfo{person}{Roger Iyengar},
  \bibinfo{person}{Padmanabhan Pillai}, {and} \bibinfo{person}{Mahadev
  Satyanarayanan}.} \bibinfo{year}{2019}\natexlab{a}.
\newblock \showarticletitle{Towards Scalable Edge-Native Applications}. In
  \bibinfo{booktitle}{\emph{Proceedings of the 4th ACM/IEEE Symposium on Edge
  Computing}} (Arlington, Virginia) \emph{(\bibinfo{series}{SEC '19})}.
  \bibinfo{publisher}{Association for Computing Machinery},
  \bibinfo{address}{New York, NY, USA}, \bibinfo{pages}{152–165}.
\newblock
\showISBNx{9781450367332}
\urldef\tempurl%
\url{https://doi.org/10.1145/3318216.3363308}
\showDOI{\tempurl}


\bibitem[\protect\citeauthoryear{Wang, Wang, Zhang, Jiang, and Chen}{Wang
  et~al\mbox{.}}{2019b}]%
        {cloudseg}
\bibfield{author}{\bibinfo{person}{Yiding Wang}, \bibinfo{person}{Weiyan Wang},
  \bibinfo{person}{Junxue Zhang}, \bibinfo{person}{Junchen Jiang}, {and}
  \bibinfo{person}{Kai Chen}.} \bibinfo{year}{2019}\natexlab{b}.
\newblock \showarticletitle{Bridging the Edge-Cloud Barrier for Real-time
  Advanced Vision Analytics}. In \bibinfo{booktitle}{\emph{11th {USENIX}
  Workshop on Hot Topics in Cloud Computing (HotCloud 19)}}.
  \bibinfo{publisher}{{USENIX} Association}, \bibinfo{address}{Renton, WA}.
\newblock
\urldef\tempurl%
\url{https://www.usenix.org/conference/hotcloud19/presentation/wang}
\showURL{%
\tempurl}


\bibitem[\protect\citeauthoryear{Wang, Qinami, Karakozis, Genova, Nair, Hata,
  and Russakovsky}{Wang et~al\mbox{.}}{2020}]%
        {wang2020towards}
\bibfield{author}{\bibinfo{person}{Zeyu Wang}, \bibinfo{person}{Klint Qinami},
  \bibinfo{person}{Ioannis~Christos Karakozis}, \bibinfo{person}{Kyle Genova},
  \bibinfo{person}{Prem Nair}, \bibinfo{person}{Kenji Hata}, {and}
  \bibinfo{person}{Olga Russakovsky}.} \bibinfo{year}{2020}\natexlab{}.
\newblock \showarticletitle{Towards fairness in visual recognition: Effective
  strategies for bias mitigation}. In \bibinfo{booktitle}{\emph{Proceedings of
  the IEEE/CVF conference on computer vision and pattern recognition}}.
  \bibinfo{pages}{8919--8928}.
\newblock


\bibitem[\protect\citeauthoryear{Xu, Zhu, Liu, Lin, and Liu}{Xu
  et~al\mbox{.}}{2018}]%
        {deepcache}
\bibfield{author}{\bibinfo{person}{Mengwei Xu}, \bibinfo{person}{Mengze Zhu},
  \bibinfo{person}{Yunxin Liu}, \bibinfo{person}{Felix~Xiaozhu Lin}, {and}
  \bibinfo{person}{Xuanzhe Liu}.} \bibinfo{year}{2018}\natexlab{}.
\newblock \showarticletitle{DeepCache: Principled Cache for Mobile Deep
  Vision}. In \bibinfo{booktitle}{\emph{Proceedings of the 24th Annual
  International Conference on Mobile Computing and Networking}} (New Delhi,
  India) \emph{(\bibinfo{series}{MobiCom '18})}.
  \bibinfo{publisher}{Association for Computing Machinery},
  \bibinfo{address}{New York, NY, USA}, \bibinfo{pages}{129–144}.
\newblock
\showISBNx{9781450359030}
\urldef\tempurl%
\url{https://doi.org/10.1145/3241539.3241563}
\showDOI{\tempurl}


\bibitem[\protect\citeauthoryear{Yang, Li, Du, Huang, and Sebe}{Yang
  et~al\mbox{.}}{2020}]%
        {yang2020embedding}
\bibfield{author}{\bibinfo{person}{Yifan Yang}, \bibinfo{person}{Guorong Li},
  \bibinfo{person}{Dawei Du}, \bibinfo{person}{Qingming Huang}, {and}
  \bibinfo{person}{Nicu Sebe}.} \bibinfo{year}{2020}\natexlab{}.
\newblock \showarticletitle{Embedding perspective analysis into multi-column
  convolutional neural network for crowd counting}.
\newblock \bibinfo{journal}{\emph{IEEE Transactions on Image Processing}}
  \bibinfo{volume}{30} (\bibinfo{year}{2020}), \bibinfo{pages}{1395--1407}.
\newblock


\bibitem[\protect\citeauthoryear{{Yifu Zhang, Peize Sun, Yi Jiang, Dongdong Yu,
  Zehuan Yuan, Ping Luo, Wenyu Liu, Xinggang Wang}}{{Yifu Zhang, Peize Sun, Yi
  Jiang, Dongdong Yu, Zehuan Yuan, Ping Luo, Wenyu Liu, Xinggang Wang}}{2021}]%
        {bytetrack}
\bibfield{author}{\bibinfo{person}{{Yifu Zhang, Peize Sun, Yi Jiang, Dongdong
  Yu, Zehuan Yuan, Ping Luo, Wenyu Liu, Xinggang Wang}}.}
  \bibinfo{year}{2021}\natexlab{}.
\newblock \bibinfo{title}{ByteTrack: Multi-Object Tracking by Associating Every
  Detection Box}.
\newblock
\newblock
\showeprint[arxiv]{2110.06864}~[cs.CV]


\bibitem[\protect\citeauthoryear{Yin, Jindal, Sekar, and Sinopoli}{Yin
  et~al\mbox{.}}{2015}]%
        {mpc}
\bibfield{author}{\bibinfo{person}{Xiaoqi Yin}, \bibinfo{person}{Abhishek
  Jindal}, \bibinfo{person}{Vyas Sekar}, {and} \bibinfo{person}{Bruno
  Sinopoli}.} \bibinfo{year}{2015}\natexlab{}.
\newblock \showarticletitle{A Control-Theoretic Approach for Dynamic Adaptive
  Video Streaming over HTTP}.
\newblock \bibinfo{journal}{\emph{SIGCOMM Comput. Commun. Rev.}}
  \bibinfo{volume}{45}, \bibinfo{number}{4} (\bibinfo{date}{aug}
  \bibinfo{year}{2015}), \bibinfo{pages}{325–338}.
\newblock
\showISSN{0146-4833}
\urldef\tempurl%
\url{https://doi.org/10.1145/2829988.2787486}
\showDOI{\tempurl}


\bibitem[\protect\citeauthoryear{Yosinski, Clune, Bengio, and Lipson}{Yosinski
  et~al\mbox{.}}{2014}]%
        {NIPS2014_375c7134}
\bibfield{author}{\bibinfo{person}{Jason Yosinski}, \bibinfo{person}{Jeff
  Clune}, \bibinfo{person}{Yoshua Bengio}, {and} \bibinfo{person}{Hod Lipson}.}
  \bibinfo{year}{2014}\natexlab{}.
\newblock \showarticletitle{How transferable are features in deep neural
  networks?}. In \bibinfo{booktitle}{\emph{Advances in Neural Information
  Processing Systems}}, \bibfield{editor}{\bibinfo{person}{Z.~Ghahramani},
  \bibinfo{person}{M.~Welling}, \bibinfo{person}{C.~Cortes},
  \bibinfo{person}{N.~Lawrence}, {and} \bibinfo{person}{K.Q. Weinberger}}
  (Eds.), Vol.~\bibinfo{volume}{27}. \bibinfo{publisher}{Curran Associates,
  Inc.}
\newblock
\urldef\tempurl%
\url{https://proceedings.neurips.cc/paper/2014/file/375c71349b295fbe2dcdca9206f20a06-Paper.pdf}
\showURL{%
\tempurl}


\bibitem[\protect\citeauthoryear{Zeiler and Fergus}{Zeiler and Fergus}{2014}]%
        {zeiler2014visualizing}
\bibfield{author}{\bibinfo{person}{Matthew~D Zeiler} {and} \bibinfo{person}{Rob
  Fergus}.} \bibinfo{year}{2014}\natexlab{}.
\newblock \showarticletitle{Visualizing and understanding convolutional
  networks}. In \bibinfo{booktitle}{\emph{Computer Vision--ECCV 2014: 13th
  European Conference, Zurich, Switzerland, September 6-12, 2014, Proceedings,
  Part I 13}}. Springer, \bibinfo{pages}{818--833}.
\newblock


\bibitem[\protect\citeauthoryear{Zhang, Yue, Shen, Zhu, Zhen, Cao, and
  Shao}{Zhang et~al\mbox{.}}{2019}]%
        {zhang2019attentional}
\bibfield{author}{\bibinfo{person}{Anran Zhang}, \bibinfo{person}{Lei Yue},
  \bibinfo{person}{Jiayi Shen}, \bibinfo{person}{Fan Zhu},
  \bibinfo{person}{Xiantong Zhen}, \bibinfo{person}{Xianbin Cao}, {and}
  \bibinfo{person}{Ling Shao}.} \bibinfo{year}{2019}\natexlab{}.
\newblock \showarticletitle{Attentional neural fields for crowd counting}. In
  \bibinfo{booktitle}{\emph{Proceedings of the IEEE/CVF international
  conference on computer vision}}. \bibinfo{pages}{5714--5723}.
\newblock


\bibitem[\protect\citeauthoryear{Zhang, Ananthanarayanan, Bodik, Philipose,
  Bahl, and Freedman}{Zhang et~al\mbox{.}}{2017a}]%
        {videostorm-nsdi17}
\bibfield{author}{\bibinfo{person}{Haoyu Zhang}, \bibinfo{person}{Ganesh
  Ananthanarayanan}, \bibinfo{person}{Peter Bodik}, \bibinfo{person}{Matthai
  Philipose}, \bibinfo{person}{Paramvir Bahl}, {and}
  \bibinfo{person}{Michael~J. Freedman}.} \bibinfo{year}{2017}\natexlab{a}.
\newblock \showarticletitle{Live Video Analytics at Scale with Approximation
  and Delay-tolerance}. In \bibinfo{booktitle}{\emph{Proceedings of the 14th
  USENIX Conference on Networked Systems Design and Implementation}} (Boston,
  MA, USA) \emph{(\bibinfo{series}{NSDI'17})}. \bibinfo{publisher}{USENIX
  Association}, \bibinfo{address}{Berkeley, CA, USA},
  \bibinfo{pages}{377--392}.
\newblock
\showISBNx{978-1-931971-37-9}
\urldef\tempurl%
\url{http://dl.acm.org/citation.cfm?id=3154630.3154661}
\showURL{%
\tempurl}


\bibitem[\protect\citeauthoryear{Zhang, Ananthanarayanan, Bodik, Philipose,
  Bahl, and Freedman}{Zhang et~al\mbox{.}}{2017b}]%
        {VideoStormNSDI2017}
\bibfield{author}{\bibinfo{person}{Haoyu Zhang}, \bibinfo{person}{Ganesh
  Ananthanarayanan}, \bibinfo{person}{Peter Bodik}, \bibinfo{person}{Matthai
  Philipose}, \bibinfo{person}{Paramvir Bahl}, {and}
  \bibinfo{person}{Michael~J. Freedman}.} \bibinfo{year}{2017}\natexlab{b}.
\newblock \showarticletitle{Live Video Analytics at Scale with Approximation
  and Delay-Tolerance}. In \bibinfo{booktitle}{\emph{14th {USENIX} Symposium on
  Networked Systems Design and Implementation ({NSDI} 17)}}.
  \bibinfo{publisher}{{USENIX} Association}, \bibinfo{address}{Boston, MA},
  \bibinfo{pages}{377--392}.
\newblock
\showISBNx{978-1-931971-37-9}
\urldef\tempurl%
\url{https://www.usenix.org/conference/nsdi17/technical-sessions/presentation/zhang}
\showURL{%
\tempurl}


\bibitem[\protect\citeauthoryear{Zhang, Chowdhery, Bahl, Jamieson, and
  Banerjee}{Zhang et~al\mbox{.}}{2015}]%
        {vigil}
\bibfield{author}{\bibinfo{person}{Tan Zhang}, \bibinfo{person}{Aakanksha
  Chowdhery}, \bibinfo{person}{Paramvir Bahl}, \bibinfo{person}{Kyle Jamieson},
  {and} \bibinfo{person}{Suman Banerjee}.} \bibinfo{year}{2015}\natexlab{}.
\newblock \showarticletitle{The Design and Implementation of a Wireless Video
  Surveillance System}. \bibinfo{pages}{426--438}.
\newblock
\urldef\tempurl%
\url{https://doi.org/10.1145/2789168.2790123}
\showDOI{\tempurl}


\bibitem[\protect\citeauthoryear{Zhao, Liu, Song, Li, and Yu}{Zhao
  et~al\mbox{.}}{2019}]%
        {zhao2019cascaded}
\bibfield{author}{\bibinfo{person}{Kun Zhao}, \bibinfo{person}{Bin Liu},
  \bibinfo{person}{Luchuan Song}, \bibinfo{person}{Weihai Li}, {and}
  \bibinfo{person}{Nenghai Yu}.} \bibinfo{year}{2019}\natexlab{}.
\newblock \showarticletitle{Cascaded residual density network for crowd
  counting}. In \bibinfo{booktitle}{\emph{2019 IEEE International Conference on
  Image Processing (ICIP)}}. IEEE, \bibinfo{pages}{2199--2203}.
\newblock


\bibitem[\protect\citeauthoryear{Zhu, Han, Mao, and Dally}{Zhu
  et~al\mbox{.}}{2016}]%
        {zhu2016trained}
\bibfield{author}{\bibinfo{person}{Chenzhuo Zhu}, \bibinfo{person}{Song Han},
  \bibinfo{person}{Huizi Mao}, {and} \bibinfo{person}{William~J Dally}.}
  \bibinfo{year}{2016}\natexlab{}.
\newblock \showarticletitle{Trained ternary quantization}.
\newblock \bibinfo{journal}{\emph{arXiv preprint arXiv:1612.01064}}
  (\bibinfo{year}{2016}).
\newblock


\bibitem[\protect\citeauthoryear{Zhu, Samajdar, Mattina, and Whatmough}{Zhu
  et~al\mbox{.}}{2018}]%
        {euphrates}
\bibfield{author}{\bibinfo{person}{Yuhao Zhu}, \bibinfo{person}{Anand
  Samajdar}, \bibinfo{person}{Matthew Mattina}, {and} \bibinfo{person}{Paul
  Whatmough}.} \bibinfo{year}{2018}\natexlab{}.
\newblock \showarticletitle{Euphrates: Algorithm-SoC Co-Design for Low-Power
  Mobile Continuous Vision}. In \bibinfo{booktitle}{\emph{Proceedings of the
  45th Annual International Symposium on Computer Architecture}} (Los Angeles,
  California) \emph{(\bibinfo{series}{ISCA '18})}. \bibinfo{publisher}{IEEE
  Press}, \bibinfo{pages}{547–560}.
\newblock
\showISBNx{9781538659847}
\urldef\tempurl%
\url{https://doi.org/10.1109/ISCA.2018.00052}
\showDOI{\tempurl}


\end{thebibliography}

\clearpage
\sloppypar
\appendix

\newpage
\twocolumn

\section{Appendix}
\label{s:appendix}

\subsection{Workloads}
 
\begin{table}[H]
\footnotesize
\centering
\begin{tabular}{|l|l|l|}
\hline
\textbf{Model} & \textbf{Object} & \textbf{Type}  \\ \hline
SSD      & people          & aggregate count       \\ \hline
Faster RCNN            & cars            & binary classification \\ \hline
SSD           & people            & count      \\ \hline
YOLOv4          & people            & detection      \\ \hline
Faster RCNN          & people            & detection      \\ \hline

\end{tabular}
\caption{Workload 1 (W1)}
\label{tab:light-workload}
\end{table}

\begin{table}[H]
\footnotesize
\centering
\begin{tabular}{|l|l|l|}
\hline
\textbf{Model} & \textbf{Object} & \textbf{Type}  \\ \hline
 YOLOv4      & people          & aggregate count       \\ \hline
Tiny YOLOv4      & people          & aggregate count       \\ \hline
Tiny YOLOv4      & people          & detection       \\ \hline
 YOLOv4      & people          &  binary classification       \\ \hline
Tiny YOLOv4      & people          & aggregate count       \\ \hline
Faster RCNN      & people          &  count       \\ \hline
Faster RCNN      & people          &  detection       \\ \hline
Faster RCNN      & car          &  count       \\ \hline
YOLOv4      & people          & aggregate count       \\ \hline
YOLOv4      & people          & detection \\ \hline
YOLOv4      & people          & count \\ \hline
 Tiny YOLOv4      & people          & aggregate count       \\ \hline
 YOLOv4      & car          &  count       \\ \hline
  YOLOv4      & car          &  detection       \\ \hline
Tiny   YOLOv4      & car          &  count       \\ \hline
 SSD     & person          &  binary classification       \\ \hline
Faster RCNN     & car          &  count    \\ \hline
SSD    & car          &  count    \\ \hline
    
\end{tabular}
\caption{Workload 2 (W2)}
\label{tab:light-workload}
\end{table}

\begin{table}[H]
\footnotesize
\centering
\begin{tabular}{|l|l|l|}
\hline
\textbf{Model} & \textbf{Object} & \textbf{Type}  \\ \hline
SSD      & car          & binary classification       \\ \hline
Faster RCNN    & people          & aggregate count        \\ \hline
Faster RCNN    & people          &  count        \\ \hline
Tiny YOLOv4    & people          &  binary classification      \\ \hline
Tiny YOLOv4   & people         &  binary classification     \\ \hline
Tiny YOLOv4    & people          & aggregate count        \\ \hline
YOLOv4    & people          &  count        \\ \hline
Faster RCNN   & people         & aggregate count        \\ \hline
SSD   & people          & binary classification        \\ \hline
Faster RCNN   & car          & count       \\ \hline
SSD   & car          & count       \\ \hline

\end{tabular}
\caption{Workload 3 (W3)}
\label{tab:light-workload}
\end{table}

\begin{table}[H]
\centering
\footnotesize
\begin{tabular}{|l|l|l|}
\hline
\textbf{Model} & \textbf{Object} & \textbf{Type}  \\ \hline
Tiny YOLOv4     & car          & count      \\ \hline
Faster RCNN     & car          & detection      \\ \hline
Faster RCNN     & people          & aggregate count      \\ \hline

\end{tabular}
\caption{Workload 4 (W4)}
\label{tab:light-workload}
\end{table}

\begin{table}[H]
\centering
\footnotesize
\begin{tabular}{|l|l|l|}
\hline
\textbf{Model} & \textbf{Object} & \textbf{Type}  \\ \hline
Tiny YOLOv4     & car          & count      \\ \hline
SSD    & car          & count      \\ \hline
Faster RCNN & people        &  aggregate count      \\ \hline
\end{tabular}
\caption{Workload 5 (W5)}
\label{tab:light-workload}
\end{table}

\begin{table}[H]
\centering
\footnotesize
\begin{tabular}{|l|l|l|}
\hline
\textbf{Model} & \textbf{Object} & \textbf{Type}  \\ \hline
Tiny YOLOv4     & people          & aggregate count      \\ \hline
Tiny YOLOv4     & people          & binary classification     \\ \hline
SSD    &  car      & count     \\ \hline
YOLOv4    &  people     & aggregate count     \\ \hline
Tiny YOLOv4    &  people     &  count     \\ \hline
Faster RCNN    &  car     &  binary classification     \\ \hline
SSD    &  people     &  detection     \\ \hline
Faster RCNN   &  car     &  detection     \\ \hline
Faster RCNN   &  people    &  aggregate count     \\ \hline
YOLOv4   &  car    &  count     \\ \hline
Tiny YOLOv4   &  people    &  aggregate count     \\ \hline
Faster RCNN  &  people    &   detection     \\ \hline
SSD  &  people    &   aggregate count     \\ \hline
YOLOv4  &  car    &   detection     \\ \hline
\end{tabular}
\caption{Workload 6 (W6)}
\label{tab:light-workload}
\end{table}

\begin{table}[H]
\centering
\footnotesize
\begin{tabular}{|l|l|l|}
\hline
\textbf{Model} & \textbf{Object} & \textbf{Type}  \\ \hline
YOLOv4     & people          & binary classification      \\ \hline
SSD    & people          & detection     \\ \hline
Tiny YOLOv4     & car          & binary classification      \\ \hline
Tiny YOLOv4     & people          & detection     \\ \hline
SSD    & people          & binary classification     \\ \hline
SSD    & people          & aggregate count    \\ \hline
Tiny YOLOv4  & people          & detection  \\ \hline
SSD  & car          & count  \\ \hline
SSD  & people          & count  \\ \hline
Faster RCNN  & people          & count  \\ \hline
YOLOv4  & people          & count  \\ \hline
Faster RCNN  & people          & binary classification  \\ \hline
Tiny YOLOv4  & people          & aggregate count  \\ \hline
Faster RCNN  & people          & aggregate count  \\ \hline
Faster RCNN  & car          & count  \\ \hline
YOLOv4  & car          & binary classification  \\ \hline

\end{tabular}
\caption{Workload 7 (W7)}
\label{tab:light-workload}
\end{table} 

\begin{table}[H]
\centering
\footnotesize
\begin{tabular}{|l|l|l|}
\hline
\textbf{Model} & \textbf{Object} & \textbf{Type}  \\ \hline
Faster RCNN  & car          &  count   \\ \hline
Tiny YOLOv4   & people          &  binary classification   \\ \hline
YOLOv4   & people          &  aggregate count  \\ \hline
YOLOv4   & car          &   count  \\ \hline
Tiny YOLOv4   & people          &  aggregate count   \\ \hline
Faster RCNN & people          &  aggregate count   \\ \hline
YOLOv4 & people          &  aggregate count   \\ \hline
Faster RCNN & car          &   count   \\ \hline
SSD & car          &   count   \\ \hline
Faster RCNN & car          &   count   \\ \hline
SSD & car          &   binary classification   \\ \hline
YOLOv4 & car          &   binary classification   \\ \hline
SSD & car          &   binary classification   \\ \hline
SSD & people          &   count  \\ \hline
YOLOv4 & people          &   count  \\ \hline
YOLOv4 & car          &   binary classification   \\ \hline
Faster RCNN & person          &   aggregate count   \\ \hline
SSD & car          &   detection   \\ \hline

\end{tabular}
\caption{Workload 8 (W8)}
\label{tab:light-workload}
\end{table}

\begin{table}[H]
\centering
\footnotesize
\begin{tabular}{|l|l|l|}
\hline
\textbf{Model} & \textbf{Object} & \textbf{Type}  \\ \hline
Tiny YOLOv4   & people          & aggregate count   \\ \hline
Faster RCNN   & people          &  count   \\ \hline
Faster RCNN   & people          &  count   \\ \hline
Tiny YOLOv4   & car          &  detection   \\ \hline
Tiny YOLOv4   & people          &  binary clasification   \\ \hline
 YOLOv4   & people          &  detection   \\ \hline
 Faster RCNN   & people          &  count   \\ \hline
  YOLOv4   & people          &  aggregate count   \\ \hline
  SSD   & people          &  aggregate count   \\ \hline

\end{tabular}
\caption{Workload 9 (W9)}
\label{tab:light-workload}
\end{table}

\begin{table}[H]
\centering
\footnotesize
\begin{tabular}{|l|l|l|}
\hline
\textbf{Model} & \textbf{Object} & \textbf{Type}  \\ \hline
Faster RCNN   & people          & aggregate count   \\ \hline
Faster RCNN   & car          &  count   \\ \hline
Faster RCNN   & people          &  count   \\ \hline

\end{tabular}
\caption{Workload 10 (W10)}
\label{tab:light-workload}
\end{table} 

\end{document}